\documentclass[useAMS,usenatbib]{mnras}
\usepackage{natbib}
\usepackage{amsmath}
\usepackage{graphicx}
\usepackage{multirow}
\usepackage{url}
\numberwithin{equation}{section}
\title[Chemical evolution of GMCs]{Chemical evolution of giant molecular clouds in simulations of galaxies}
\author[A. J. Richings and J. Schaye]{Alexander J. Richings$^{1, 2}$\thanks{Email: a.j.richings@northwestern.edu} and Joop Schaye$^{2}$\\
$^{1}$Center for Interdisciplinary Exploration and Research in Astrophysics (CIERA) and Department of Physics and Astronomy,\\ 
Northwestern University, 2145 Sheridan Road, Evanston, IL 60208, USA\\
$^{2}$Leiden Observatory, Leiden University, PO Box 9513, 2300 RA Leiden, the Netherlands}

\begin{document}

\date{Accepted 2016 May 10. Received 2016 May 09; in original form 2016 April 01}

\pagerange{\pageref{firstpage}--\pageref{lastpage}} \pubyear{2016}

\maketitle

\label{firstpage}

\begin{abstract} 

We present an analysis of Giant Molecular Clouds (GMCs) within hydrodynamic simulations of isolated, low-mass ($M_{\ast} \sim 10^{9} \, \rm{M}_{\odot}$) disc galaxies. We study the evolution of molecular abundances and the implications for CO emission and the $X_{\rm{CO}}$ conversion factor in individual clouds. We define clouds either as regions above a density threshold $n_{\rm{H,} \, \rm{min}} = 10 \, \rm{cm}^{-3}$, or using an observationally motivated CO intensity threshold of $0.25 \, \rm{K} \, \rm{km} \, \rm{s}^{-1}$. Our simulations include a non-equilibrium chemical model with $157$ species, including 20 molecules. We also investigate the effects of resolution and pressure floors (i.e. Jeans limiters). We find cloud lifetimes up to $\approx 40 \, \rm{Myr}$, with a median of $13 \, \rm{Myr}$, in agreement with observations. At one tenth solar metallicity, young clouds ($\la 10 - 15 \, \rm{Myr}$) are underabundant in H$_{2}$ and CO compared to chemical equilibrium, by factors of $\approx 3$ and $1 - 2$ orders of magnitude, respectively. At solar metallicity, GMCs reach chemical equilibrium faster (within $\approx 1 \, \rm{Myr}$). We also compute CO emission from individual clouds. The mean CO intensity, $I_{\rm{CO}}$, is strongly suppressed at low dust extinction, $A_{\rm{v}}$, and possibly saturates towards high $A_{\rm{v}}$, in agreement with observations. The $I_{\rm{CO}} - A_{\rm{v}}$ relation shifts towards higher $A_{\rm{v}}$ for higher metallicities and, to a lesser extent, for stronger UV radiation. At one tenth solar metallicity, CO emission is weaker in young clouds ($\la 10 - 15 \, \rm{Myr}$), consistent with the underabundance of CO. Consequently, $X_{\rm{CO}}$ decreases by an order of magnitude from $0$ to $15 \, \rm{Myr}$, albeit with a large scatter. 

\end{abstract}


\section{Introduction}

Molecular hydrogen is the main constituent of Giant Molecular Clouds (GMCs), making up most of their mass. However, cold H$_{2}$ is difficult to observe in emission, as the lowest rotational transition of the H$_{2}$ molecule has an excitation energy of $E / k_{\rm{B}} = 510 \, \rm{K}$ \citep{dabrowski84}. It is therefore difficult to excite H$_{2}$ at the cold temperatures typical of GMCs ($\sim 10 \, \rm{K}$). 

CO is typically the next most abundant molecule in GMCs. It is also much easier to excite the rotational and vibrational levels of the CO molecule at low temperatures. For example, the lowest rotational transition of CO ($J = 1 - 0$) has an excitation energy of $E / k_{\rm{B}} = 5.53 \, \rm{K}$. CO emission is therefore commonly used as a tracer of molecular gas in GMCs \citep[e.g.][]{solomon87, dame01, heyer01}. The velocity-integrated CO intensity, $I_{\rm{CO}}$, is then converted to an H$_{2}$ column density, $N_{\rm{H_{2}}}$, using a conversion factor $X_{\rm{CO}}$, defined as: 

\begin{equation}\label{xco_equation}
X_{\rm{CO}} = \frac{N_{\rm{H_{2}}}}{I_{\rm{CO}}} \, \rm{cm}^{-2} \, ( \rm{K} \, \rm{km} \, \rm{s}^{-1})^{-1}. 
\end{equation} 
To accurately determine the molecular content of a GMC in this way, we therefore require a detailed understanding of the $X_{\rm{CO}}$ factor, including how it depends on the physical conditions in the GMC, such as its metallicity and the radiation field. 

There have been many studies, both observational and theoretical, to determine the $X_{\rm{CO}}$ factor (see \citealt{bolatto13} for a recent review). Observational studies use various methods to determine the total molecular content, which can then be compared to the CO emission to determine the $X_{\rm{CO}}$ factor. For example, virial techniques assume that the GMC is in virial equilibrium, which allows one to measure the total mass of a GMC from its size and velocity dispersion, which is assumed to be the molecular mass \citep[e.g.][]{scoville87, solomon87}. Other studies estimate the dust content of GMCs, either by mapping the extinction towards background stars \citep[e.g.][]{frerking82, lombardi06, pineda08}, or by measuring dust emission in the far-infrared \citep[e.g.][]{dame01, planck11}. This can then be converted into a total gas column density, assuming a dust-to-gas ratio. Diffuse gamma-ray emission arising from interactions between cosmic rays and nucleons can also be used to estimate the total gas column density \citep[e.g.][]{strong96, abdo10, ackermann12}. 

Some theoretical studies of the $X_{\rm{CO}}$ factor use models of photodissociation regions (PDRs), where a cloud of gas is illuminated from one side by an external UV radiation field. \citet{tielens85} use PDR models to determine the chemical and temperature structure of such clouds for various gas densities and radiation fields. \citet{vandishoeck88} and \citet{visser09} focus on the chemistry and photodissociation of CO in PDR models, and they use these models to determine how the CO column density varies with dust extinction. \citet{sternberg14} recently presented a detailed study of the H\textsc{i}-to-H$_{2}$ transition in clouds, using both analytic theory and numerical PDR models. These PDR models assume that the abundances of molecules and atoms are in chemical equilibrium, or a `steady state', and that the clouds have a constant density profile. 

These PDR models can then be used to study how the $X_{\rm{CO}}$ factor depends on the physical conditions. For example, \citet{bell06} use PDR models to explore how $X_{\rm{CO}}$ varies in different environments. They find that, at low dust extinction, $A_{\rm{v}}$, $X_{\rm{CO}}$ decreases with increasing $A_{\rm{v}}$, until it reaches a minimum and subsequently increases with $A_{\rm{v}}$ once the CO line becomes optically thick. They show that the $X_{\rm{CO}} - A_{\rm{v}}$ profile depends on cloud properties, including gas density, radiation field strength, metallicity and turbulent velocity dispersion. 

Other theoretical studies of the $X_{\rm{CO}}$ factor use hydrodynamic simulations of a turbulent interstellar medium (ISM) to study the environmental dependence of the $X_{\rm{CO}}$ factor, which account for more realistic cloud geometries \citep[e.g.][]{glover11, shetty11a, shetty11b, clark15}. \citet{narayanan11, narayanan12} combine hydrodynamic simulations of isolated and merging galaxies, using a subgrid model for cold gas below $10^{4} \, \rm{K}$, with radiative transfer calculations of dust and molecular line emission to explore how galaxy mergers and the galactic environment affect the $X_{\rm{CO}}$ factor. \citet{feldmann12} combine the results of sub-parsec resolution simulations from \citet{glover11} with gas distributions from the cosmological simulations of \citet{gnedin11} to model the $X_{\rm{CO}}$ factor, finding a metallicity dependence of $X_{\rm{CO}}$ (averaged over kpc scales) of $X_{\rm{CO}} \propto Z^{-\gamma}$, where $\gamma \approx 0.5 - 0.8$. 

Theoretical models of $X_{\rm{CO}}$ need to determine the abundances of CO and H$_{2}$ under various conditions. The simplest approach is to assume that these abundances are in chemical equilibrium \citep[e.g.][]{narayanan11, narayanan12}. However, this assumption may not be valid if the formation time-scale of molecules is comparable to the lifetimes of GMCs, particularly in young clouds. Observational estimates have suggested a wide range of GMC lifetimes, from a few Myr \citep[e.g.][]{elmegreen00}, to $\approx 20 - 40 \, \rm{Myr}$ \citep[e.g.][]{bash77, kawamura09, murray11, miura12}, to hundreds of Myr \citep[e.g.][]{scoville79}. 

\citet{bell06} include time-dependent chemistry of H$_{2}$ and CO in their PDR models, with metallicities $0.01 \leq Z / \rm{Z}_{\odot} \leq 1.0$\footnote{Throughout this paper we use a solar metallicity of $\rm{Z}_{\odot} = 0.0129$ \citep{wiersma09}, although other studies that we quote in this paragraph use different definitions of $\rm{Z}_{\odot}$.}, and they consider various cloud ages. They find significant evolution in the $X_{\rm{CO}}$ factor at times $\la 1 \, \rm{Myr}$, with less evolution for cloud ages $1 - 10 \, \rm{Myr}$, and no notable evolution beyond $10 \, \rm{Myr}$, even though it takes up to $100 \, \rm{Myr}$ for the chemical abundances to reach steady-state in their models. \citet{glover11} and \citet{shetty11a, shetty11b} also include time-dependent chemistry in their simulations of a turbulent ISM, with metallicities $0.03 \leq Z / \rm{Z}_{\odot} \leq 1.0$ and $0.1 \leq Z / \rm{Z}_{\odot} \leq 1.0$, respectively. However, since they include only a region of the ISM in their simulations, and not an entire galaxy, they may be missing some aspects of the evolution of GMCs in a galactic environment. Indeed, \citet{dobbs13} explore GMC evolution in simulations of isolated disc galaxies, with solar metallicity, and they find complex evolutionary histories. GMCs in their simulations often form by assembling from smaller clouds and ambient ISM material, or by breaking off from larger clouds, while they are dispersed by stellar feedback and shear, or are accreted onto larger clouds. It would therefore be useful to explore the chemical evolution of GMCs within a realistic galactic environment. 

In this paper we investigate how the molecular abundances of GMCs evolve, and under what conditions these abundances are out of chemical equilibrium. We consider the effects of cloud age, metallicity and the radiation field. We can then determine how the conditions affect the $X_{\rm{CO}}$ factor. We study clouds of dense gas ($n_{\rm{H}} > 10 \, \rm{cm}^{-3}$) in the high-resolution Smoothed Particle Hydrodynamics (SPH) simulations of isolated disc galaxies presented in \citet{richings16}, hereafter Paper I. These simulations include a treatment for the non-equilibrium chemistry of $157$ species, including 20 different molecules \citep{richings14a, richings14b}. We also run radiative transfer calculations on these simulations in post-processing to determine the $^{12}$CO $J = 1 - 0$ line emission\footnote{For the remainder of this paper, we will use `CO' to refer to $^{12}$CO, unless stated otherwise.} from individual GMCs, and hence compute their $X_{\rm{CO}}$ factors. 

The remainder of this paper is organised as follows. In section~\ref{simulations} we summarise the simulations and initial conditions from paper I. In section~\ref{methods_section} we describe the methods that we use to analyse GMCs in these simulations, including how we identify clouds, how we link clouds in previous and subsequent snapshots to identify their progenitors and descendants, and how we create maps of CO emission from individual clouds in post-processing. In section~\ref{scaling_relations_section} we investigate the scaling relations of these clouds and compare them to observations. In section~\ref{chemical_evolution_section} we look at the H$_{2}$ and CO abundances of our simulated GMCs as a function of cloud age to explore their chemical evolution. In section~\ref{co_emission_section} we use the CO $J = 1 - 0$ line emission from simulated clouds to investigate the $X_{\rm{CO}}$ factor, and we summarise our main results in section~\ref{conclusions}. Finally, in Appendix~\ref{res_appendix} we explore how our results are affected by changing the resolution of our simulations, and in Appendix~\ref{pfloor_appendix} we explore how our results are affected by the pressure floor that we impose in our simulations to ensure that the Jeans mass is always well-resolved. 

\section{Simulations}\label{simulations}

We study GMCs in the suite of hydrodynamic simulations of isolated disc galaxies that were first presented in paper I. The details of how these simulations were run, along with properties of the galaxies such as their star formation histories and their outflow rates and velocities, can be found in paper I. Here we summarise the main features of these simulations. 

The simulations were run using a modified version of the tree/SPH code \textsc{gadget3}, last described in \citet{springel05}. The hydrodynamics solver has been replaced with the suite of hydrodynamical methods collectively known as \textsc{anarchy}, which incorporates many of the latest improvements on `classical' SPH methods, including the pressure-entropy formulation of SPH, as derived by \citet{hopkins13}; a switch for artificial conduction, similar to the one used by \citet{price08}; a switch for artificial viscosity, from \citet{cullen10}; the time-step limiters from \citet{durier12}; and the $C^{2}$ \citet{wendland95} kernel, for which we use 100 neighbours. \textsc{anarchy} will be described in more detail in Dalla Vecchia (in preparation); see also Appendix A of \citet{schaye15} for a full description of our version of \textsc{anarchy}. 

\subsection{Chemistry and subgrid models}\label{models_section} 

We follow the chemical evolution of the abundances of ions and molecules in the gas using the chemical model of \citet{richings14a, richings14b}. This model includes all ionisation states of the 11 elements that contribute most to the cooling rate\footnote{H, He, C, N, O, Ne, Mg, Si, S, Ca, Fe}, along with 20 molecular species\footnote{H$_{2}$, H$_{2}^{+}$, H$_{3}^{+}$, OH, H${_2}$O, C$_{2}$, O$_{2}$, HCO$^{+}$, CH, CH$_{2}$, CH$_{3}^{+}$, CO, CH$^{+}$, CH$_{2}^{+}$, OH$^{+}$, H$_{2}$O$^{+}$, H$_{3}$O$^{+}$, CO$^{+}$, HOC$^{+}$, O$_{2}^{+}$}, most importantly H$_{2}$ and CO. This gives us a chemical network of 157 species in total. The chemical species evolve via collisional ionisation, radiative and di-electronic recombination, charge transfer reactions, photoionisation (including Auger ionisation), cosmic ray ionisation (parameterised by an H\textsc{i} cosmic ray ionisation rate of $2.5 \times 10^{-17} \, \rm{s}^{-1}$; \citealt{williams98}), and various molecular reactions, including the formation of H$_{2}$ on dust grains \citep{cazaux02} and in the gas phase. 

The photoionisation, photoheating and photoelectric dust heating rates are computed assuming a constant, uniform UV radiation field, either the local interstellar radiation field (ISRF) of \citet{black87}, or ten per cent of this ISRF. We also use a self-shielding prescription to account for the attenuation of photochemical rates by dust and gas \citep{richings14b}. This prescription includes self-shielding of H$_{2}$ and CO, and shielding of CO by H$_{2}$. We assume that the shielding occurs locally, which allows us to express the column density of each particle as the density, $\rho$, multiplied by a local shielding length, $L$. For the shielding length, we use a local Sobolev-like approximation, $L = \rho / \lvert 2 \nabla \rho \rvert$ \citep[e.g.][]{gnedin09}. 

From the chemical network we obtain a system of 158 differential equations (157 chemical rate equations and the thermal equation for the temperature evolution), which we integrate for each gas particle over each hydrodynamic time-step using the implicit differential equation solver \textsc{Cvode}, from the \textsc{sundials}\footnote{\url{https://computation.llnl.gov/casc/sundials/main.html}} suite. This enables us to follow the non-equilibrium evolution of ion and molecule abundances, and also to evolve the temperature using cooling rates computed from these abundances, without needing to assume chemical equilibrium. 

Gas particles are allowed to form stars if their hydrogen number density, $n_{\rm{H}}$, exceeds a threshold of $1 \, \rm{cm}^{-3}$ and their temperature is below $1000 \, \rm{K}$. If a particle meets these criteria, it forms stars at a rate per unit volume given by the gas density over the local free fall time, multiplied by an efficiency factor $\epsilon_{\rm{SF}}$, which we take to be $0.005$. Gas particles are then stochastically converted into star particles according to a probability that is determined from the particle's star formation rate and the hydrodynamic time-step. The value of the efficiency, $\epsilon_{\rm{SF}}$ (and the value of the heating temperature, $\Delta T$, used in the stellar feedback model; see below) were chosen to reproduce the observed Kennicutt-Schmidt relation (see fig. 3 in paper I). 

We include feedback from star formation using a thermal supernova prescription similar to that of \citet{dallavecchia12}, with some modifications. As each star particle is treated as a simple stellar population (rather than an individual star), we can calculate the number of supernovae that explode from each star particle in a given time-step, using the stellar lifetimes of \citet{portinari98} and assuming a \citet{chabrier03} initial mass function (IMF). When supernovae explode, their energy is injected into the gas thermally by stochastically selecting neighbouring gas particles to be heated by $\Delta T = 10^{7.5} \, \rm{K}$. 

\begin{table*}
\footnotesize
\centering
\begin{minipage}{168mm}
\caption{Properties of the galaxy simulations used in this paper: total mass $M_{200}$ within the radius $R_{200, \rm{crit}}$ enclosing a mean density of 200 times the critical density of the Universe at redshift zero, NFW concentration $c_{200}$ of the dark matter halo, initial stellar mass $M_{\ast, \, \rm{init}}$, initial gas mass $M_{\rm{gas, \, init}}$, disc gas mass fraction $f_{\rm{d, \, gas}}$, mass per gas or star particle $m_{\rm{baryon}}$, gravitational softening length $\epsilon_{\rm{soft}}$, gas metallicity $Z$, and UV radiation field. Parameters highlighted in bold have been changed from the ref model.}
\centering
\begin{tabular}{lccccccccc}
\hline
Model & $M_{200}$ & $c_{200}$ & $M_{\ast, \, \rm{init}}$ & $M_{\rm{gas, \, init}}$ & $f_{\rm{d, \, gas}}$ & $m_{\rm{baryon}}$ & $\epsilon_{\rm{soft}}$ & $Z$ & UV Field \\
 & ($\rm{M}_{\odot}$) & & ($\rm{M}_{\odot}$) & ($\rm{M}_{\odot}$) & & ($\rm{M}_{\odot}$) & ($\rm{pc}$) & ($\rm{Z}_{\odot}$) & \\
\hline
ref & $10^{11}$ & 8.0 & $1.4 \times 10^{9}$ & $4.8 \times 10^{8}$ & 0.3 & 750 & 3.1 & 0.1 & ISRF\footnote{ISRF of \citet{black87}} \\
hiZ & $10^{11}$ & 8.0 & $1.4 \times 10^{9}$ & $4.8 \times 10^{8}$ & 0.3 & 750 & 3.1 & \textbf{1.0} & ISRF \\
lowISRF & $10^{11}$ & 8.0 & $1.4 \times 10^{9}$ & $4.8 \times 10^{8}$ & 0.3 & 750 & 3.1 & 0.1 & \textbf{10\% ISRF} \\
\hline
\vspace{-0.15in}
\label{ic_parameters}
\end{tabular}
\end{minipage}
\end{table*}

By imposing a minimum heating temperature, we ensure that we reduce artificial radiative losses due to our finite resolution, which might otherwise make the stellar feedback unrealistically inefficient. We are unable to resolve individual supernovae. Instead, at the fiducial resolution used in our simulations ($750 \, \rm{M}_{\odot}$ per gas particle) each heating event corresponds to approximately ten supernovae exploding simultaneously. The probability of stochastically selecting a gas particle to be heated is computed such that, when averaged over time and over all particles in the simulation, the expectation value for the total injected thermal energy is equal to the total available energy from supernovae.

The difference between our stellar feedback model and that of \citet{dallavecchia12} is that we distribute the total available supernova energy from each star particle over time, according to the lifetimes of massive stars, rather than injecting it all at $30 \, \rm{Myr}$ after the birth of the star particle. 

To ensure that the Jeans mass is always resolved, we impose a density-dependent pressure floor, $P_{\rm{floor}}$, in the hydrodynamic equations, such that the Jeans mass will always be at least a factor $N_{\rm{J, \, m}}$ times the mass within the SPH kernel. This is similar to the methods used by e.g. \citet{robertson08, schaye08, hopkins11}. The pressure floor is given by equation 2.12 of paper I: 

\begin{equation}\label{floor_mass}
P_{\rm{floor, \, m}} = \left( \frac{36}{\pi^{5}} \right)^{1/3} \frac{G}{\gamma} (N_{\rm{J, \, m}} N_{\rm{ngb}}^{\rm{SPH}} m_{\rm{gas}})^{2/3} \rho^{4/3}, 
\end{equation}
where $\gamma = 5 / 3$ is the ratio of specific heats, $N_{\rm{ngb}}^{\rm{SPH}}$ is the number of SPH neighbours, $m_{\rm{gas}}$ is the mass per SPH particle, and $\rho$ is the gas density. We use a conservative fiducial value of $N_{\rm{J, \, m}} = 4$ in our simulations, but see Appendix~\ref{pfloor_appendix} for the effects of lowering this pressure floor. We impose this Jeans limiter as a pressure floor rather than a temperature floor (as used by \citealt{schaye08}) so that gas particles can continue to cool below the temperature corresponding to the pressure floor, and thus will evolve towards thermal and chemical equilibrium for the given density. 

\subsection{Initial conditions} 

We ran simulations of isolated disc galaxies using initial conditions based on the model of \citet{springeletal05}. These initial conditions were generated using a modified version of a code that was kindly provided to us by Volker Springel. Each galaxy has a total mass within $R_{200, \rm{crit}}$ (i.e. the radius enclosing $200$ times the critical density) of $M_{200} = 10^{11} \, \rm{M}_{\odot}$. The galaxies initially consist of a rotating disc of gas and stars and a central stellar bulge, embedded in a dark matter halo. The initial stellar mass is $M_{\ast} = 1.4 \times 10^{9} \, \rm{M}_{\odot}$, which is consistent with the abundance matching results of \citet{moster13} corrected for baryonic effects according to the prescription of \citet{sawala15}. Twenty per cent of the initial stellar mass is in the bulge, with the remainder in the stellar disc. We use a gas mass fraction in the disc of 30 per cent, which gives an initial gas mass of $4.8 \times 10^{8} \, \rm{M}_{\odot}$. 

The gas and stellar discs initially have an exponential surface density profile with a radial scale length of $2.0 \, \rm{kpc}$. The vertical structure of the stellar disc has an isothermal profile with a scale height of ten per cent of the radial scale length, while the gas is initially in chemical equilibrium with a constant temperature of $10^{4} \, \rm{K}$ and a vertical structure set up in hydrostatic equilibrium using an iterative procedure. At this temperature, most of the hydrogen is in H\textsc{ii} in chemical equilibrium. The stellar bulge has a \citet{hernquist90} density profile, and the dark matter halo follows a \citet{hernquist90} profile that is scaled to match a \citet{navarro96} (NFW) profile in the inner regions with a concentration $c_{200} = 8.0$, which agrees with the redshift zero mass-concentration relation of \citet{duffy08}. 

We use a fiducial resolution of $750 \, \rm{M}_{\odot}$ per gas or star particle, with $100$ SPH neighbours, and a gravitational softening length of $3.1 \, \rm{pc}$ (but see Appendix~\ref{res_appendix} for runs with a factor four higher/lower mass resolution), and we model the dark matter halo using a static potential. Each simulation initially contains $6.45 \times 10^{5}$ gas particles and $1.88 \times 10^{6}$ star particles. 

We include a constant, uniform UV radiation field, along with a local self-shielding prescription, and the gas metallicity is held fixed, with dust-to-gas mass ratios of $2.4 \times 10^{-3} \, Z / \rm{Z}_{\odot}$ and $4.0 \times 10^{-3} \, Z / \rm{Z}_{\odot}$ for graphite and silicate grain species, respectively. These dust-to-gas ratios were taken from the `ISM' grain abundances used by the photoionisation code \textsc{cloudy}\footnote{\url{http://nublado.org/}} version $13.01$ \citep{ferland13}, and we assume that they scale linearly with metallicity, $Z$. However, this assumption of a linear scaling between dust-to-gas ratio and metallicity may not be accurate, particularly at low metallicity. For example, the dust content of low-metallicity dwarf galaxies is found to be less than what one would expect from a linear scaling \citep[e.g.][]{remyruyer14}. 

In paper I, we ran six simulations with different combinations of metallicity and UV radiation field. Each simulation was repeated twice, once with the full non-equilibrium chemical model of \citet{richings14a, richings14b}, and once using cooling rates computed assuming chemical equilibrium. In this paper, we focus on three of these simulations: ref (ten per cent solar metallicity and the local ISRF of \citealt{black87}), hiZ (solar metallicity and the \citealt{black87} ISRF), and lowISRF (ten per cent solar metallicity and ten per cent of the \citealt{black87} ISRF), all evolved using the full non-equilibrium chemical model. We focus on these as they are the most relevant for conditions in molecular clouds in low-mass galaxies. Of the three remaining simulations in paper I, lowZ (with one per cent solar metallicity) did not form dense clouds, as the gas was mostly unable to cool to a cold ($\sim 100 \, \rm{K}$) phase; UVB (evolved with the redshift zero UV background of \citealt{haardt01}) used an extragalactic UV radiation field that is more relevant for the circum- and inter-galactic medium than for molecular clouds; and UVBthin neglected self-shielding of UV radiation, which is necessary for the formation of molecules. The properties of our simulations are summarised in Table~\ref{ic_parameters}. 

Our simulations all use a constant H\textsc{i} cosmic ray ionisation rate of $2.5 \times 10^{-17} \, \rm{s}^{-1}$ \citep{williams98}. By keeping the cosmic ray rate fixed as we vary the strength of the UV radiation field, we can isolate the effects of the UV radiation alone on the molecular clouds. However, in reality, it is likely that both the UV radiation and the cosmic rays are produced in star forming regions. Therefore, we would expect both to vary in proportion to the local star formation rate (see e.g. \citealt{clark15} for examples of varying both the UV radiation field and the cosmic ray ionisation rate together in simulations of molecular clouds). 

\section{Analysis methods}\label{methods_section} 

In this section we describe the methods that we use to analyse gas clouds in our simulations, including the algorithm that we use to identify clouds (\S\ref{clump_finder}), how we link clouds to their progenitors and descendants to define their mass evolution (\S\ref{evolution_section}), and how we create maps of CO emission from individual clouds (\S\ref{CO_map_methods}). 

\subsection{Clump finding algorithm}\label{clump_finder} 

Observationally, molecular clouds are typically identified as regions detected in emission from a molecular tracer (often CO) above an intensity threshold \citep[e.g.][]{larson81, solomon87}. This is approximately equivalent to selecting regions above a molecular gas surface density threshold. However, as we are interested in the atomic to molecular transition, we do not want to select only clouds that are already molecular. We therefore need a criterion that is based on the total gas content, and not just on the molecular content. 

Furthermore, \citet{dobbs15} found that using a grid-based approach to identify clouds above a surface density threshold in simulations can create problems for studying the cloud evolution. They found that clouds that are identified with such a method appear to evolve on shorter time-scales than is seen in the three-dimensional particle distribution. These errors arise due to the projection onto a two-dimensional grid, as the gas moves relative to the grid. 

\begin{figure}
\centering
\mbox{
	\includegraphics[width=80mm]{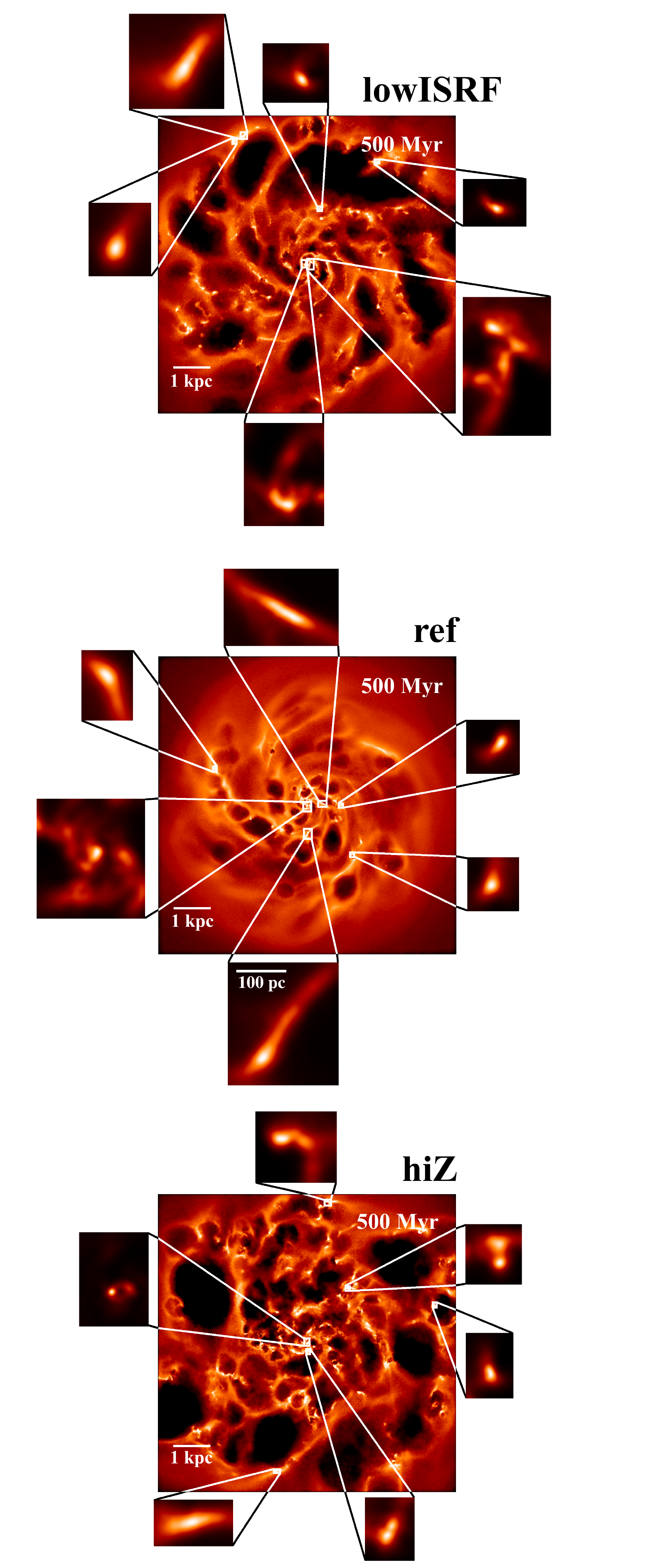}}
\caption{Maps of gas surface density after $500 \, \rm{Myr}$ from simulations lowISRF \textit{(0.1 Z$_{\odot}$, ten per cent of the \citet{black87} ISRF; top)}, ref \textit{(0.1 Z$_{\odot}$, \citet{black87} ISRF; centre)} and hiZ \textit{(Z$_{\odot}$, \citet{black87} ISRF; bottom)}. Each map is $8 \, \rm{kpc}$ across and views the disc of the galaxy face-on. We also zoom in (by a factor 13) on the most massive clouds in each simulation. We see a wide range of morphologies, including nearly spherical clouds, clouds that have been sheared into long filaments, and clouds with multiple density peaks that are indicative of cloud mergers.} 
\label{gasMapsFig}
\end{figure}

We therefore base our clump finding algorithm on the particle-based approach used by \citet{dobbs15}. This is a Friends-of-Friends (FoF) algorithm that acts on dense gas particles. We first select gas particles with a hydrogen number density, $n_{\rm{H}}$, above a threshold $n_{\rm{H,} \, \rm{min}}$. We then link together nearby dense particles by taking each particle in turn and identifying particles that lie within a linking length, $l$. 

There are two parameters in this method, $n_{\rm{H,} \, \rm{min}}$ and $l$. However, as noted by \citet{dobbs15}, they are degenerate, as denser particles will be closer together. We use a density threshold of $n_{\rm{H,} \, \rm{min}} = 10 \, \rm{cm}^{-3}$, which is comparable to the density at which we expect the transition from atomic to molecular hydrogen to occur \citep[e.g.][]{schaye01, gnedin09}. This ensures that we focus on clouds that are likely to become molecular. We then use a linking length $l = 10 \, \rm{pc}$, which corresponds approximately to the mean spacing between gas particles at the density threshold, $n_{\rm{H,} \, \rm{min}}$, for our resolution of $750 \, \rm{M}_{\odot}$ per particle. 

Fig.~\ref{gasMapsFig} shows maps of the gas surface density in each of our three simulations after $500 \, \rm{Myr}$. Each map is $8 \, \rm{kpc}$ across and views the disc face-on. We also zoom in (by a factor 13) on the six most massive clouds in each simulation. We see that these clouds show a wide range of morphologies. Some are approximately spherical, while others have been stretched into long, thin filaments by shear in the rotating disc. We also see some clouds with two or more density peaks, which suggests that they consist of multiple clumps that are in the process of merging. 

\begin{figure}
\centering
\mbox{
	\includegraphics[width=84mm]{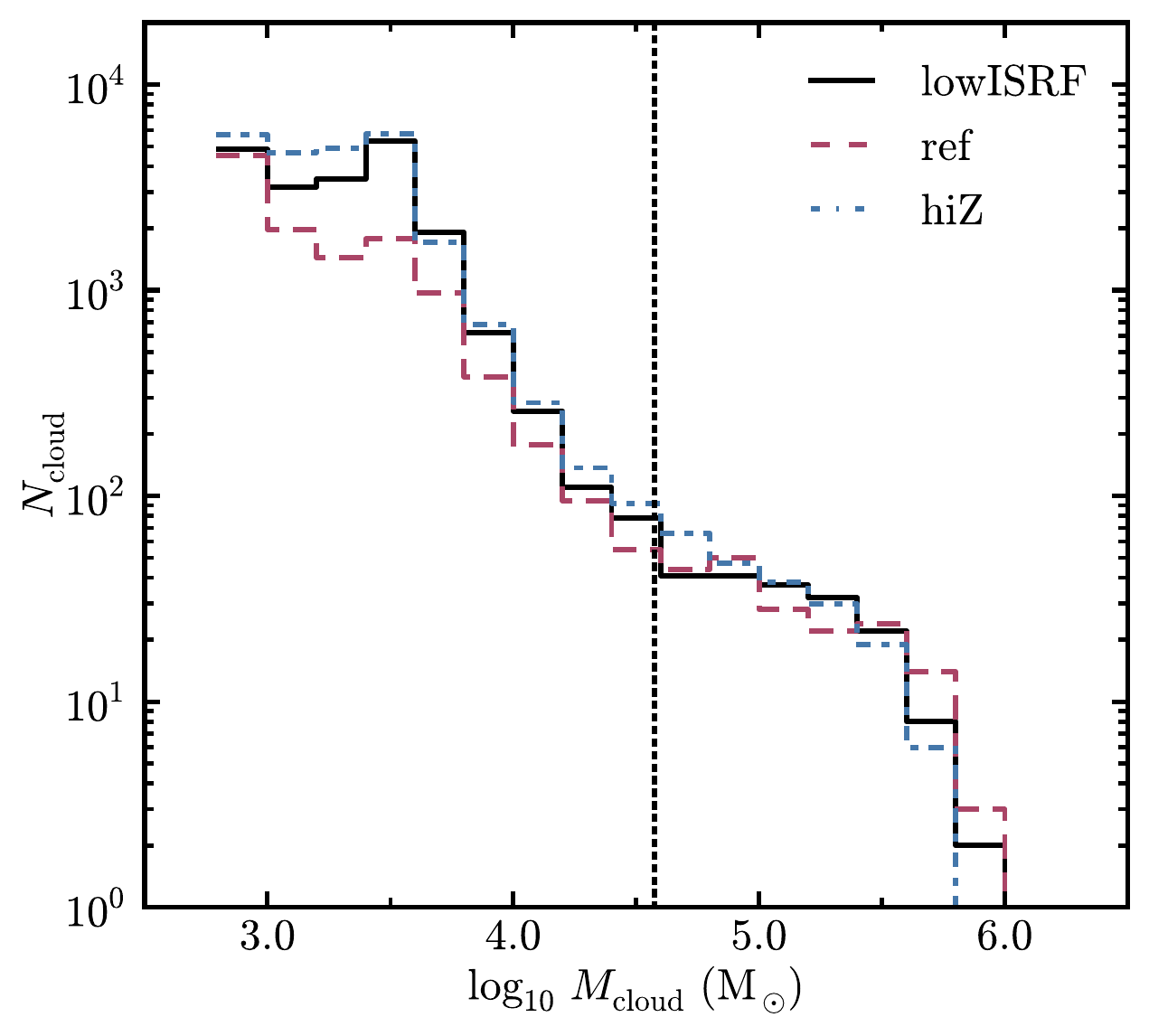}}
\caption{Cloud mass functions from simulations lowISRF \textit{(black solid curve)}, ref \textit{(red dashed curve)} and hiZ \textit{(blue dot-dashed curve)}, taken from snapshots at $100 \, \rm{Myr}$ intervals from $100 \, \rm{Myr}$ to $900 \, \rm{Myr}$. The vertical dotted line shows a mass of $3.75 \times 10^{4} \, \rm{M}_{\odot}$, which corresponds to the mass of a cloud containing 50 particles. We consider only clouds above this mass for the remainder of this study. All three simulations show similar cloud mass functions.} 
\label{massFunctionFig}
\end{figure}

Fig.~\ref{massFunctionFig} shows the cloud mass functions for each simulation. Here we have identified clouds in snapshots at $100 \, \rm{Myr}$ intervals, from $100 \, \rm{Myr}$ to $900 \, \rm{Myr}$, and combined the different snapshots into a single mass function for each simulation. We will show in the next section that the clouds have lifetimes $< 100 \, \rm{Myr}$ if we define lifetimes using the particles originally in the cloud when it is identified, so we do not double-count cloud mass by combining snapshots in this way. However, if we follow the total mass of cloud progenitors/descendants, we find that individual cloud structures can survive for $> 100 \, \rm{Myr}$, although new gas has cycled through them. Such long-lived cloud structures will appear multiple times in Fig.~\ref{massFunctionFig}. 

All three simulations show similar cloud mass functions. For the remainder of this study, we shall focus on clouds that contain at least $50$ gas particles to avoid poorly resolved clouds in our analysis (but see Appendix~\ref{pfloor_appendix} for the effects of the pressure floor on low-mass clouds). This corresponds to a mass of $3.75 \times 10^{4} \, \rm{M}_{\odot}$, shown by the vertical dotted line in Fig.~\ref{massFunctionFig}. 

We also need to define the radius of each cloud, which will be important for comparing to the observed molecular cloud scaling relations (see \S\ref{scaling_relations_section}). We determine the radius by finding the 3-dimensional ellipsoid that approximately encloses the particles in the cloud. First, we compute the moment of inertia tensor, $I$: 

\begin{equation}
I_{ij} = \sum_{k = 1}^{N} m_{k} (\lvert \mathbf{r} \rvert^{2} \delta_{ij} - r_{k, \, i} r_{k, \, j}), 
\end{equation}
where $m_{k}$ is the mass of the $k^{\rm{th}}$ particle, $\mathbf{r}_{k}$ is the position vector of the $k^{\rm{th}}$ particle in the cloud's centre of mass frame, the summation is over the $N$ particles in the cloud, $i$ and $j$ index the Cartesian directions ($i, j = 1, 2, 3$ in 3d), and $\delta_{ij}$ is the Kronecker delta function. The eigenvectors of $I$ give the directions of the principle axes of the cloud. We then determine the maximum extent of the particle distribution along each principle axis to obtain the semi-major, intermediate and minor axes, $a$, $b$ and $c$ respectively, of the ellipsoid that approximately encloses the particles in the cloud. Finally, we define the cloud radius, $R_{\rm{mean}}$, to be the geometric mean of these three axes, i.e.:

\begin{equation}\label{r_mean_eqn}
R_{\rm{mean}} = (abc)^{1/3}. 
\end{equation}

The above cloud definition is based on a density threshold. However, observations define molecular clouds based on a CO intensity threshold. We therefore also consider an alternative cloud definition, based on the CO emission, which we discuss in section~\ref{CO_map_methods}. 

\subsection{Cloud mass evolution}\label{evolution_section} 

To follow the mass evolution of individual clouds, and hence determine their ages and lifetimes, we first ran the clump finding algorithm described above on all snapshots, taken at intervals of $1 \, \rm{Myr}$. We then took each massive cloud (containing at least $50$ particles) in a given snapshot and traced back its main progenitor in preceding snapshots, and its main descendant in subsequent snapshots. 

There are a couple of different ways in which we can link a given cloud to its progenitors and descendants. In the first method that we use, we first take all particles in a given cloud in snapshot $i$. We then look for these particles in the preceding snapshot, $i - 1$, and we identify the cloud that contains the most of these particles. This cloud is selected as the main progenitor. We then take all of the particles in the main progenitor, and we look for these particles in snapshot $i - 2$. We repeat this process to find the main progenitor in each preceding snapshot until we can no longer identify a progenitor. Finally, we repeat the above procedure in the snapshots following $i$ to identify the main descendants. 

The above method allows us to follow the evolution of the total mass of the cloud. In this way, we can trace coherent cloud structures through time. However, gas will cycle through individual clouds, with new gas being added to the cloud via smooth accretion or mergers, while existing gas can break off into smaller clouds or disperse into the ISM. Therefore, after some time, it is possible that a cloud will no longer contain any of the material that was originally in the cloud in snapshot $i$. 

We therefore also considered an alternative method to link clouds with their progenitors and descendants, in which we consider only gas particles that were originally in the cloud in snapshot $i$. This is similar to how \citet{dobbs13} trace the evolution of GMCs in their simulations of an isolated disc galaxy. We first take the particles in the cloud in snapshot $i$ and identify the main progenitor in snapshot $i - 1$ that contains the most of these particles, as before. However, we then take only the particles in the main progenitor that were originally in the cloud in snapshot $i$ (and not all of the main progenitor's particles), and we trace these back to snapshot $i - 2$ to find the preceding main progenitor, and so on. We then repeat this procedure in later snapshots to identify the main descendants. In this way we can trace the evolution of only gas that was originally in the cloud in snapshot $i$. 

\begin{figure}
\centering
\mbox{
	\includegraphics[width=84mm]{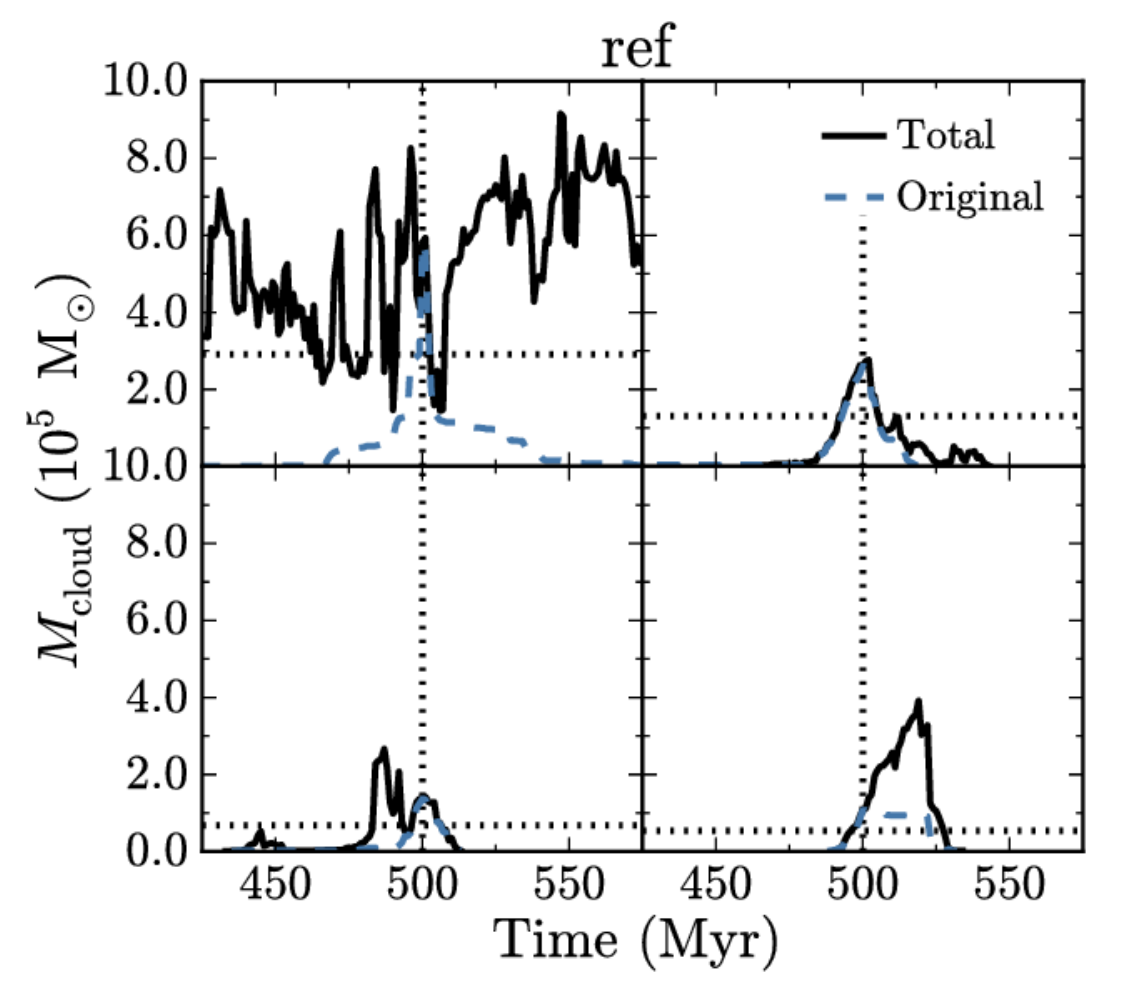}}
\caption{Mass evolution of four clouds selected at $500 \, \rm{Myr}$ from the ref simulation. We show the evolution of the total cloud mass \textit{(black solid curves)} and of the mass of particles originally in the cloud at $500 \, \rm{Myr}$ \textit{(blue dashed curves)}. The horizontal dotted lines indicate half of the mass of the cloud at $500 \, \rm{Myr}$, which we use to define the cloud ages and lifetimes (see text).} 
\label{massEvolutionFig}
\end{figure}

In Fig.~\ref{massEvolutionFig} we show examples of the mass evolution of individual clouds selected at $500 \, \rm{Myr}$ from the ref simulation. For each cloud we show the evolution of the total cloud mass (black solid curves), using the first method described above, and of the mass of original particles that were in the cloud at $500 \, \rm{Myr}$ (blue dashed curves), using the second method described above. The horizontal dotted line in each panel indicates half of the mass of the cloud at $500 \, \rm{Myr}$, which we use to define the age and lifetime of the cloud (see below). 

For some clouds, the evolution of the total and original mass are similar (e.g. the top right panel). These are clouds that reach their peak mass close to $500 \, \rm{Myr}$, and have fairly simple evolutionary histories, for example with no significant cloud mergers bringing in new material at later times. 

In many other clouds, the evolution is more complex. For example, in the bottom right panel, the cloud is still growing at $500 \, \rm{Myr}$. The total mass of the cloud therefore quadruples over the following $19 \, \rm{Myr}$, after which it rapidly declines. However, by definition, the original mass is a maximum in the original snapshot (at $500 \, \rm{Myr}$ in this example). We see that the original mass in this example remains nearly constant over the same period. 

\begin{figure}
\centering
\mbox{
	\includegraphics[width=84mm]{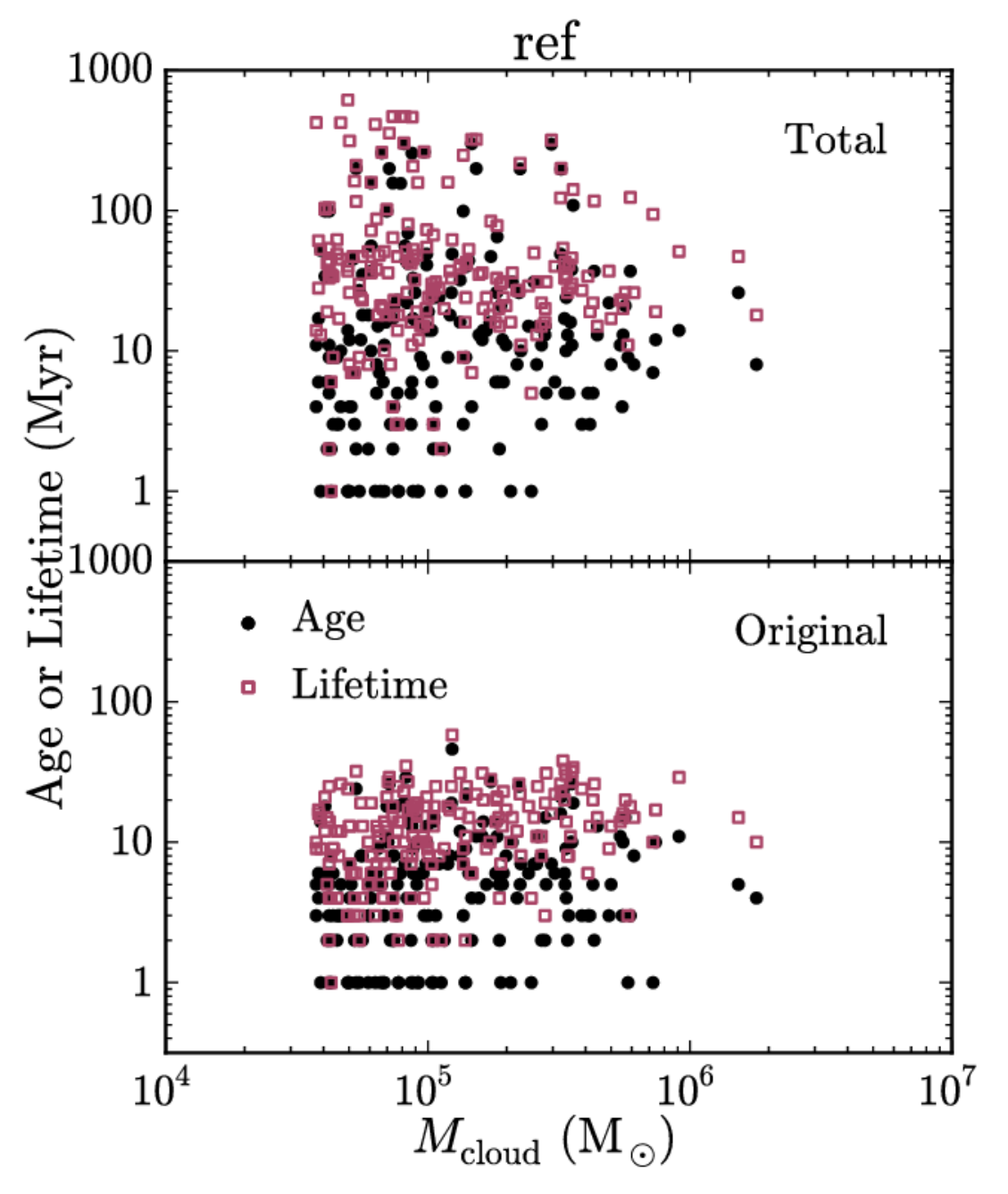}}
\caption{Cloud ages \textit{(black circles)}, defined as the time since the cloud's main progenitor had half of its current mass, and cloud lifetimes \textit{(red squares)}, defined as the period from when the main progenitor had half of its current mass to when the main descendant is reduced to half of its current mass. We define ages and lifetimes using either the total progenitor/descendant mass \textit{(top panel)}, or the mass of particles that were originally part of the cloud, i.e. at the time the cloud was identified \textit{(bottom panel)}. We show all clouds with at least $50$ particles, selected in snapshots at $100 \, \rm{Myr}$ intervals from $100 \, \rm{Myr}$ to $900 \, \rm{Myr}$, from the ref simulation. Our other two simulations (lowISRF and hiZ; not shown) have similar distributions of cloud ages and lifetimes. The horizontal stripes of points at low ages/lifetimes are due to the finite interval ($1 \, \rm{Myr}$) between snapshots.
\vspace{-0.1 in}} 
\label{cloudAgeFig}
\end{figure}

Finally, in some clouds (e.g. the top left panel), we find that the progenitors and descendants traced by the total mass extend over a much longer time period than those traced only by the original particles at $500 \, \rm{Myr}$. These are clouds that are constantly cycling through new gas, via accretion and cloud mergers, while existing gas breaks away or is blown away and disperses. The cloud in the top left panel is located at the centre of the galaxy. We saw in Fig.~\ref{gasMapsFig} that there is more dense gas near the centre, with several clouds packed closely together within the central few hundred parsecs. This explains why we see a strong cycling of gas through individual clouds in this region. 

We can now use the mass evolution of a cloud's progenitors and descendants to determine the age and lifetime of the cloud, based on either the total mass of the cloud or the mass of original particles. We define the age of the cloud as the time since the mass was half of its current value, and we define its lifetime to be the total period over which its mass is greater than half of its current value. For example, suppose we identify a cloud at time $t_{\rm{now}}$. In the past, its main progenitor had half of its current mass at time $t_{\rm{past}}$, and in the future, its main descendant is reduced to half of its current mass at time $t_{\rm{future}}$. The age is then $t_{\rm{now}} - t_{\rm{past}}$, and the lifetime is $t_{\rm{future}} - t_{\rm{past}}$. The ages and lifetimes will depend on the mass fraction that we use to define them. For example, if we use the time when the main progenitor/descendant was a quarter of the cloud's current mass, rather than half, the median lifetimes are increased by $\approx 50$ per cent. However, by using a factor of half, there can only be one `main' progenitor/descendant in each snapshot over the cloud's lifetime, and we avoid ambiguities arising from multiple progenitors/descendants with equal mass. 

In Fig.~\ref{cloudAgeFig} we plot the ages (black circles) and lifetimes (red squares) of clouds from the ref simulation versus their current mass, using the total progenitor/descendant mass (top panel) or the mass of original particles (bottom panel). Our other two simulations (lowISRF and hiZ; not shown) have similar distributions of cloud ages and lifetimes. We show all clouds with at least $50$ particles identified in snapshots at $100 \, \rm{Myr}$ intervals, from $100 \, \rm{Myr}$ to $900 \, \rm{Myr}$. Note that, while we only show clouds with at least $50$ particles, we still identify clouds with as few as $25$ particles, so we can trace the clouds in Fig.~\ref{cloudAgeFig} to the time when they had, or will have, half of their current mass. 

If we use the total cloud mass (top panel), we find a median cloud age of $12 \, \rm{Myr}$, and a median lifetime of $33 \, \rm{Myr}$. There is a lot of scatter in cloud ages and lifetimes, with many having ages and lifetimes of a few hundred $\rm{Myr}$. Note that, since we combine snapshots at $100 \, \rm{Myr}$ intervals in this figure, evolutionary tracks will appear multiple times if they have a lifetime longer than this interval. 

In the bottom panel, we see that cloud ages and lifetimes defined using only the original particles are shorter than those defined from the total mass. Using this definition, we find a median age of $5 \, \rm{Myr}$ and a median lifetime of $13 \, \rm{Myr}$. There is again a lot of scatter in cloud ages and lifetimes, but we find that most clouds have an age $\la 30 \, \rm{Myr}$ and a lifetime $\la 40 \, \rm{Myr}$. Observational estimates, typically based on associated signatures of star formation such as young stellar clusters and H\textsc{ii} regions, find GMC lifetimes $\approx 20 - 40 \, \rm{Myr}$ \citep[e.g.][]{bash77, kawamura09, murray11, miura12}, although \citet{elmegreen00} and \citet{scoville79} find lifetimes of a few Myr and hundreds of Myr, respectively. We find no clear trend of age or lifetime with the current mass of the cloud. 

Since we run each simulation for $1 \, \rm{Gyr}$, we follow the evolution of the galaxy for many cloud lifetimes. This is important as it ensures that the evolution of individual clouds is not strongly affected by the initial chemical state of the gas at the beginning of the simulation, when most of the hydrogen was in H\textsc{ii}. 

An important caveat to note is that the stellar feedback model used in these simulations is for feedback from supernovae. This means that we do not explicitly model feedback processes that act on shorter time-scales than supernovae, such as stellar winds and photoheating from H\textsc{ii} regions. Such processes might disrupt clouds before the supernovae explode, thus shortening their ages and lifetimes. 

For the remainder of this paper, we will use cloud ages and lifetimes defined via the mass of original particles (i.e. the bottom panel of Fig.~\ref{cloudAgeFig}). This definition gives a better indication of how long the current material has been in the cloud. However, both age/lifetime definitions that we have considered here (using total or original mass) involve tracing individual particles through time in the simulations, which is not possible in observations. Observational estimates are typically based on nearby signatures of star formation, such as young stellar clusters and H\textsc{ii} regions \citep[e.g.][]{kawamura09}. It is not clear which of our definitions is likely to correspond more closely with these observational definitions, so we need to be careful when comparing to observed GMC lifetimes. 

\vspace{-0.1 in}

\subsection{CO emission maps}\label{CO_map_methods} 

We computed CO emission from the $J = 1 - 0$ line in our simulations in post-processing, using the publicly available Monte-Carlo radiative transfer code \textsc{radmc-3d}\footnote{\url{http://www.ita.uni-heidelberg.de/~dullemond/software/radmc-3d/}} (version 0.38), written by Cornelis Dullemond. This code follows emission from user-specified molecular and atomic lines, and also includes thermal emission, absorption and scattering from dust grains. We used molecular CO data from the \textsc{lamda} database\footnote{\url{http://home.strw.leidenuniv.nl/~moldata/}} \citep{schoier05}, including collisional excitation rates of CO by ortho- and para-H$_{2}$ \citep{yang10}. We assumed an ortho-to-para ratio of 3:1 for H$_{2}$. We included two species of dust grains, graphite and silicate, with dust opacities from \citet{martin90}, who used the power-law size distribution of dust grains from \citet{mathis77}. 

Line emission from CO depends on the level populations of the CO molecule. The simplest method is to assume that the level populations are in Local Thermodynamic Equilibrium (LTE). However, this assumption may not always be valid. We therefore computed the level populations in non-LTE using the Local Velocity Gradient (LVG) method, also known as the \citet{sobolev57} approximation. This method assumes that, due to gas motions, photons emitted from transitions in the CO molecule will become sufficiently Doppler shifted after travelling some distance that the photon can no longer be absorbed by the same transition that produced it. This allows us to define an escape probability for these photons based on their velocity gradient. We can then determine the level populations, including radiative excitation by line photons, from local quantities alone. A more detailed description of the LVG method, as implemented in \textsc{radmc-3d}, can be found in \citet{shetty11a}. 

In addition to thermal broadening of the emission line, \textsc{radmc-3d} also allows the inclusion of doppler broadening by unresolved microturbulence. In our simulations, we impose a density-dependent pressure floor, $P_{\rm{floor}}$, on the gas to ensure that the Jeans mass is resolved by at least 4 SPH kernel masses, to prevent artificial fragmentation (see section~\ref{models_section}). While the implementation of this pressure floor was motivated by numerical reasons, its effect on the cloud will be similar to a pressure term from unresolved turbulence. We can attribute a one-dimensional velocity dispersion, $\sigma_{\rm{floor, \, 1D}}$, to the pressure floor according to $P_{\rm{floor}} = \rho \sigma_{\rm{floor, \, 1D}}^{2}$. Using equation~\ref{floor_mass} for $P_{\rm{floor}}$, with our fiducial parameters $N_{\rm{J, \, m}} = 4$, $m_{\rm{gas}} = 750 \, \rm{M}_{\odot}$ and $N_{\rm{ngb}}^{\rm{SPH}} = 100$, we find:

\begin{equation}\label{sigma_floor_eqn}
\sigma_{\rm{floor, \, 1D}} = 1.18 \left( \frac{\rho}{10^{-24} \, \rm{g} \, \rm{cm}^{-3}} \right)^{1/6} \, \rm{km} \, \rm{s}^{-1}. 
\end{equation} 
We therefore include microturbulent broadening due to this pressure floor when computing the CO line emission, with a velocity dispersion given by equation~\ref{sigma_floor_eqn}. The Doppler broadening due to this microturbulence is then added to the thermal broadening in quadrature. 

For each cloud in our simulations, we extracted a region around the cloud and interpolated the gas density, temperature and velocities, along with the densities of CO and H$_{2}$, onto a 3d Cartesian grid with a resolution of $1 \, \rm{pc}$, using the same $C^{2}$ \citet{wendland95} kernel with $100$ SPH neighbours as was used in the simulations. We used \textsc{radmc-3d} to compute the total emission from the $J = 1 - 0$ line and thermal dust emission in $480$ wavelength bins covering a velocity range $\pm 60 \, \rm{km} \, \rm{s}^{-1}$ centred on the line, which we projected onto a plane parallel to the galactic disc. We then repeated this without line emission to create a map of the thermal dust emission only, which we finally subtracted from the total emission to produce a continuum-subtracted map of CO $J = 1 - 0$ line emission. 

\begin{figure}
\centering
\mbox{
	\includegraphics[width=84mm]{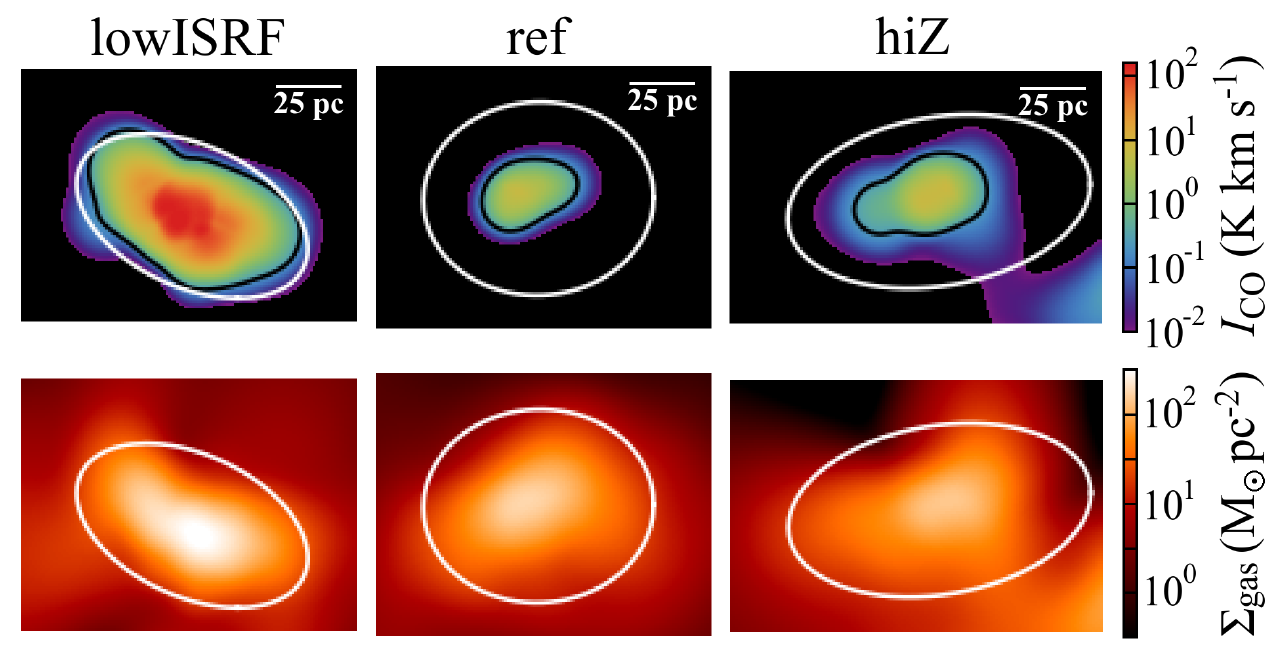}}
\caption{CO $J = 1 - 0$ line emission maps \textit{(top row)} and gas surface density maps \textit{(bottom row)} of molecular clouds from the lowISRF \textit{(left)}, ref \textit{(centre)} and hiZ \textit{(right)} simulations. The white ellipse in each panel indicates the boundary of the cloud, defined by gas particles with a density above a threshold $n_{\rm{H}, \, \rm{min}} = 10 \, \rm{cm}^{-3}$. The black contours in the top row show $I_{\rm{CO}} = 0.25 \, \rm{K} \, \rm{km} \, \rm{s}^{-1}$, which corresponds to the $3 \sigma$ intensity threshold for the Small Magellanic Cloud in the observations of \citet{leroy11}. In the examples from the ref and hiZ simulations, only the centres of the clouds would be detectable in a typical CO survey.}
\label{co_maps_fig}
\end{figure}

In Fig.~\ref{co_maps_fig} we show examples of CO emission maps from individual molecular clouds in the lowISRF, ref and hiZ simulations (left, centre and right columns, respectively). The top and bottom rows show maps of the CO emission and gas surface density, respectively. 

The white ellipse in each panel shows the boundary of the cloud, defined by gas particles with a density above a threshold $n_{\rm{H}, \, \rm{min}} = 10 \, \rm{cm}^{-3}$. This ellipse was computed by projecting the 3d ellipsoid that approximately encloses the particles in the cloud onto the image plane, where the 3d ellipsoid is based on the principle axes of the moment of inertia tensor, as described in section~\ref{clump_finder}. For comparison, the black contours in the top row show $I_{\rm{CO}} = 0.25 \, \rm{K} \, \rm{km} \, \rm{s}^{-1}$. This corresponds to the $3 \sigma$ intensity threshold for the observations of the Small Magellanic Cloud in \citet{leroy11}. In the examples from the ref and hiZ simulations (centre and right columns, respectively), only the centres of the clouds are above this detection threshold. 

We thus see that our standard definition of a cloud, based on a fixed density threshold, includes a larger region than if we had defined clouds based on the observable CO emission. We therefore also consider an alternative cloud definition, in which we only include regions in the 2d maps of CO emission with $I_{\rm{CO}} > 0.25 \, \rm{K} \, \rm{km} \, \rm{s}^{-1}$. For this alternative cloud definition, we also compute the projected cloud mass, $M_{\rm{proj}}$, and size, $R_{\rm{proj}} = (A / \pi)^{1/2}$, from the 2d maps, rather than from the 3d particle distribution, where $M_{\rm{proj}}$ and $A$ are the total mass and area, respectively, of pixels above the CO intensity threshold. This alternative cloud definition provides a fairer comparison with observations. 

It is important to note that these CO emission maps may be sensitive to resolution. In particular, high-resolution simulations of dense clouds find that most CO is concentrated in compact structures, with sizes $\sim 1 \, \rm{pc}$ and densities $\sim 10^{3} \, \rm{cm}^{-3}$ \citep[e.g.][]{glover12}. Such structures are poorly resolved in our simulations, even in our high resolution simulation in Appendix~\ref{res_appendix}, which may make the predicted CO emission uncertain. 

\section{Cloud scaling relations}\label{scaling_relations_section} 

Observations of molecular clouds find strong relations between their properties such as size, velocity dispersion and mass, both in Milky Way GMCs \citep[e.g.][]{larson81, solomon87, heyer09} and in extragalactic GMCs \citep[e.g.][]{bolatto08}. For example, building on the original relations identified by \citet{larson81}, \citet{solomon87} studied a sample of GMCs in the Milky Way, and found that the line of sight velocity dispersion, $\sigma$, follows a power law relation with cloud radius, $R$: 

\begin{equation}\label{scaling_eqn1} 
\sigma = 0.72 \left( \frac{R}{\rm{pc}} \right)^{0.5} \, \rm{km} \, \rm{s}^{-1}. 
\end{equation} 
By assuming that the clouds are in virial equilibrium, they estimated the cloud masses, $M$, and they found a power-law relation between $M$ and $\sigma$: 

\begin{equation}\label{scaling_eqn2} 
M = 2000 \left( \frac{\sigma}{\rm{km} \, \rm{s}^{-1}} \right)^{4} \, \rm{M}_{\odot}. 
\end{equation} 
Combining equations~\ref{scaling_eqn1} and \ref{scaling_eqn2} produces the following relation between $M$ and $R$: 

\begin{equation}\label{scaling_eqn3}
M = 540 \left( \frac{R}{\rm{pc}} \right)^{2} \, \rm{M}_{\odot}, 
\end{equation}
which implies that the clouds in their sample have a constant mean surface density of $\Sigma = 170 \, \rm{M}_{\odot} \, \rm{pc}^{-2}$. Later studies have corrected this value to $200 \, \rm{M}_{\odot} \, \rm{pc}^{-2}$ to account for an updated estimate for the Sun's galactocentric radius of $8.5 \, \rm{kpc}$, rather than $10 \, \rm{kpc}$ as originally used \citep[see e.g.][]{heyer09}. The corrected mass-size relation is then: 

\begin{equation}\label{scaling_eqn4}
M = 625 \left( \frac{R}{\rm{pc}} \right)^{2} \, \rm{M}_{\odot}. 
\end{equation}
The updated galactocentric radius of the Sun will also affect the size-linewidth relation in equation~\ref{scaling_eqn1}. However, the correction for this relation is smaller than the errors. 

\citet{heyer09} re-examined the sample of \citet{solomon87}, and used the $^{13}$CO luminosities of the clouds to estimate their molecular hydrogen masses. They found a median molecular surface density of $42 \, \rm{M}_{\odot} \, \rm{pc}^{-2}$, lower than the value determined by \citet{solomon87} from the virial mass. Furthermore, they found that $\Sigma$ is not constant, and that $\sigma / R^{0.5}$ varies systematically with surface density as $\Sigma^{0.5}$. 

\begin{figure}
\centering
\mbox{
	\includegraphics[width=84mm]{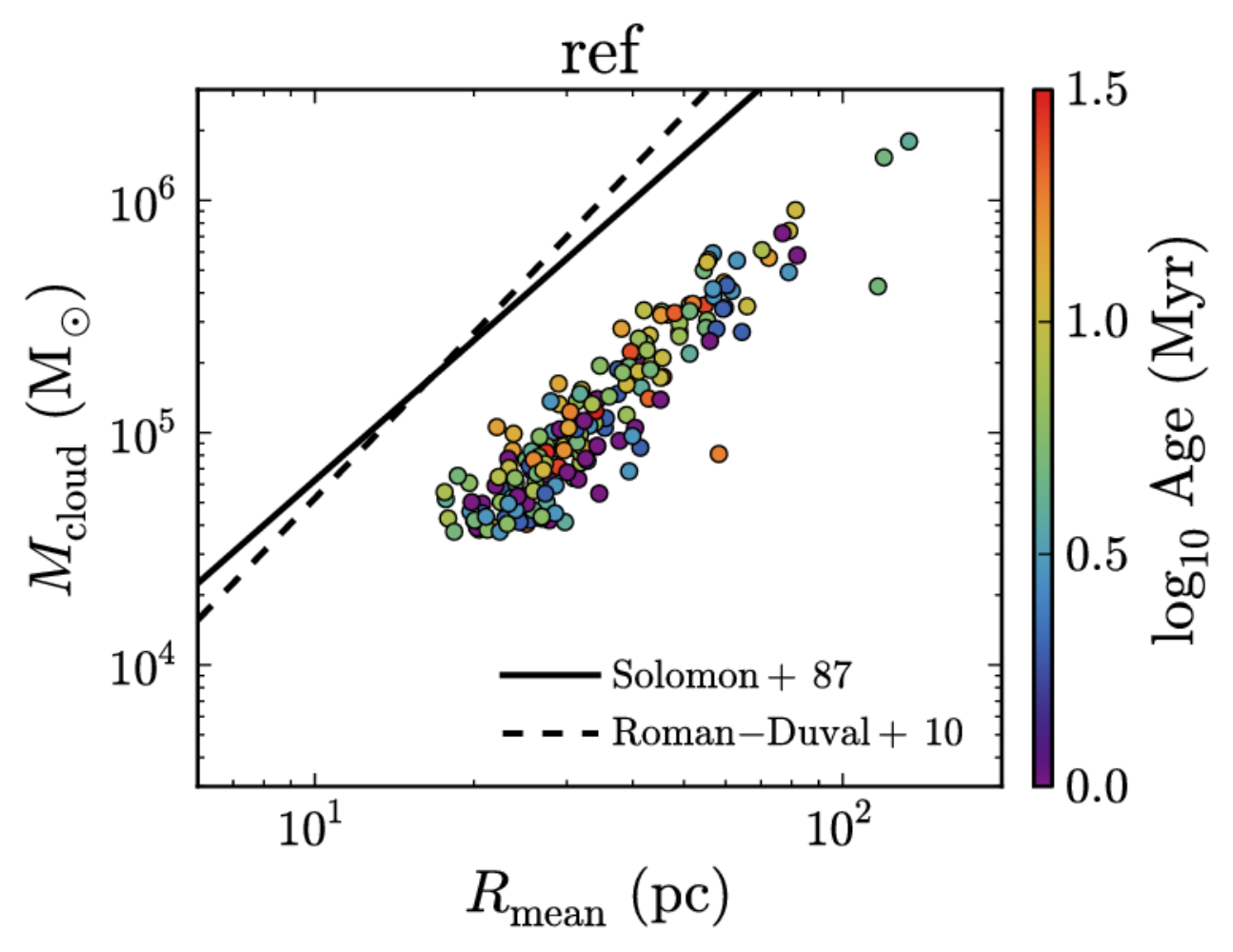}}
\caption{Cloud mass, $M_{\rm{cloud}}$, versus cloud radius, $R_{\rm{mean}} = (abc)^{1/3}$, for all clouds with at least $50$ particles identified in snapshots at $100 \, \rm{Myr}$ intervals from $100 \, \rm{Myr}$ to $900 \, \rm{Myr}$ in the ref simulation. Our other two simulations (not shown) have very similar cloud mass-size relations. The colour scale indicates the cloud age. The black lines show the observed relations of \citet{solomon87} \textit{(solid; our equation~\ref{scaling_eqn4})} and \citet{romanduval10} \textit{(dashed; our equation~\ref{scaling_eqn5})}. Our simulated clouds follow a similar slope to the observed relations, but the normalisation, which depends on the cloud definition, is a factor $\approx 4$ lower than is observed.}
\label{massSizeFig}
\end{figure}

\citet{romanduval10} studied the properties of molecular clouds in the BU-FCRAO Galactic Ring Survey \citep{jackson06} and the UMSB survey \citep{clemens86, sanders86} in the Milky Way. They found the following mass-size relation: 

\begin{equation}\label{scaling_eqn5}
M = 228 \left( \frac{R}{\rm{pc}} \right)^{2.36} \, \rm{M}_{\odot},
\end{equation}
based on $^{13}$CO line emission. 

Fig.~\ref{massSizeFig} shows the cloud mass-size relation from the ref simulation, where the cloud radius, $R_{\rm{mean}}$, was calculated from the 3d particle distribution (see equation~\ref{r_mean_eqn}). Our other two simulations (lowISRF and hiZ; not shown) have very similar cloud mass-size relations. We show all clouds with at least $50$ particles identified in snapshots at $100 \, \rm{Myr}$ intervals from $100 \, \rm{Myr}$ to $900 \, \rm{Myr}$. The colour scale indicates the age of the cloud, defined from the particles originally in the cloud in the current snapshot, as described in section~\ref{evolution_section}. The black solid and dashed lines show the observed relations of \citet{solomon87} and \citet{romanduval10}, respectively, i.e. equations~\ref{scaling_eqn4} and \ref{scaling_eqn5}. 

Our simulated clouds follow a similar slope to the observed relations, but the normalisation is a factor $\approx 4$ lower than is observed. The lower normalisation is determined by the density threshold that we use to define a cloud, and suggests that our definition includes a larger region around the cloud than would be included in a typical observational survey based on CO emission. Indeed, we saw in Fig.~\ref{co_maps_fig} that only the central regions of our simulated clouds have velocity-integrated CO intensities above a threshold of $0.25 \, \rm{K} \, \rm{km} \, \rm{s}^{-1}$. For a fairer comparison with observations, we therefore also considered an alternative cloud definition in which we include only regions above this CO intensity threshold, and compute cloud properties in projection, as described in section~\ref{CO_map_methods}. 

\begin{figure}
\centering
\mbox{
	\includegraphics[width=84mm]{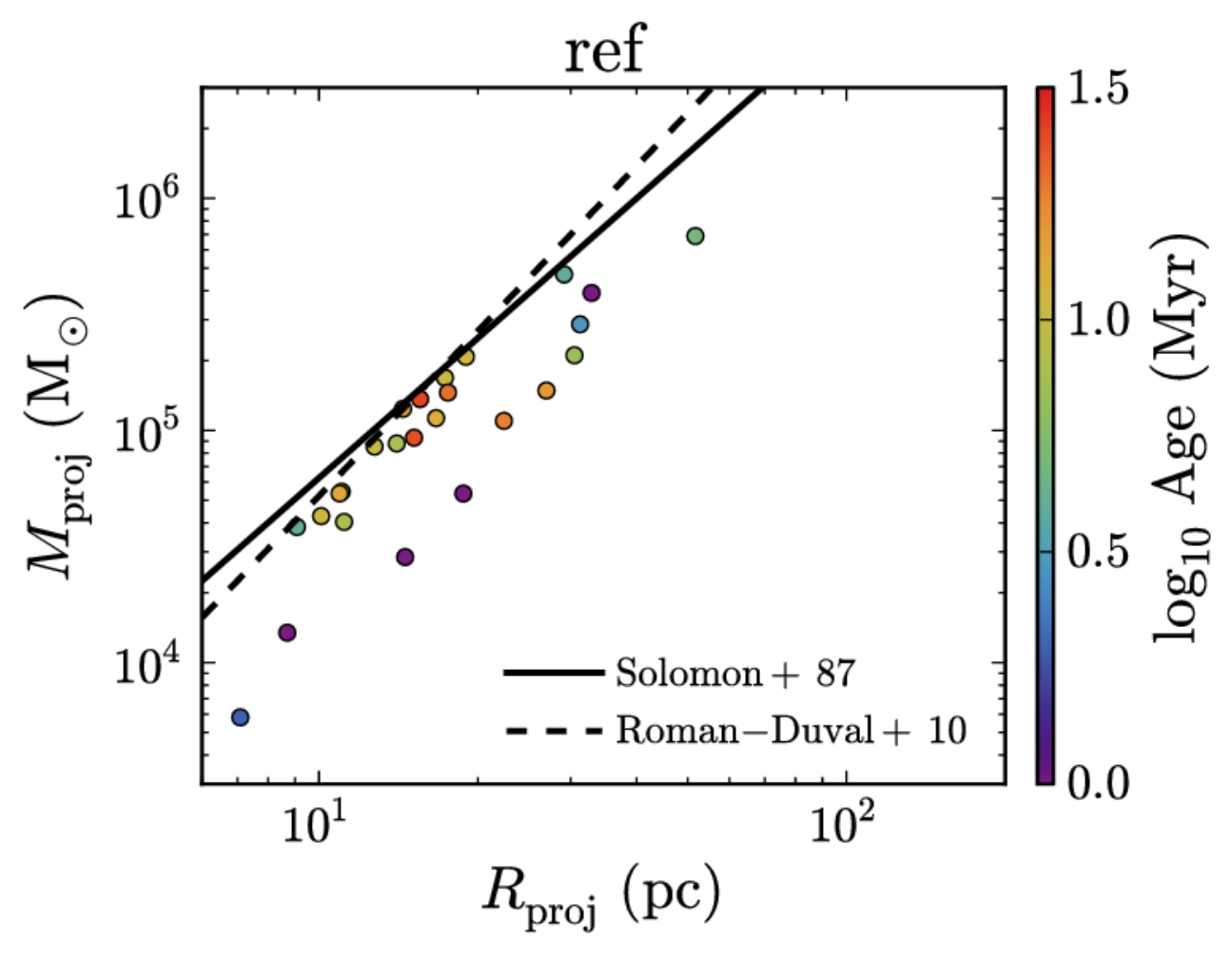}}
\caption{As Fig.~\ref{massSizeFig}, but for cloud masses and projected sizes determined from CO-detectable ($I_{\rm{CO}} > 0.25 \, \rm{K} \, \rm{km} \, \rm{s}^{-1}$) regions only. The normalisation of this relation in our simulations is a factor $\approx 2$ lower than the observed relations (\citealt{solomon87} and \citealt{romanduval10}; black solid and dashed lines respectively) when we include only CO-detectable regions, compared to a factor of four lower than is observed when we used a density-based cloud definition (see Fig.~\ref{massSizeFig}).}
\label{massSizeThreshFig}
\end{figure}

\begin{figure}
\centering
\mbox{
	\includegraphics[width=84mm]{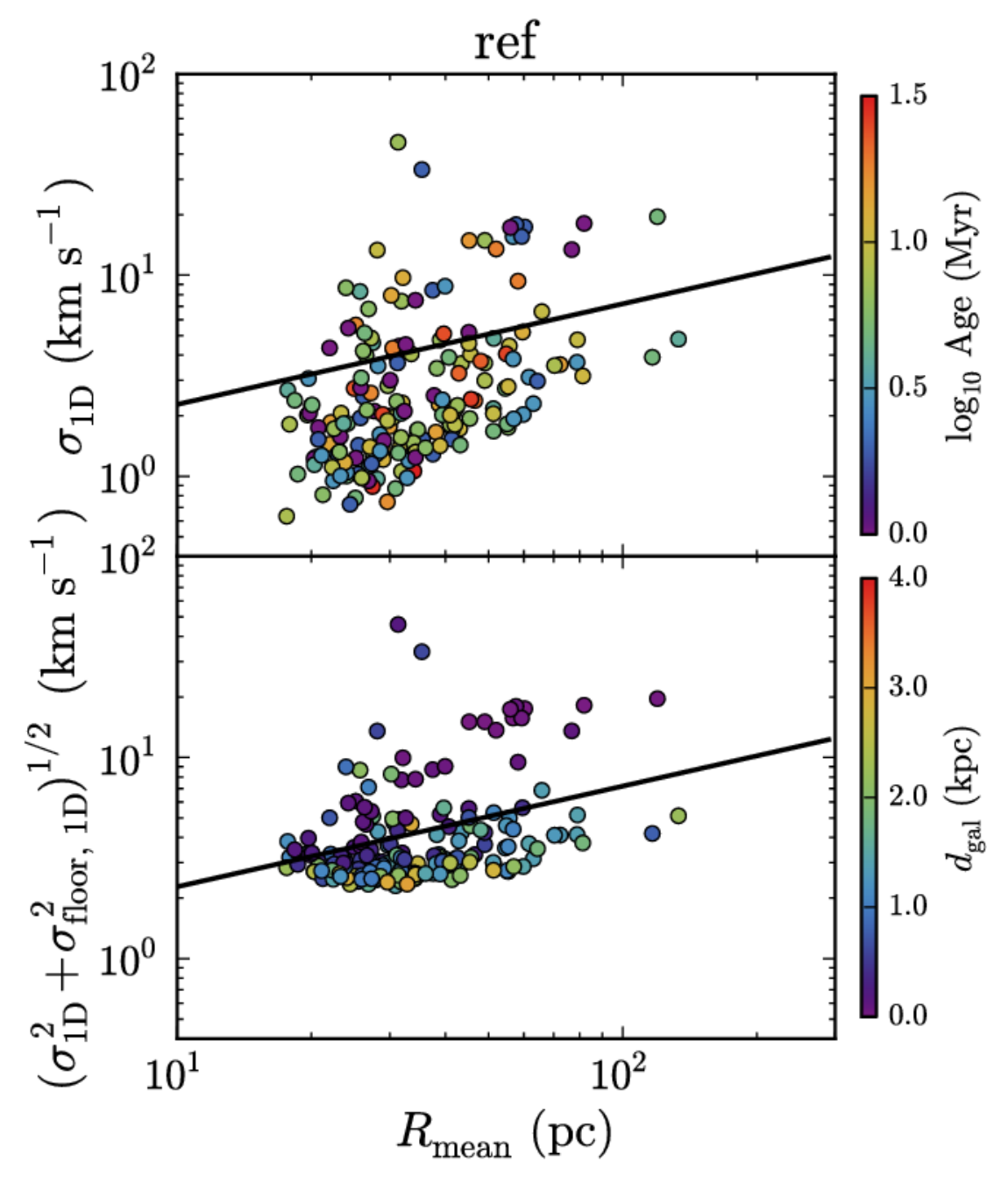}}
\caption{One-dimensional velocity dispersion, measured from gas particle motions only \textit{($\sigma_{1\rm{D}}$; top panel)}, and including the contribution from the pressure floor \textit{($\sigma_{\rm{floor, \, 1D}}$; bottom panel)}, plotted against cloud radius. The colour scale in the top panel indicates the cloud age, while in the bottom panel it indicates the distance of the cloud's centre of mass from the galaxy centre. The black lines indicate the observed relation of \citet{solomon87}. We show only the ref simulation here, although the relations in our other two simulations are very similar. If we do not account for the pressure floor (top panel), we find unrealistically low velocity dispersions ($< 1 \, \rm{km} \, \rm{s}^{-1}$). When we include the pressure floor as an unresolved turbulence term (bottom panel), we find velocity dispersions $> 2.3 \, \rm{km} \, \rm{s}^{-1}$ for all clouds, although there remains a large scatter in this relation. The clouds with the highest velocity dispersion tend to be close to the galaxy centre.} 
\label{velocityDispersionFig}
\end{figure}

Fig.~\ref{massSizeThreshFig} shows the mass-size relation computed using only CO-detectable regions, for the ref simulation. Compared to Fig.~\ref{massSizeFig}, for a density-based cloud definition, the clouds lie much closer to the observed relations of \citet{solomon87} and \citet{romanduval10} (black solid and dashed lines, respectively), although the normalisation of this relation in our simulations is still a factor $\approx 2$ lower than is observed (compared to a factor $\approx 4$ in Fig.~\ref{massSizeFig}). However, even our CO-based cloud definition is not identical to the definitions used in these two observational studies. Our criterion is based on a minimum velocity-integrated CO intensity in the projected, two-dimensional position-position space of the CO emission maps. However, \citet{solomon87} define clouds as regions above a minimum CO brightness temperature of $1 \, \rm{K}$ in the three-dimensional position-position-velocity (PPV) space. \citet{romanduval10} use a minimum velocity-integrated intensity of $4 \sigma = 0.23 \sqrt{N_{\nu}} \, \rm{K} \, \rm{km} \, \rm{s}^{-1}$, where $N_{\nu}$ is the number of velocity channels, but for $^{13}$CO line emission, rather than $^{12}$CO as used by us. Additionally, when measuring $^{13}$CO column densities to compute the cloud mass, they include only velocity channels above a $^{13}$CO brightness temperature of $4\sigma = 1 \, \rm{K}$. Therefore, the remaining discrepancy in the normalisation of the cloud mass-size relation is likely due to the different CO thresholds that we use. Given that observational studies use a range of cloud definitions, with different clump finding algorithms and using different molecular emission lines, we keep our CO-based definition general, rather than try to match a particular observational study. 

\begin{figure}
\centering
\mbox{
	\includegraphics[width=84mm]{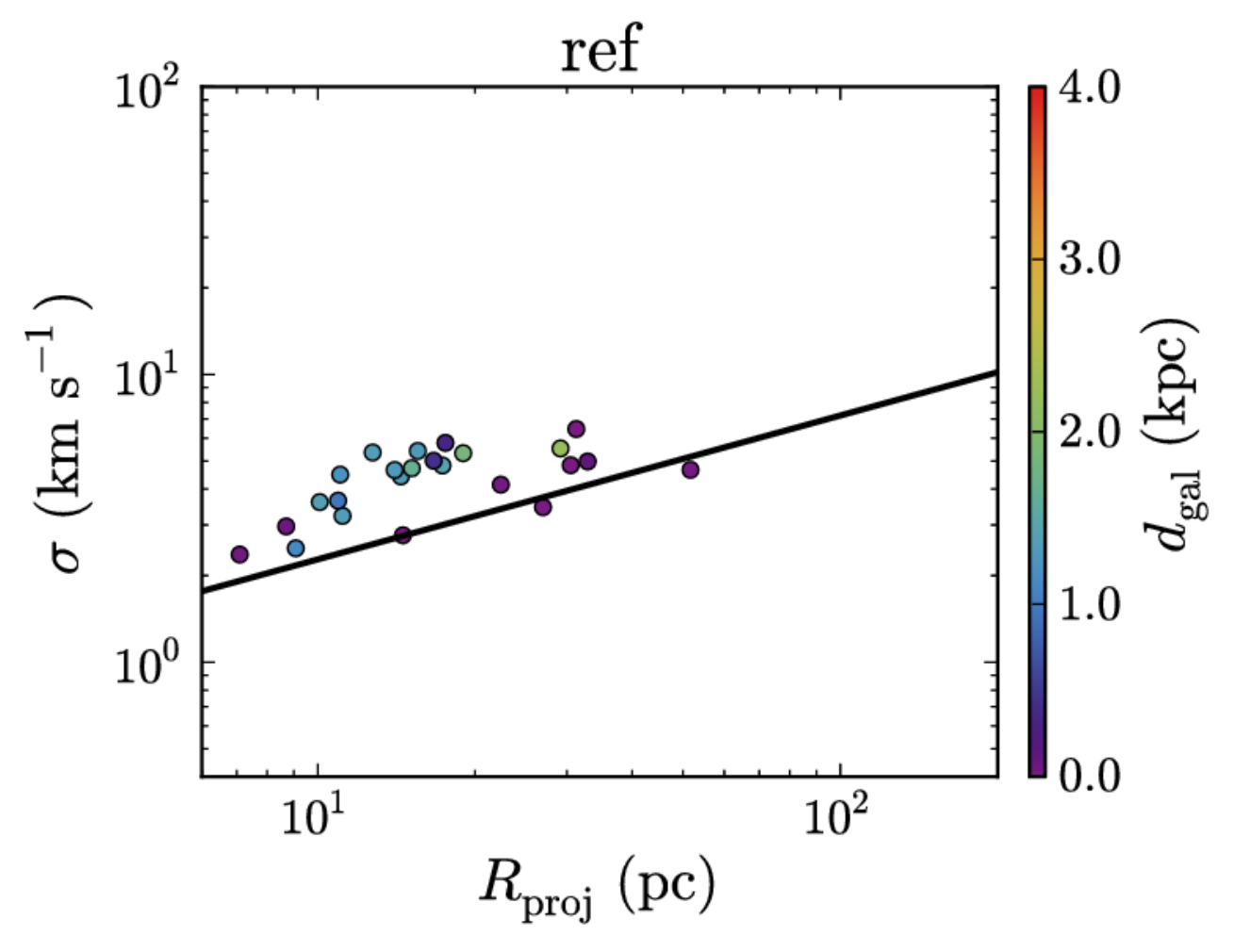}}
\caption{As Fig.~\ref{velocityDispersionFig}, but measured from CO-detectable ($I_{\rm{CO}} > 0.25 \, \rm{K} \, \rm{km} \, \rm{s}^{-1}$) regions of each cloud only. Also, we use projected sizes and compute the velocity dispersion by fitting a single Gaussian component to the CO spectrum. We exclude all clouds that show multiple peaks in their CO spectrum, as they cannot be fit with a single velocity component, and likely consist of multiple clouds that are undergoing mergers. We find better agreement with the observed relation of \citet{solomon87} \textit{(black line)} than we saw in Fig.~\ref{velocityDispersionFig}.}
\label{velocityDispersionThreshFig}
\end{figure}

\citet{pan16} used a hydrodynamic simulation of a barred spiral galaxy to investigate how the definition of GMCs in position-position-position (PPP) or PPV space affects their properties. They found that the power law indices of the cloud scaling relations vary with cloud definition and, for a PPV-based definition, with disc inclination. \citet{duartecabral16} also explored different definitions of GMCs in PPP or PPV space, based on either H$_{2}$ or CO, using a high-resolution ($3.85 \, \rm{M}_{\odot}$ per SPH particle) re-simulation of a section of a spiral galaxy simulation. They found that a PPV-based definition tends to blend clouds that would be physically separated in PPP space, although this effect is less significant if the PPV emission maps have high spatial resolution and high sensitivity. They also found that CO densities tend to trace only the high-density H$_{2}$ gas, rather than all molecular gas. 

Fig.~\ref{velocityDispersionFig} compares the velocity dispersion-cloud size relation from the ref simulation (coloured points) to the observed relation from \citet{solomon87}, i.e. our equation~\ref{scaling_eqn1} (black solid line), for our standard density-based cloud definition. In the top panel of Fig.~\ref{velocityDispersionFig}, we plot the one-dimensional velocity dispersion, $\sigma_{1\rm{D}} = \sigma_{3\rm{D}} / \sqrt{3}$, where the three-dimensional velocity dispersion is $\sigma_{3\rm{D}}^{2} = \sigma_{x}^{2} + \sigma_{y}^{2} + \sigma_{z}^{2}$, and $\sigma_{i}^{2} = \left< v_{i}^{2} \right> - \left< v_{i} \right>^{2}$ for the $i^{\rm{th}}$ component of the particle velocities, $v_{i}$, where angular brackets indicate a mass-weighted average over all particles in the cloud. The colour scale in the top panel of Fig.~\ref{velocityDispersionFig} indicates the cloud age. 

We see that the relation between $\sigma_{1\rm{D}}$ and $R_{\rm{mean}}$ is steeper than observed. In particular, we find very low velocity dispersions, $\sigma_{1\rm{D}} < 1 \, \rm{km} \, \rm{s}^{-1}$, far below the observed relation of \citet{solomon87}. However, these measurements of $\sigma_{1\rm{D}}$ in the simulations do not account for unresolved turbulence. As noted in section~\ref{CO_map_methods}, the pressure floor that we impose on the gas to ensure that the Jeans mass is always well-resolved will have a similar effect on the cloud as a pressure term from unresolved turbulence, with a turbulent velocity dispersion, $\sigma_{\rm{floor, \, 1D}}$, given by equation~\ref{sigma_floor_eqn}. We therefore need to include the effects of this pressure floor. 

In the bottom panel of Fig.~\ref{velocityDispersionFig}, we compute $\sigma_{\rm{floor, \, 1D}}$ for each cloud using its mean density, add this to $\sigma_{1\rm{D}}$ in quadrature, and plot the total velocity dispersion against cloud radius. By accounting for the pressure floor in this way, we avoid unrealistically low velocity dispersions (the lowest value is now $\approx 2.3 \, \rm{km} \, \rm{s}^{-1}$). This will be important for computing CO emission in our simulated clouds, as the CO $J = 1 - 0$ line is often optically thick in molecular clouds, so the line intensity will depend on the line width. 

The observed velocity dispersions will also include a component due to the thermal broadening of the molecular lines that are used to measure $\sigma_{1 \rm{D}}$. For CO, with a mean molecular weight $\mu = 28$, the thermal velocity is $\sigma_{\rm{th, \, 1D}} = \sqrt{k_{\rm{B}} T / \mu m_{\rm{p}}} = 0.17 \, \rm{km} \, \rm{s}^{-1}$ at a temperature $T = 100 \, \rm{K}$, where $k_{\rm{B}}$ and $m_{\rm{p}}$ are the Boltzmann constant and proton mass, respectively. Thus, $\sigma_{\rm{th, \, 1D}}$ is small compared to $\sigma_{\rm{floor, \, 1D}}$ in our simulations, so we do not include $\sigma_{\rm{th, \, 1D}}$ in Fig.~\ref{velocityDispersionFig}, although we do account for thermal broadening when we compute CO line emission, as described in section~\ref{CO_map_methods}. 

Even accounting for the pressure floor, we still find a lot of scatter in this relation in our simulations, with some clouds showing velocity dispersions $> 10 \, \rm{km} \, \rm{s}^{-1}$. In the top panel of Fig.~\ref{velocityDispersionFig}, we found no trend in this relation with the cloud age. However, in the bottom panel, the colour scale indicates the distance of the cloud's centre of mass from the centre of the galaxy. We see that clouds with the highest velocity dispersions ($\ga 10 \, \rm{km} \, \rm{s}^{-1}$) are generally found within the central $\approx 1 \, \rm{kpc}$ of the galaxy. We also find that many of these high velocity dispersion clouds contain multiple density peaks that indicate substructures within the cloud. Therefore, some of the scatter towards high velocity dispersions is likely to be caused by motions of substructures within the cloud, possibly created by ongoing cloud-cloud mergers, which are more common in the centre of the galaxy. Interestingly, observations of molecular clouds in the centre of the Milky Way also find higher velocity widths compared to the linewidth-size relation of nearby molecular clouds in the Galactic disc \citep[e.g.][]{oka01}. 

Fig.~\ref{velocityDispersionThreshFig} shows the velocity dispersion-size relation in the ref simulation using our CO-based cloud definition, i.e. restricted to CO-detectable regions and with cloud sizes computed in projection. The 1d velocity dispersion of each cloud was measured by fitting a single Gaussian component to the CO spectrum extracted from pixels above the $I_{\rm{CO}}$ threshold. We visually inspected each spectrum and excluded those with multiple peaks, which cannot be fit with a single velocity component. These systems are likely multiple clouds that are undergoing mergers. The velocity dispersion was then obtained from the width of the best-fitting Gaussian. The width of the CO spectrum includes microturbulent Doppler broadening by the pressure floor. By defining clouds above a CO intensity threshold, measuring the velocity dispersion from the width of the CO spectrum rather than motions of the gas particles, and excluding merging systems with multiple velocity components, we find better agreement with the observed relation of \citet{solomon87} than we saw in Fig.~\ref{velocityDispersionFig}. 

\begin{figure}
\centering
\mbox{
	\includegraphics[width=84mm]{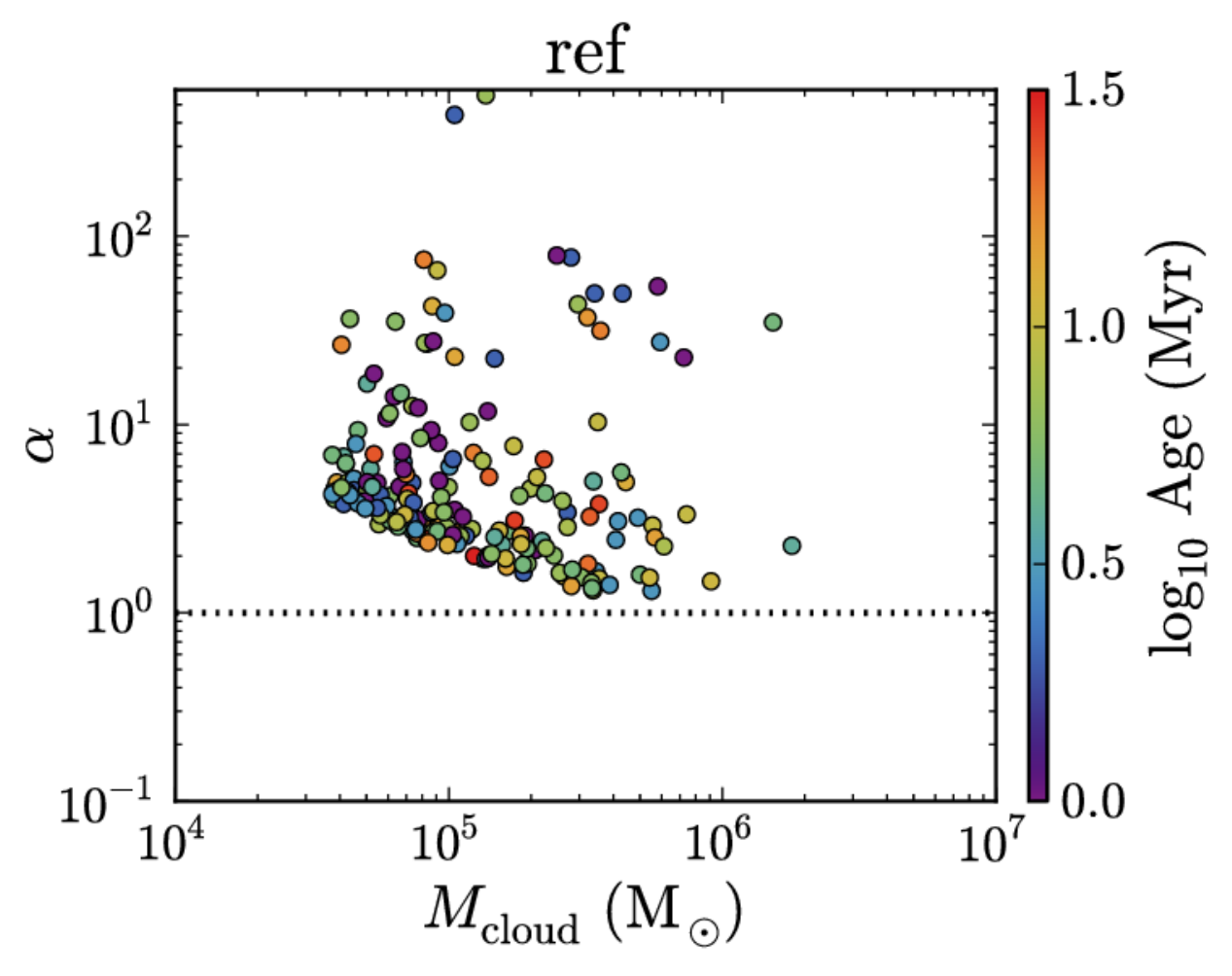}}
\caption{Virial parameter, $\alpha$, versus cloud mass for all clouds with more than $50$ particles in snapshots at $100 \, \rm{Myr}$ intervals from $100 \, \rm{Myr}$ to $900 \, \rm{Myr}$, from the ref simulation. Clouds in our other simulations (lowISRF and hiZ; not shown) have similar distributions of $\alpha$. The colour scale indicates the cloud age. The horizontal dotted line shows $\alpha = 1$, which corresponds to virial equilibrium. We see that all of our simulated clouds have $\alpha > 1$, i.e. they are not virialised.  However, the lower envelope of points is caused by the pressure floor, which prevents low-mass clouds ($\la 3 \times 10^{5} \, \rm{M}_{\odot}$) from becoming strongly gravitationally bound.} 
\label{virialFig}
\end{figure}

In Fig.~\ref{virialFig} we plot the virial parameter, $\alpha$, against $M_{\rm{cloud}}$, from the ref simulation, where: 

\begin{equation}
\alpha = \frac{5 \sigma^{2} R_{\rm{mean}}}{G M_{\rm{cloud}}} 
\end{equation}
\citep[e.g.][]{bertoldi92, dobbs11},  and we include the pressure floor in the velocity dispersion, i.e. $\sigma^{2} = \sigma_{1\rm{D}}^{2} + \sigma_{\rm{floor, \, 1D}}^{2}$. The numerical factor on the right hand side depends weakly on the density profile of the cloud. The value of $5$ that we use here corresponds to a cloud with constant density; for comparison, in a cloud with a power-law density profile $\rho \propto r^{-2}$, this numerical factor would be $3$. 

The horizontal dotted line shows $\alpha = 1$, which corresponds to virial equilibrium, with $2 K + W = 0$, where $K$ and $W$ are the kinetic and gravitational potential energies, respectively. A cloud that is gravitationally bound, with $K + W < 0$, requires $\alpha < 2$. While we do find clouds with $\alpha \approx 1 - 2$ in our simulations, which are marginally bound (but not virialised), most have $\alpha > 2$, and thus are unbound. \citet{dobbs11} similarly found that most (but not all) GMCs in their simulations of isolated disc galaxies are gravitationally unbound. They attributed this to cloud-cloud collisions and stellar feedback, which regulate the velocity dispersion within the clouds. However, in our simulations, the lack of clouds with low virial parameters is partially due to the pressure floor, at least for masses $\la 3 \times 10^{5} \, \rm{M}_{\odot}$. We find a lower envelope of $\alpha \propto M^{-2/3}$ in Fig.~\ref{virialFig}, whereas observations find that $\alpha$ is approximately constant with mass \citep[e.g.][]{rosolowsky07}. This scaling of $\alpha$ with cloud mass is what we would expect when the virial parameter is determined by the pressure floor, with $\sigma_{\rm{floor, \, 1D}} \propto \rho^{1/6}$ (equation~\ref{sigma_floor_eqn}) and $R_{\rm{mean}} \propto M^{1/2}$ (as seen in Fig.~\ref{massSizeFig}). It is therefore apparent that the pressure floor prevents the low-mass clouds from becoming gravitationally bound in our simulations. 

Some observational studies also suggest that molecular clouds may be gravitationally unbound \citep[e.g.][]{heyer01}. \citet{dobbs11} also demonstrated that many of the GMCs in the sample of \citet{heyer09} have $\alpha > 2$ (see, for example, the centre bottom panel of fig. 1 of \citealt{dobbs11}). However, there are still some GMCs in this sample with $\alpha < 1$, which we do not see in our simulations. Furthermore, other studies suggest that molecular clouds may be marginally gravitationally bound, with $\alpha \approx 1$ \citep[see e.g.][]{mckee07}. 

\begin{figure}
\centering
\mbox{
	\includegraphics[width=84mm]{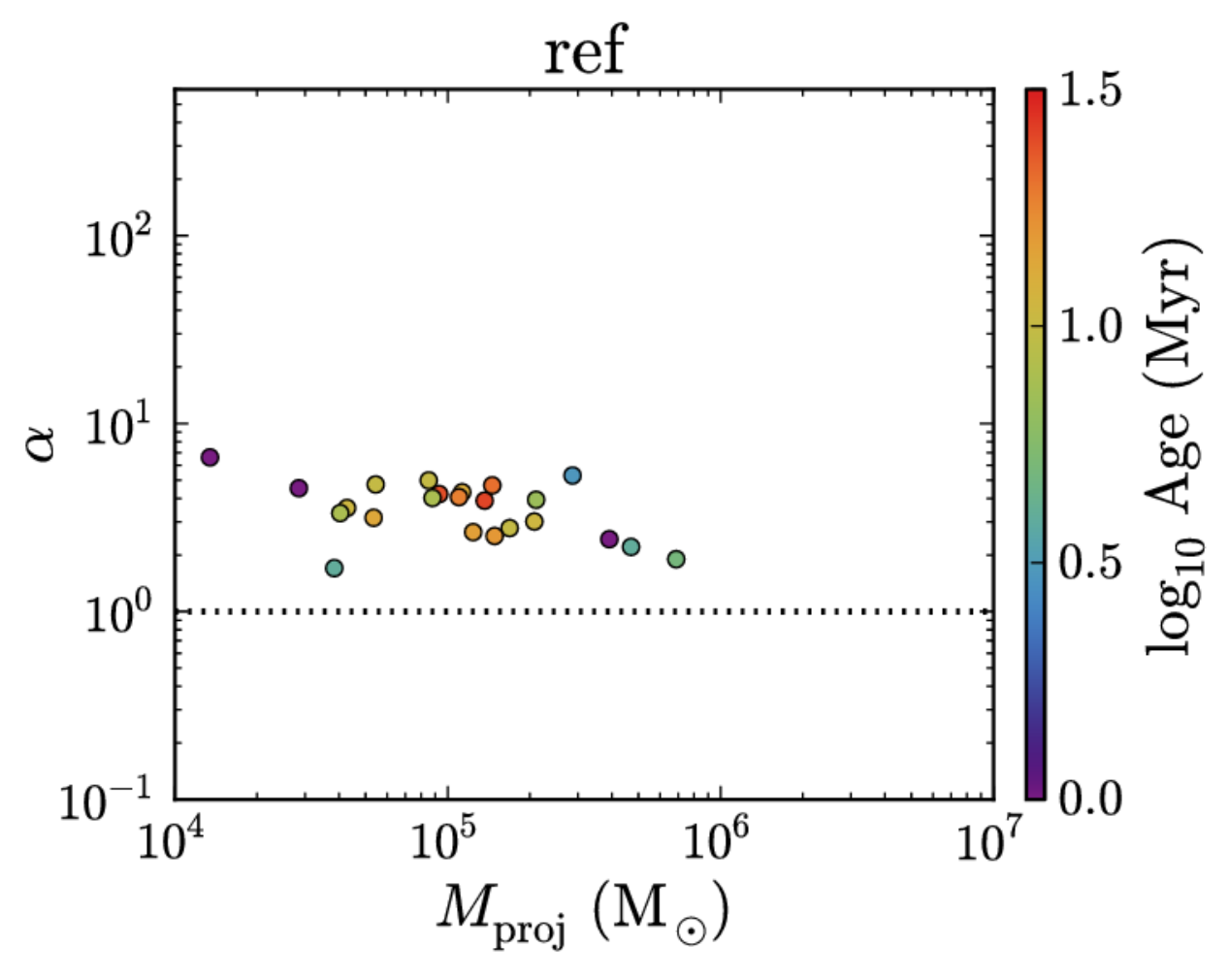}}
\caption{As Fig.~\ref{virialFig}, but measured from CO-detectable regions only, and with $\alpha$ calculated from velocity dispersions measured from the simulated CO spectra. Compared to Fig.~\ref{virialFig}, for a density-based cloud definition, the dependence of the lower envelope of $\alpha$ on cloud mass is weaker. However, we still find that all clouds are unvirialised ($\alpha > 1$), even for a CO-based cloud definition.} 
\label{virialThreshFig}
\end{figure}

To test how the pressure floor affects our results, we repeated the ref model twice, with the pressure floor lowered by factors of $4$ and $16$ in terms of the Jeans mass, corresponding to $N_{\rm{J, \, m}} = 1$ and $0.25$, respectively. We present these comparisons in Appendix~\ref{pfloor_appendix}, and we also present resolution tests, with the mass resolution increased/decreased by a factor of four, in Appendix~\ref{res_appendix}. We summarise the main results here. 

\begin{figure*}
\centering
\mbox{
	\includegraphics[width=134mm]{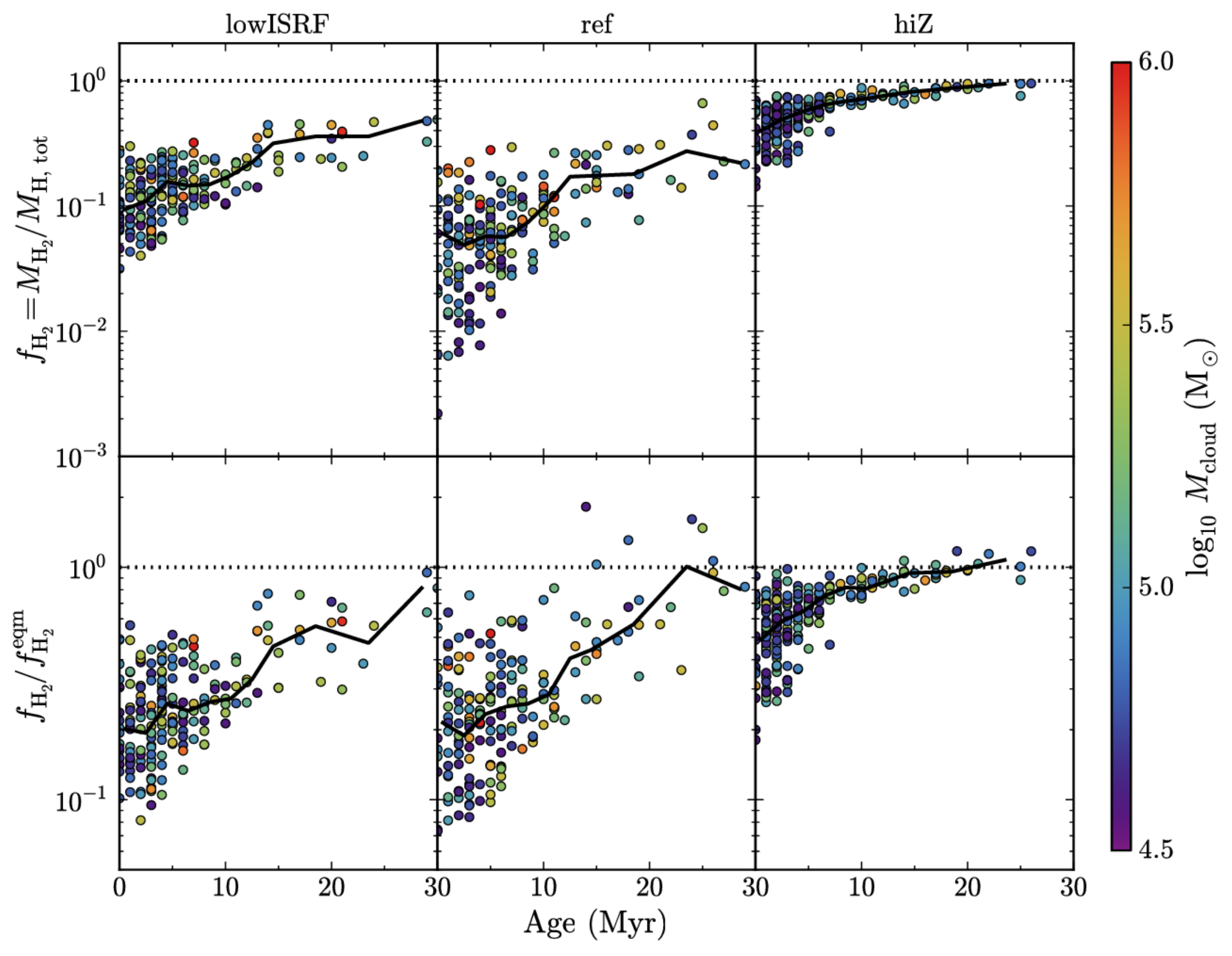}}
\caption{Molecular hydrogen fraction, $f_{\rm{H_{2}}}$ \textit{(top row)}, and the ratio of the H$_{2}$ fraction to the H$_{2}$ fraction in chemical equilibrium, $f_{\rm{H_{2}}} / f_{\rm{H_{2}}}^{\rm{eqm}}$ \textit{(bottom row)}, plotted against cloud age for all clouds with at least $50$ particles identified in snapshots at $100 \, \rm{Myr}$ intervals from $100 \, \rm{Myr}$ to $900 \, \rm{Myr}$ in our three simulations: lowISRF \textit{(left column)}, ref \textit{(centre column)} and hiZ \textit{(right column)}. The dotted horizontal lines indicate a value of unity, and the colour scale indicates cloud mass. We also show the median $f_{\rm{H_{2}}}$ or $f_{\rm{H_{2}}} / f_{\rm{H_{2}}}^{\rm{eqm}}$ in bins of age \textit{(solid curves)}. We see that $f_{\rm{H_{2}}}$ increases, and moves closer to chemical equilibrium, with increasing cloud age, and does so faster for higher metallicity.}  
\label{cloudH2Fraction}
\end{figure*}

As we lower the pressure floor, the low-mass ($\la 3 \times 10^{5} \, \rm{M}_{\odot}$), most poorly resolved clouds become more compact. They extend to lower values of $\alpha$ and can become strongly gravitationally bound, with $\alpha < 1$, and thus they can survive for longer, with ages up to $\approx 50 \, \rm{Myr}$. However, clouds with higher masses than this are unaffected by the pressure floor. This also means that the mass-size relation becomes flatter when we lower the pressure floor, and no longer agrees with the observed slope of this relation. Furthermore, these trends are not seen in our resolution tests. In particular, our highest resolution run reproduces the observed slope of the mass-size relation, and there are no clouds with $\alpha < 1$. Therefore, it is likely that the compact, long-lived clouds with $\alpha < 1$ that we find when we lower the pressure floor are artifacts of spurious fragmentation and collapse that may arise when we do not fully resolve the Jeans scale \citep[e.g.][]{bate97, truelove97}. 

Despite the differences that arise from lowering the pressure floor, we find that the median relations of molecule abundances with cloud age, and of CO intensity and $X_{\rm{CO}}$ factor with dust extinction, which we present for our fiducial simulations in the next two sections, are insensitive to the pressure floor, although the scatter in these relations does increase as we lower the pressure floor. These relations are also insensitive to resolution, although the scatter is higher at higher resolution. 

Fig.~\ref{virialThreshFig} shows the virial parameter plotted against cloud mass for our CO-based cloud definition, including only regions above the $I_{\rm{CO}}$ threshold, and using velocity dispersions measured by fitting a single Gaussian component to the simulated CO spectra, as described above. We show clouds from the ref simulation (using our fiducial pressure floor, with $N_{\rm{J, \, m}} = 4$), and we exclude those with multiple peaks in the CO spectrum, which cannot be fit by a single Gaussian component. Compared to Fig.~\ref{virialFig} (for a density-based cloud definition), the dependence of the lower envelope of $\alpha$ on cloud mass is weaker, which suggests that the impact of the pressure floor on the virial parameter is less severe when we use a CO-based cloud definition. However, we still find that all clouds in our simulations are unvirialised, with $\alpha > 1$. 

\section{Chemical evolution}\label{chemical_evolution_section}

\begin{figure*}
\centering
\mbox{
	\includegraphics[width=134mm]{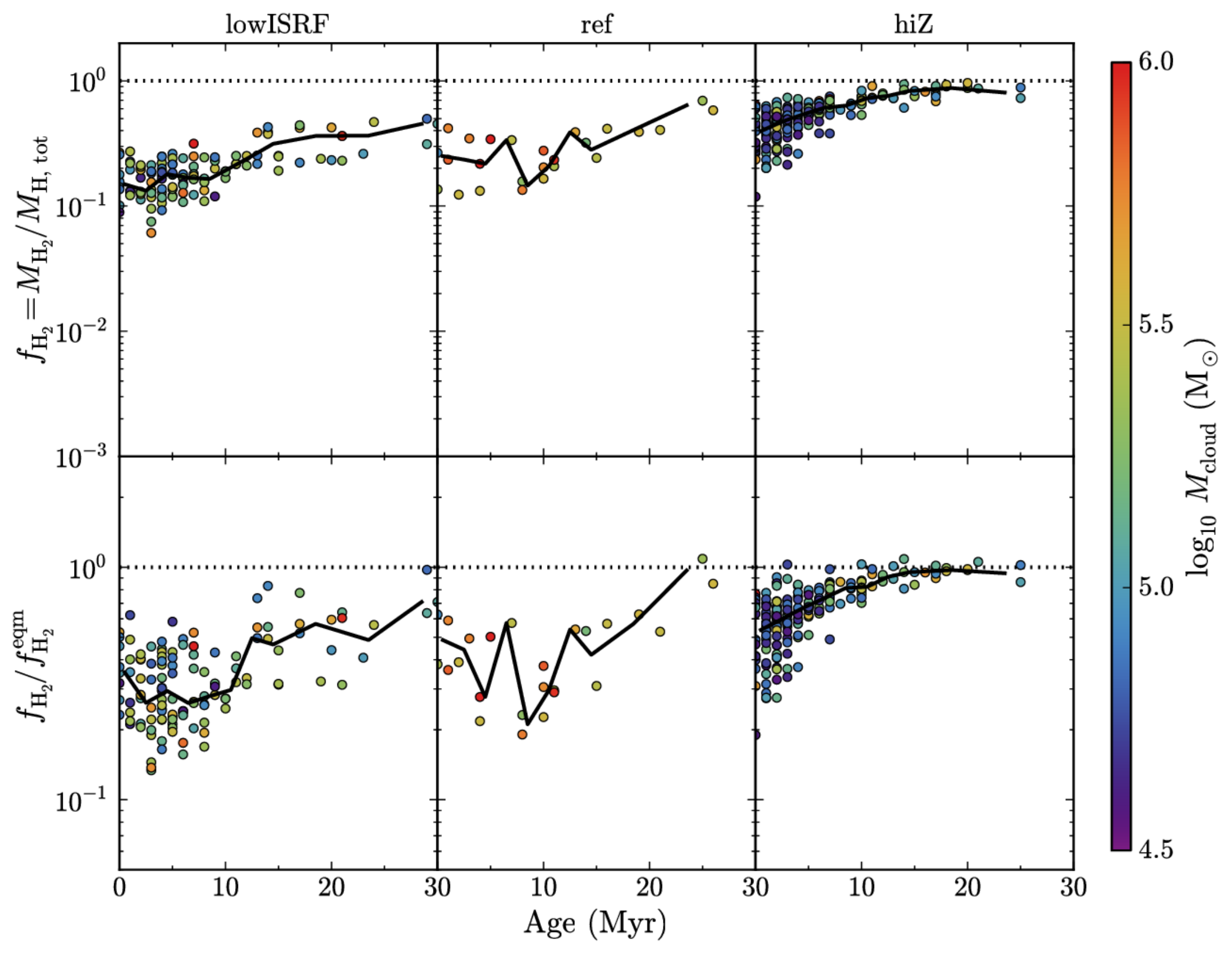}}
\caption{As Fig.~\ref{cloudH2Fraction}, but only for the region within each cloud with $I_{\rm{CO}} > 0.25 \, \rm{K} \, \rm{km} \, \rm{s}^{-1}$. By restricting the cloud definition to CO-detectable regions, we find less scatter in the H$_{2}$ fraction for the lowISRF and ref simulations, while the hiZ simulation is mostly unaffected.} 
\label{cloudH2FractionThresh}
\end{figure*}

We now look at the evolution of molecular abundances within the dense clouds that we have identified in our simulations. In particular, we investigate the time-scales over which these clouds become fully molecular, and whether they remain close to chemical equilibrium throughout their evolution. As in the previous section, we consider two cloud definitions: one based on a minimum density threshold ($n_{\rm{H}} > 10 \, \rm{cm}^{-3}$), and one based on a minimum velocity-integrated CO intensity threshold ($I_{\rm{CO}} > 0.25 \, \rm{K} \, \rm{km} \, \rm{s}^{-1}$). We consider two molecular species: H$_{2}$, which is the most prevalent molecule in interstellar gas, and CO, which is the most easily observed molecule. 

\subsection{Molecular hydrogen} 

Fig.~\ref{cloudH2Fraction} shows the molecular hydrogen fraction, $f_{\rm{H_{2}}} = M_{\rm{H_{2}}} / M_{\rm{H}, \, \rm{tot}}$ (where $M_{\rm{H_{2}}}$ is the mass of H$_{2}$ and $M_{\rm{H}, \, \rm{tot}}$ is the total mass of hydrogen in the cloud), for all clouds with at least $50$ particles identified in snapshots at $100 \, \rm{Myr}$ intervals from $100 \, \rm{Myr}$ to $900 \, \rm{Myr}$, using our density-based cloud definition. In the top row of Fig.~\ref{cloudH2Fraction} we plot $f_{\rm{H_{2}}}$ against the age of the cloud, and in the bottom row we plot the ratio $f_{\rm{H_{2}}} / f_{\rm{H_{2}}}^{\rm{eqm}}$, where $f_{\rm{H_{2}}}^{\rm{eqm}}$ is the molecular hydrogen fraction if all gas particles are set to chemical equilibrium. The three columns in Fig.~\ref{cloudH2Fraction} correspond to our three simulations (lowISRF, ref and hiZ), and the colour of each point indicates the mass of the cloud. The black curve in each panel shows the median in bins of age. 

The top row of Fig.~\ref{cloudH2Fraction} shows that the H$_{2}$ fraction increases with age, while there does not appear to be any significant trend with cloud mass. The simulation using solar metallicity (hiZ, right column) shows the highest H$_{2}$ fractions for a given cloud age. This is as expected, as we assume that the formation rate of H$_{2}$ on dust grains scales linearly with metallicity. In contrast, our reference simulation (ref, centre column), with a metallicity of ten per cent solar, shows the lowest H$_{2}$ fractions. 

In run hiZ, the median cloud H$_{2}$ fraction reaches $f_{\rm{H_{2}}} = 0.5$ after $\approx 3 \, \rm{Myr}$. In the ref simulation, with a factor of ten lower metallicity than hiZ, molecular hydrogen takes longer to build up, and only reaches $f_{\rm{H_{2}}} \approx 0.2$ after $\approx 30 \, \rm{Myr}$. Finally, in run lowISRF (left column), with ten per cent solar metallicity and ten per cent of the ISRF used in the other two simulations, the median H$_{2}$ fraction is always a factor $\approx 2$ higher than for ref. Thus, the time-scale for forming molecular H$_{2}$ in dense clouds is shorter at higher metallicity and (to a lesser extent) in the presence of a weaker UV radiation field. 

In the bottom row of Fig.~\ref{cloudH2Fraction}, we see that the H$_{2}$ fraction in young clouds is below what we would expect in chemical equilibrium. The clouds in the run at solar metallicity (hiZ) reach chemical equilibrium the fastest, with the median $f_{\rm{H_{2}}} / f_{\rm{H_{2}}}^{\rm{eqm}}$ already at $50$ per cent after $\approx 1 \, \rm{Myr}$ (which is the smallest time-scale that we show here, as we only have snapshots at $1 \, \rm{Myr}$ intervals). After $\approx 13 \, \rm{Myr}$, $f_{\rm{H_{2}}}$ has reached $90$ per cent of its equilibrium value. 

At lower metallicity, clouds take longer to reach chemical equilibrium. For example, clouds in the ref and lowISRF simulations reach $50$ per cent of the equilibrium H$_{2}$ fraction after $\approx 16 \, \rm{Myr}$, and they reach $90$ per cent after $\approx 22 \, \rm{Myr}$ and $\approx 30 \, \rm{Myr}$ respectively. In the ref simulation, clouds still have a low H$_{2}$ fraction ($f_{\rm{H_{2}}} \approx 0.2$) after $30 \, \rm{Myr}$, although they have reached chemical equilibrium by this time. In other words, these clouds are still not fully molecular, even in chemical equilibrium. This suggests that, in the reference simulation, the H\textsc{i}-to-H$_{2}$ transition, which depends on both metallicity and radiation field, lies further above the density threshold that we use to define our clouds ($n_{\rm{H}, \, \rm{min}} = 10 \, \rm{cm}^{-3}$) than in the other two simulations. In the ref simulation our definition of a dense cloud therefore includes a greater proportion of the H\textsc{i} envelope. 

In Fig.~\ref{cloudH2FractionThresh} we repeat Fig.~\ref{cloudH2Fraction}, but for our CO-based cloud definition, i.e. including only regions with $I_{\rm{CO}} > 0.25 \, \rm{K} \, \rm{km} \, \rm{s}^{-1}$. We compute $f_{\rm{H_{2}}}$ by projecting the H$_{2}$ column density onto the same image grid as was used for the CO emission maps, and selecting pixels above the $I_{\rm{CO}}$ threshold. 

\begin{figure*}
\centering
\mbox{
	\includegraphics[width=134mm]{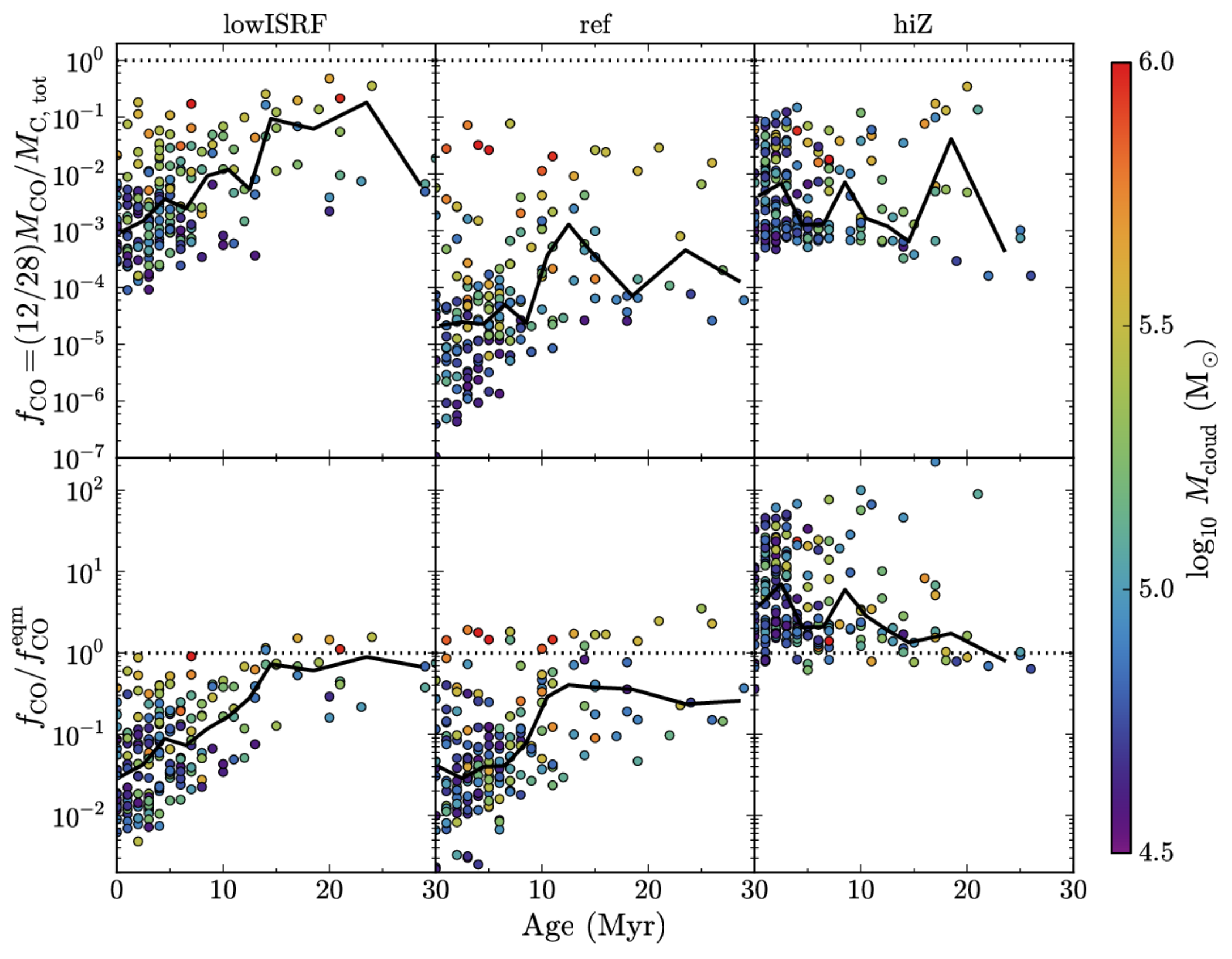}}
\caption{As Fig.~\ref{cloudH2Fraction}, but for the CO fraction, $f_{\rm{CO}} = (12/28) M_{\rm{CO}} / M_{\rm{C}, \, \rm{tot}}$ \textit{(top row)}, and the ratio of the CO fraction to the equilibrium value, $f_{\rm{CO}} / f_{\rm{CO}}^{\rm{eqm}}$. We define $f_{\rm{CO}}$ as the fraction, by mass, of carbon in CO molecules (hence the factor of (12/28)). The lowISRF and ref runs show increasing $f_{\rm{CO}}$ and $f_{\rm{CO}} / f_{\rm{CO}}^{\rm{eqm}}$ with cloud age and mass, while the hiZ run shows no strong trends. For lowISRF and ref the CO fraction in young clouds is typically lower than in equilibrium, whereas for hiZ it is typically higher than in equilibrium.}
\label{cloudCOFraction}
\end{figure*}

\begin{figure*}
\centering
\mbox{
	\includegraphics[width=134mm]{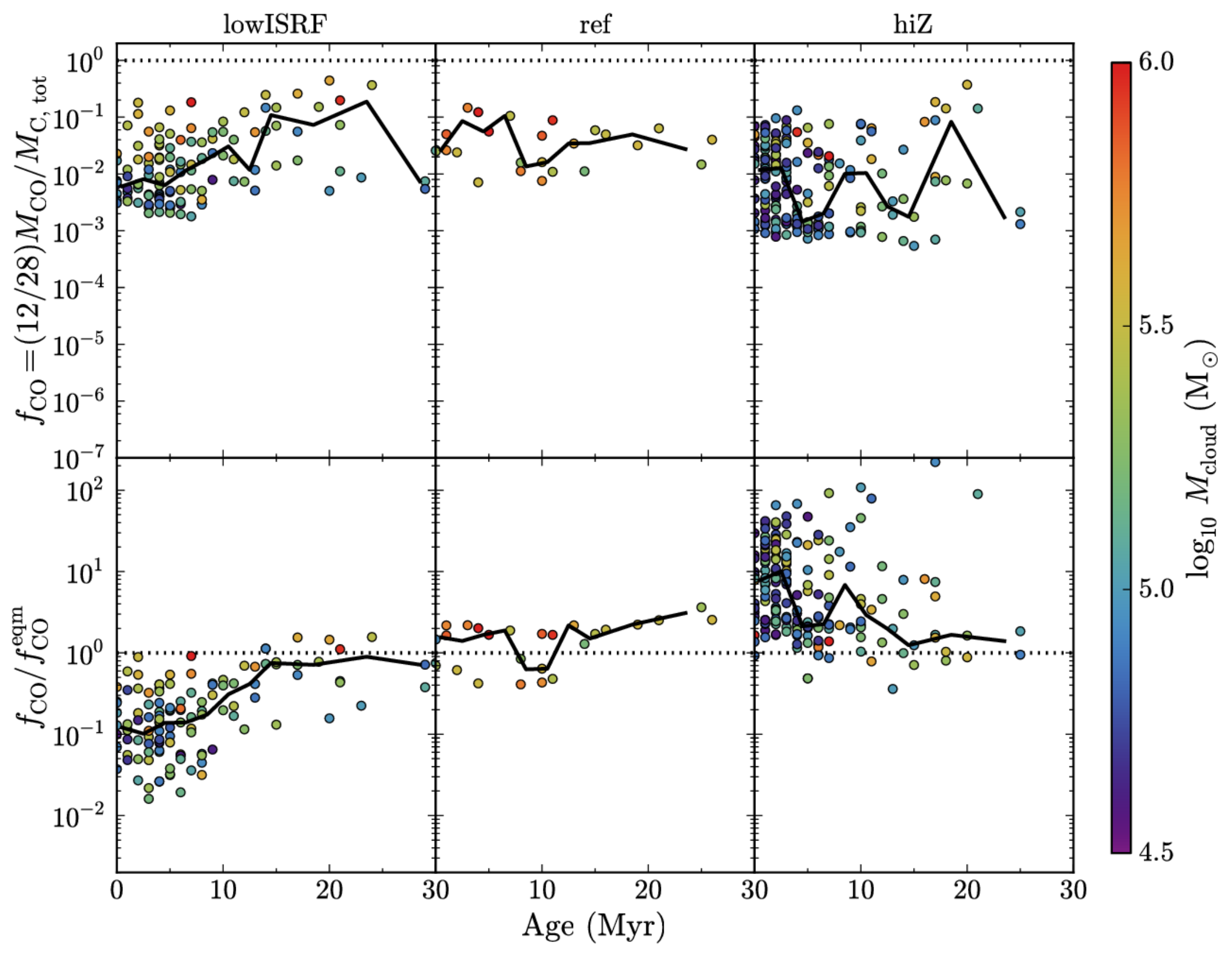}}
\caption{As Fig.~\ref{cloudCOFraction}, but only for the region within each cloud with $I_{\rm{CO}} > 0.25 \, \rm{K} \, \rm{km} \, \rm{s}^{-1}$. The CO fraction in CO-detectable regions in the ref simulation is now close to equilibrium, even in young clouds, whereas with our previous cloud definition it was an order of magnitude below equilibrium in clouds younger than $\approx 10 \, \rm{Myr}$. However, for lowISRF and hiZ the median CO fraction in young clouds remains an order of magnitude lower and higher than in equilibrium, respectively.} 
\label{cloudCOFractionThresh}
\end{figure*}

In the top row of Fig.~\ref{cloudH2FractionThresh}, the H$_{2}$ fraction in the lowISRF and ref simulations shows less scatter than we previously saw in Fig.~\ref{cloudH2Fraction}, and the values of $f_{\rm{H_{2}}}$ are higher, as we exclude the outer atomic envelope of the cloud. The ratio of H$_{2}$ fraction in non-equilibrium and H$_{2}$ fraction in equilibrium in the bottom row of Fig.~\ref{cloudH2FractionThresh} also shows less scatter. 

The lowISRF run shows similar trends with cloud age as previously, whereas the ref run shows weaker evolution with age, with H$_{2}$ fractions closer to equilibrium (within a factor $\approx 2 - 5$) in young clouds when we include only CO-detectable regions. The ref simulation contained the most CO-faint pixels, because its combination of low metallicity and high radiation field resulted in the lowest CO fractions (see Fig.~\ref{cloudCOFraction} in the next section). Therefore, restricting our cloud definition to CO-detectable regions has the strongest effect for the ref run. The CO-detectable regions are located in the dense cores of the clouds, which we would expect to reach chemical equilibrium faster, since collisional reaction rates typically scale with $n^{2}$, where $n$ is the density. This likely explains why the H$_{2}$ fraction in the ref simulation reaches equilibrium faster when we select only CO-detectable regions. 

In the hiZ simulation, the H$_{2}$ fraction is mostly unaffected by restricting the cloud definition to CO-detectable regions, with similar scatter and trends with age as in Fig.~\ref{cloudH2Fraction}.

One caveat to note is that these conclusions on the formation time-scale of H$_{2}$ may be sensitive to resolution. In particular, small-scale turbulence makes the gas form dense clumps, which increases the formation rate of H$_{2}$ in turbulent clouds \citep[e.g.][]{glover07b, micic12}. However, our simulations do not resolve this small-scale turbulence, even in the high resolution test in Appendix~\ref{res_appendix}, so it is likely that we underestimate the formation rate of H$_{2}$. 

\citet{krumholz11} compared the equilibrium H$_{2}$ model of \citet{krumholz08, krumholzetal09} and \citet{mckee10} to the non-equilibrium H$_{2}$ model of \citet{gnedin11}, applied to cosmological zoom-in simulations of a Milky Way progenitor galaxy and to a simulation of a cosmological box, $25 h^{-1} \, \rm{Mpc}$ on a side, run with the Adaptive Refinement Tree (ART) code \citep{kravtsov03}. They found excellent agreement at metallicities $\ga 10^{-2} \, \rm{Z}_{\odot}$, suggesting that non-equilibrium effects are unimportant for H$_{2}$ at the metallicities that we consider here. However, their simulations were run at a lower resolution than we use. For example, their cosmological zoom-in simulations had a maximum resolution of $65 \, \rm{pc}$, compared to a gravitational softening of $3.1 \, \rm{pc}$ in our simulations. 

\citet{hu16} explored the effects of non-equilibrium chemistry on the H$_{2}$ fraction in hydrodynamic simulations of dwarf ($M_{200} = 2 \times 10^{10} \, \rm{M}_{\odot}$) galaxies at a much higher resolution than we use ($4 \, \rm{M}_{\odot}$ per SPH particle). They find that the H$_{2}$ mass is out of equilibrium throughout their simulations, which agrees with our results. 

\subsection{Carbon monoxide} 

The top row of Fig.~\ref{cloudCOFraction} shows the mass fraction of carbon in CO, $f_{\rm{CO}} = \frac{12}{28} \frac{M_{\rm{CO}}}{M_{\rm{C,} \, \rm{tot}}}$, for each cloud as a function of cloud age, where $M_{\rm{CO}}$ and $M_{\rm{C,} \, \rm{tot}}$ are the CO and total carbon masses respectively, while the bottom row shows $f_{\rm{CO}} / f_{\rm{CO}}^{\rm{eqm}}$, where $f_{\rm{CO}}^{\rm{eqm}}$ is the CO mass fraction in chemical equilibrium. The colour scale indicates cloud mass, the black curves show the median in bins of age, and the left, centre and right columns show runs lowISRF, ref and hiZ respectively. 

In runs lowISRF and ref, which both assume $0.1 \, \rm{Z}_{\odot}$, we see that $f_{\rm{CO}}$ tends to increase with cloud age. However, there is more scatter in $f_{\rm{CO}}$ at fixed age than we saw for $f_{\rm{H_{2}}}$ in Fig.~\ref{cloudH2Fraction}. A handful of clouds reach $f_{\rm{CO}} \approx 0.5$ in the lowISRF run, but many are several orders of magnitude below unity. 

In the bottom left and bottom centre panels, we see that $f_{\rm{CO}}$ in young clouds ($\la 10 - 15 \, \rm{Myr}$) tends to be below equilibrium by $1-2$ orders of magnitude in the lowISRF and ref runs, while clouds older than this are typically closer to equilibrium, although for ref the median $f_{\rm{CO}} / f_{\rm{CO}}^{\rm{eqm}}$ is still only $0.2 - 0.5$. However, these non-equilibrium effects do not fully explain the very low CO fractions that we find in the top row. These low values of $f_{\rm{CO}}$ are partly due to the density threshold, $n_{\rm{H,} \, \rm{min}} = 10 \, \rm{cm}^{-3}$, that we use to define a cloud. This is close to the density of the H\textsc{i}-to-H$_{2}$ transition, which can occur once H$_{2}$ becomes self-shielded. However, CO forms once it becomes shielded from dissociating radiation by dust, which typically occurs at higher densities. 

The lowISRF and ref runs also show trends of $f_{\rm{CO}}$ with cloud mass, with more massive clouds showing higher CO fractions that are closer to equilibrium. This is because massive clouds are more likely to contain higher density regions where dust shielding is sufficient to form CO. 

The simulation using solar metallicity (hiZ; right column) has higher CO fractions than the ref simulation. This is due to the higher dust abundance at higher metallicity, and hence stronger dust shielding from dissociating radiation. We see no strong trend of $f_{\rm{CO}}$ with cloud age in the hiZ simulation. In the bottom right panel, we see that the CO fraction is either close to equilibrium or enhanced, by up to two orders of magnitude in some cases. The enhanced CO fractions that we see in the hiZ run are due to fluctuations in the dust extinction seen by individual particles within the cloud. Particles with enhanced CO abundances had $A_{\rm{v}} \ga 1$ within the previous few Myr, but $A_{\rm{v}}$ has since declined. Since the photodissociation rate of CO decreases exponentially with $A_{\rm{v}}$, a small decrease in $A_{\rm{v}}$ can produce a large increase in photodissociation rate. However, it takes a finite time for the CO to be destroyed, thus we see enhanced CO abundances. We see much less enhancement of CO at lower metallicity (lowISRF and ref) because, in these runs, $A_{\rm{v}}$ rarely exceeds unity, thus CO rarely becomes fully shielded from dissociating radiation. 

\begin{figure*}
\centering
\mbox{
	\includegraphics[width=134mm]{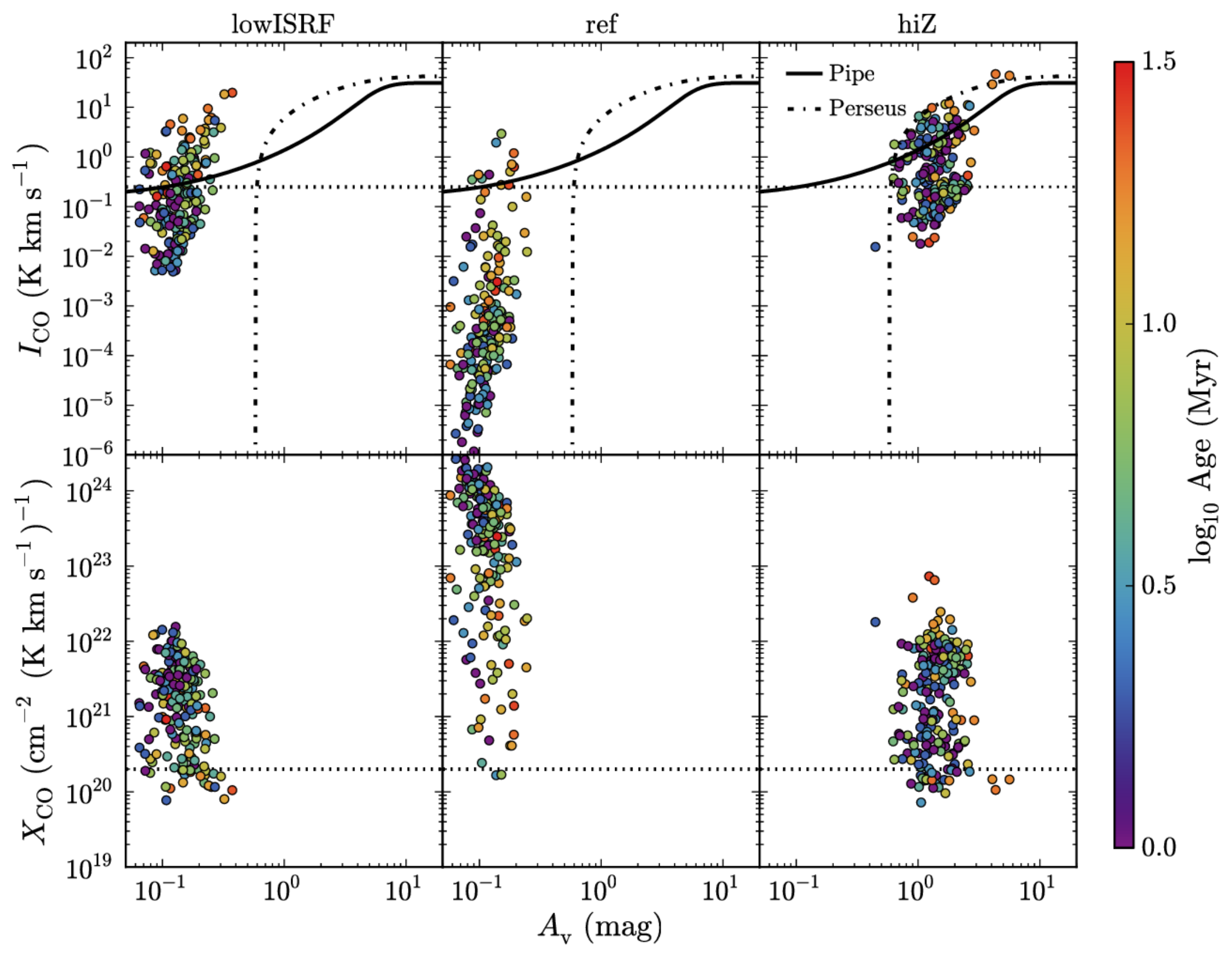}}
\caption{Mean velocity-integrated CO intensity, $I_{\rm{CO}}$ \textit{(top row)}, and $X_{\rm{CO}}$ factor \textit{(bottom row)}, plotted against mean dust extinction, $A_{\rm{v}}$, for clouds from the lowISRF \textit{(left)}, ref \textit{(centre)} and hiZ \textit{(right)} simulations. We include all clouds with at least $50$ particles in snapshots from $100 \, \rm{Myr}$ to $900 \, \rm{Myr}$, in $100 \, \rm{Myr}$ intervals. The colour scale indicates cloud age. In the top row, we also show the $I_{\rm{CO}} - A_{\rm{v}}$ relations observed in the Pipe nebula (\citealt{lombardi06}; solid curves) and the Perseus cloud (\citealt{pineda08}; dot-dashed curves) in the Milky Way. The horizontal dotted line in the top row indicates $I_{\rm{CO}} = 0.25 \, \rm{K} \, \rm{km} \, \rm{s}^{-1}$, which corresponds to the $3 \sigma$ intensity threshold for the Small Magellanic Cloud in the observations of \citet{leroy11}. In the bottom row, the horizontal dotted line indicates the typical value measured in molecular clouds in the Milky Way, $X_{\rm{CO}} = 2 \times 10^{20} \, \rm{cm}^{-2} \, (\rm{K} \, \rm{km} \, \rm{s}^{-1})^{-1}$ \citep[e.g.][]{bolatto13}.}
\label{co_Av_fig}
\end{figure*}

Fig.~\ref{cloudCOFractionThresh} shows CO fractions of clouds using a CO-based cloud definition, i.e. including only regions above the $I_{\rm{CO}}$ threshold. The effects of limiting our cloud definition to CO-detectable regions on CO fractions are similar to the effects it had on H$_{2}$ fractions that we saw in the previous section. In the lowISRF and ref runs, CO fractions are higher and show less scatter. The trends with cloud age in the lowISRF run are similar to those for a density-based cloud definition, while the ref simulation shows weaker evolution and is close to equilibrium, even in young clouds. The CO fractions in the hiZ run are mostly unaffected by the choice of cloud definition, and hence for young clouds they remain strongly enhanced compared to equilibrium. 

\section{CO emission and the $X_{\rm{CO}}$ factor}\label{co_emission_section}

Observations of molecular clouds often use CO emission as a tracer of molecular gas. The H$_{2}$ column density is then determined from the CO intensity using a conversion factor, $X_{\rm{CO}}$, as given in equation~\ref{xco_equation} (see \citealt{bolatto13} for a recent review). If the abundances of H$_{2}$ and CO are out of equilibrium in young clouds, as we found to be the case in the previous section, then this may affect the $X_{\rm{CO}}$ factor. To investigate this, we used our maps of CO emission from the clouds in our simulations to measure $X_{\rm{CO}}$. 

The CO properties of a cloud are expected to depend on the dust extinction, as dust shields the cloud from photodissociating radiation, enabling the formation of CO \citep[e.g.][]{lombardi06, pineda08, feldmann12, lee15}. Fig.~\ref{co_Av_fig} shows the mean velocity-integrated CO intensity ($I_{\rm{CO}}$; top row) and the mean $X_{\rm{CO}}$ factor (bottom row) in each cloud (using our density-based cloud definition) from our three simulations (lowISRF, ref and hiZ, in the left, centre and right columns respectively), as a function of the mean dust extinction, $A_{\rm{v}}$. In each cloud, we average $I_{\rm{CO}}$, $A_{\rm{v}}$ and the H$_{2}$ column density, $N_{\rm{H_{2}}}$, over all pixels within the projected ellipse containing the cloud particles, i.e. the white ellipses in Fig.~\ref{co_maps_fig}. The mean $X_{\rm{CO}}$ factor is then $\left< N_{\rm{H_{2}}} \right> / \left< I_{\rm{CO}} \right>$. The horizontal dotted line in the top row indicates $I_{\rm{CO}} = 0.25 \, \rm{K} \, \rm{km} \, \rm{s}^{-1}$, which corresponds to the $3 \sigma$ intensity threshold for the Small Magellanic Cloud in the observations of \citet{leroy11}, and is the minimum $I_{\rm{CO}}$ threshold that we use in our CO-based cloud definition. In the bottom row, the horizontal dotted line shows a value of $X_{\rm{CO}} = 2 \times 10^{20} \, \rm{cm}^{-2} \, (\rm{K} \, \rm{km} \, \rm{s}^{-1})^{-1}$, typical of molecular clouds in the Milky Way \citep{bolatto13}. The colour scale in both rows indicates cloud age. 

In the top row, we see that $I_{\rm{CO}}$ increases steeply with $A_{\rm{v}}$, particularly in the lowISRF and ref simulations. For comparison, we also show the observed $I_{\rm{CO}} - A_{\rm{v}}$ relations seen in the Pipe nebula \citep{lombardi06} and the Perseus cloud \citep{pineda08} in the Milky Way. The observations of \citet{pineda08} in particular find that $I_{\rm{CO}}$ cuts off at low $A_{\rm{v}}$, below a threshold $A_{\rm{v}} = 0.58$. This is unsurprising, as CO typically relies on dust to become shielded from dissociating radiation before it can form. Therefore, the steep $I_{\rm{CO}} - A_{\rm{v}}$ relation that we find in our simulations is likely due to this threshold effect, with most clouds lying close to the threshold. Since $A_{\rm{v}}$ depends on the column density, along with metallicity, this strong $I_{\rm{CO}} - A_{\rm{v}}$ relation reflects the fact that it is the column density, rather than the volume density, of a cloud that determines the molecular properties of the cloud, as this determines whether the cloud is shielded from dissociating radiation. 

At high $A_{\rm{v}}$, observations find that $I_{\rm{CO}}$ becomes saturated as the CO line becomes optically thick \citep[e.g.][]{lombardi06, pineda08}. The lowISRF and ref simulations do not extend above $A_{\rm{v}} \approx 0.4$ and so do not saturate, but the hiZ run contains clouds up to $A_{\rm{v}} \approx 6$. These high-$A_{\rm{v}}$ clouds in the hiZ run do suggest a much shallower relation than at lower $A_{\rm{v}}$ in the same simulation, and are consistent with the observed saturation in \citet{pineda08}, although we only have a few high-$A_{\rm{v}}$ clouds, so it is not clear if this relation is truely saturated in our simulations at high $A_{\rm{v}}$. 

Comparing the different panels in the top row, we see that the threshold $A_{\rm{v}}$ below which $I_{\rm{CO}}$ is strongly suppressed increases with metallicity (centre versus right) and, to a lesser extent, with the intensity of the radiation field (left versus centre). The dependence on radiation field is understandable, as a stronger radiation field requires a higher dust extinction before CO can become shielded. 

However, the reason for the dependence on metallicity is more complicated. If CO is shielded only by dust, then the dissociation rate decreases $\propto \exp(- \gamma_{\rm{d}} A_{\rm{v}})$, where $\gamma_{\rm{d}} = 3.53$ \citep{vandishoeck06}. The threshold, $A_{\rm{v}}^{\rm{thresh}}$, then arises from the exponential cut off due to shielding. The formation of CO proceeds via a series of reactions, so the overall formation rate, $R_{\rm{form}}$, will be determined by the rate-limiting step. These reactions are typically two-body interactions, so $R_{\rm{form}}$ scales with density squared. It also depends on the availability of carbon and oxygen, with densities $n_{\rm{C, tot}}$ and $n_{\rm{O, tot}}$ respectively, so $R_{\rm{form}} \propto n_{\rm{C, tot}} n_{\rm{O, tot}} \propto Z^{2} n_{\rm{H, tot}}^{2}$, where $Z$ is the metallicity and $n_{\rm{H, tot}}$ is the total hydrogen number density. However, the rate-limiting step may depend on only Oxygen or Carbon, and not both (e.g. if the formation of an intermediate species such as CH$_{2}^{+}$ is the slowest step), in which case $R_{\rm{form}} \propto Z n_{\rm{H, tot}}^{2}$. If we define $A_{\rm{v}}^{\rm{thresh}}$ to be when the CO fraction is some value, say $f_{\rm{CO}} = 0.1$, then $\exp(- \gamma_{\rm{d}} A_{\rm{v}}^{\rm{thresh}}) \propto Z^{i} n_{\rm{H, tot}}^{2}$, where $i = 1$ or $2$, depending on the rate-limiting step in the formation of CO. Since the clouds in all three of our simulations follow the same mass-size relation, and have the same distribution of cloud masses (Fig.~\ref{massFunctionFig}), the average cloud surface density and volume density are independent of $Z$. Thus, the metallicity dependence of $A_{\rm{v}}^{\rm{thresh}}$ is given by: 

\begin{equation}
A_{\rm{v}}^{\rm{thresh}} = -\frac{i}{\gamma_{\rm{d}}} \ln(Z) + \rm{constant}. 
\end{equation}
We therefore expect $A_{\rm{v}}^{\rm{thresh}}$ to decrease weakly with increasing metallicity, if the attenuation of CO photodissociation is due to dust shielding. However, this is opposite to what is seen in Fig.~\ref{co_Av_fig}. 

\begin{figure*}
\centering
\mbox{
	\includegraphics[width=134mm]{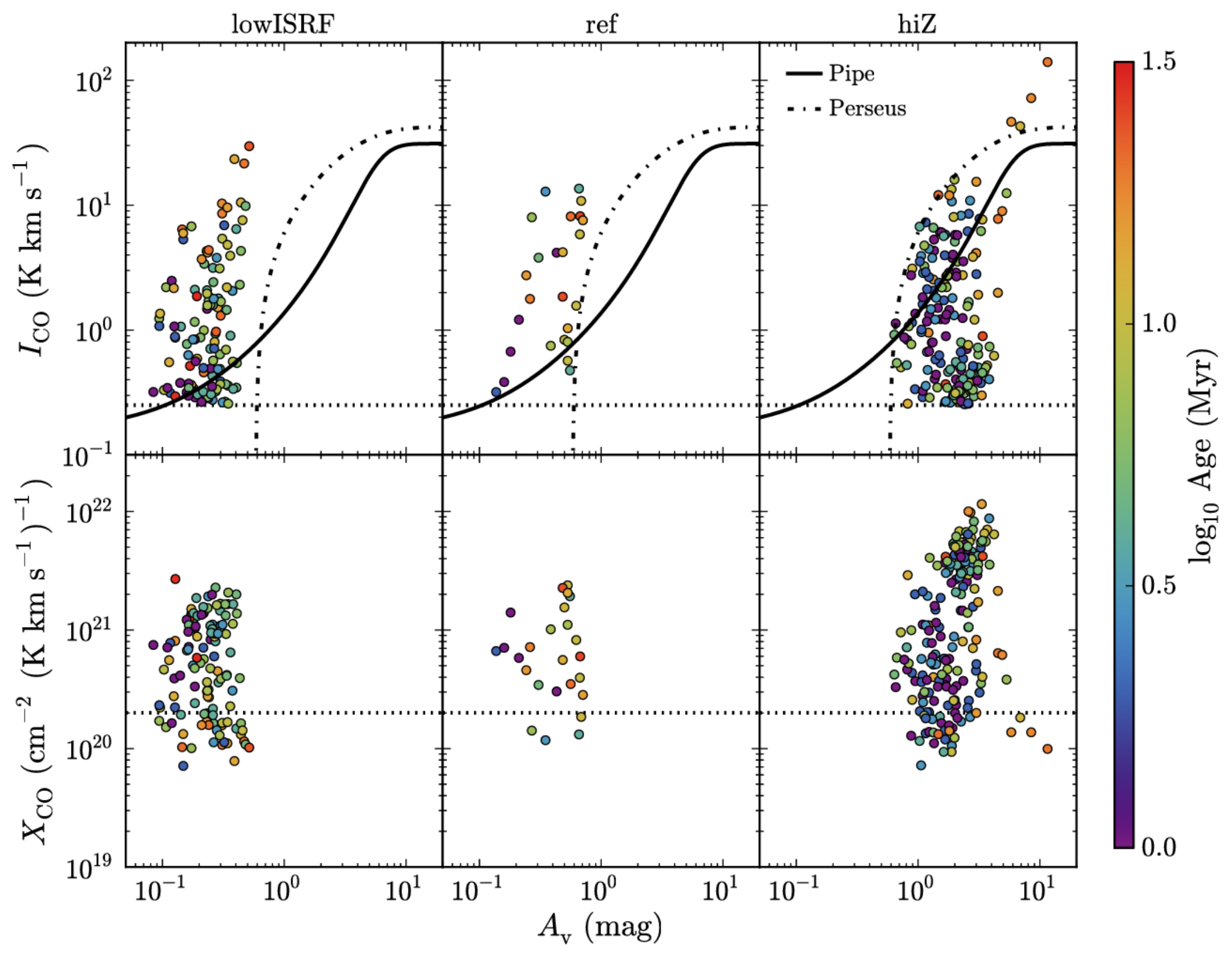}}
\caption{As Fig.~\ref{co_Av_fig}, but for $I_{\rm{CO}}$, $X_{\rm{CO}}$ and $A_{\rm{v}}$ averaged over only pixels with $I_{\rm{CO}} > 0.25 \, \rm{K} \, \rm{km} \, \rm{s}^{-1}$. Note that the ranges of the y-axes are much smaller than in Fig.~\ref{co_Av_fig}. There is much less scatter in $I_{\rm{CO}}$ and $X_{\rm{CO}}$ at fixed $A_{\rm{v}}$ compared to Fig.\ref{co_Av_fig}, as we exclude CO-faint pixels that would be undetectable in typical CO surveys of molecular clouds.}
\label{co_Av_thresh_fig}
\end{figure*}

The reason for this discrepancy is that, in the ref simulation, with ten per cent solar metallicity, the shielding of CO is primarily due to H$_{2}$, and not dust. H$_{2}$ shielding will cut off the CO dissociation rate at a threshold H$_{2}$ column density, $N_{\rm{H_{2}}}^{\rm{thresh}} = f_{\rm{H_{2}}} N_{\rm{H, tot}}^{\rm{thresh}}$, where $f_{\rm{H_{2}}}$ is the H$_{2}$ fraction of the cloud and $N_{\rm{H, tot}}^{\rm{thresh}}$ is the total hydrogen column density at the threshold. Then $A_{\rm{v}}^{\rm{thresh}} \propto Z N_{\rm{H, tot}}^{\rm{thresh}}$. If $N_{\rm{H, tot}}^{\rm{thresh}}$ is constant, $A_{\rm{v}}^{\rm{thresh}}$ will increase linearly with $Z$. However, $N_{\rm{H_{2}}}^{\rm{thresh}}$ decreases logarithmically with $Z$ due to the increased CO formation rate, as described above, and $f_{\rm{H_{2}}}$ is generally higher in the hiZ run (Fig.~\ref{cloudCOFraction}). Additionally, dust shielding becomes more important at high metallicity, which further reduces $A_{\rm{v}}^{\rm{thresh}}$, and CO self-shielding also plays a role in some clouds. We thus find a sub-linear increase in $A_{\rm{v}}^{\rm{thresh}}$ with $Z$. 

\citet{pineda08} find that the $I_{\rm{CO}} - A_{\rm{v}}$ relation in separate regions of the Perseus cloud also varies, suggesting that this relation depends on the physical conditions in the cloud. Using the Meudon PDR code \citep{lepetit06}, they find that the variations in the $I_{\rm{CO}} - A_{\rm{v}}$ relation that they observe can be explained by variations in physical conditions, particularly volume density and internal gas motions. They also find that the $I_{\rm{CO}} - A_{\rm{v}}$ relation moves to higher $A_{\rm{v}}$ in the presence of stronger radiation fields in their models, consistent with what we see in our simulations. However, they do not consider variations in metallicity, which we find to be more important. \citet{lee15} measure the $I_{\rm{CO}} - A_{\rm{v}}$ relation in the Large and Small Magellanic Clouds, and compare these to the Milky Way. They find that, at fixed $A_{\rm{v}}$, $I_{\rm{CO}}$ decreases with increasing dust temperature, suggesting a dependence on radiation field strength that is consistent with our simulations. However, they find that the $I_{\rm{CO}} - A_{\rm{V}}$ relations in these three galaxies are similar, despite their different metallicities. 

The bottom row of Fig.~\ref{co_Av_fig} shows a large range in $X_{\rm{CO}}$, spanning from two (lowISRF) to four (ref) orders of magnitude. We find no strong trends of $X_{\rm{CO}}$ with $A_{\rm{v}}$. However, if we look at the highest-$A_{\rm{v}}$ clouds in the hiZ run ($A_{\rm{v}} > 4$), the scatter in $X_{\rm{CO}}$ is much smaller, and the clouds lie within a factor two of the Milky Way value. We see a similar trend at high $A_{\rm{v}}$ when we lower the pressure floor in the ref run (see the bottom right panel of Fig.~\ref{IcoApp}). As we discuss further in Appendix~\ref{pfloor_appendix}, when we lower the pressure floor the clouds become more compact and extend to higher $A_{\rm{v}}$. In the run with the lowest pressure floor ($N_{\rm{J, \, m}} = 0.25$), the scatter in $X_{\rm{CO}}$ at $A_{\rm{v}} > 0.3$ is reduced by a factor four compared to the whole sample, and at $A_{\rm{v}} > 0.6$ the clouds are consistent with the Milky Way value of $X_{\rm{CO}}$. This suggests that the large scatter arises because the clouds are diffuse, with low $A_{\rm{v}}$, and the scatter is greatly reduced at high $A_{\rm{v}}$. However, this is based on a small number of clouds. 

\citet{bell06} find a strong relation between $X_{\rm{CO}}$ and $A_{\rm{v}}$ in their one-dimensional PDR models. However, the $X_{\rm{CO}} - A_{\rm{v}}$ relations that they consider show how $X_{\rm{CO}}$ varies with depth in a given cloud, whereas in Fig.~\ref{co_Av_fig} we show the mean $X_{\rm{CO}}$ and $A_{\rm{v}}$ for individual clouds. Indeed, \citet{bell06} show that their $X_{\rm{CO}} - A_{\rm{v}}$ relation varies with cloud properties such as density and turbulent velocity. The scatter that we find in $X_{\rm{CO}}$ in our simulations is therefore likely driven by the wide range of cloud properties in our sample. 

In the two runs at $0.1 \, \rm{Z}_{\odot}$ (lowISRF and ref), we find a trend of increasing $I_{\rm{CO}}$ with cloud age. This is consistent with the trend of increasing $f_{\rm{CO}}$ with age that we saw in Fig.~\ref{cloudCOFraction}. These two runs also show a trend of decreasing $X_{\rm{CO}}$ with increasing cloud age. The median $X_{\rm{CO}}$ factor in bins of age decreases by more than an order of magnitude for $0$ to $15 \, \rm{Myr}$, although there is a large scatter (two and four orders of magnitude in lowISRF and ref, respectively) in $X_{\rm{CO}}$ at fixed age. The trend in $X_{\rm{CO}}$ at ages $> 15 \, \rm{Myr}$ is uncertain, as there are few clouds at high ages. We see no strong trends of $I_{\rm{CO}}$ or $X_{\rm{CO}}$ with age at solar metallicity (hiZ), as the time-scales to reach equilibrium are shorter at higher metallicity, as we saw in Figs.~\ref{cloudH2Fraction} and \ref{cloudCOFraction}. 

In Fig.~\ref{co_Av_thresh_fig} we plot $I_{\rm{CO}}$ (top row) and $X_{\rm{CO}}$ (bottom row) versus $A_{\rm{v}}$ for our CO-based cloud definition, where these three quantities are now averaged over only pixels within the cloud with $I_{\rm{CO}} > 0.25 \, \rm{K} \, \rm{km} \, \rm{s}^{-1}$. Note that the ranges of the y-axes are much smaller than in Fig.~\ref{co_Av_fig}. In the lowISRF and ref simulations (left and centre columns), there is much less scatter in both $I_{\rm{CO}}$ and $X_{\rm{CO}}$ than we saw in Fig.~\ref{co_Av_fig}, for a density-based cloud definition. The reduction in scatter in the hiZ simulation is more modest. We also see more clearly how the $I_{\rm{CO}} - A_{\rm{v}}$ relation shifts towards higher $A_{\rm{v}}$ at higher metallicity and, to a lesser extent, at higher radiation field strength. While most high-$A_{\rm{v}}$ clouds in the hiZ run remain consistent with the observed saturation of $I_{\rm{CO}}$, there are now two clouds that lie a factor of $\approx 2-3$ above the observed relation. We again find no strong trends of $X_{\rm{CO}}$ with $A_{\rm{v}}$, except that clouds with $A_{\rm{v}} > 6$ in the hiZ run show much less scatter and are within a factor two of the Milky Way value. In the lowISRF run, the trend of median $X_{\rm{CO}}$ with cloud age is similar to that using a density-based cloud definition, while the trend of $X_{\rm{CO}}$ with age in the ref run is much weaker when we use a CO-based cloud definition. 

It is important to note that the values of the $X_{\rm{CO}}$ factor that we have presented in this section may be sensitive to resolution. In particular, as we noted in section~\ref{CO_map_methods}, high-resolution simulations of dense clouds \citep[e.g.][]{glover12} have found that most CO is concentrated in dense ($\sim 10^{3} \, \rm{cm}^{-3}$), compact ($\sim 1 \, \rm{pc}$) structures that are poorly resolved in our simulations. This will make the predicted CO emission, and hence the $X_{\rm{CO}}$ factor, uncertain. Our resolution tests in Appendix~\ref{res_appendix} do not show any significant change in the $I_{\rm{CO}} - A_{\rm{v}}$ or $X_{\rm{CO}} - A_{\rm{v}}$ relations, except that there is more scatter at higher resolution. However, even our highest resolution run, with a gravitational softening of $2 \, \rm{pc}$, cannot resolve the $\sim 1 \, \rm{pc}$ structures that are likely to dominate the CO emission. 

\citet{glover16} studied the emission of CO and C\textsc{i} in high resolution ($0.005 \, \rm{M}_{\odot}$ per SPH particle) simulations of individual molecular clouds with a range of metallicities, dust-to-gas ratios, UV radiation fields and cosmic ray ionisation rates. They found that, in low metallicity clouds ($Z \la 0.2 \, \rm{Z}_{\odot}$), $X_{\rm{CO}}$ decreases with time, while there is no strong evolution at high metallicity. This behaviour is similar to what we find in our simulations. They also demonstrated that C\textsc{i} is a better tracer of molecular gas than CO at low metallicity. 

\citet{smith14} studied the molecular gas and CO properties in high resolution simulations of Milky Way-type galaxies, run with the moving mesh code \textsc{arepo} \citep{springel10} and with time-dependent chemistry based on \citet{glover07a, glover07b} and \citet{nelson97}. They considered four simulations of galaxies with different surface densities and radiation field strengths, all assuming solar metallicity. The mean $X_{\rm{CO}}$ factor averaged over the galaxy in their four simulations was $3.89 - 13.1 \times 10^{20} \, \rm{cm}^{-2} \, (\rm{K} \, \rm{km} \, \rm{s}^{-1})^{-1}$ when all gas was included. However, $42 - 85$ per cent of the molecular mass in their simulations was `CO-dark', with a CO intensity $\leq 0.1 \, \rm{K} \, \rm{km} \, \rm{s}^{-1}$. When they included only gas with $I_{\rm{CO}} > 0.1 \, \rm{K} \, \rm{km} \, \rm{s}^{-1}$, the mean $X_{\rm{CO}}$ factor in their simulations was $1.48 - 2.28 \times 10^{20} \, \rm{cm}^{-2} \, (\rm{K} \, \rm{km} \, \rm{s}^{-1})^{-1}$. In other words, when they included only CO-bright regions, the mean $X_{\rm{CO}}$ factor decreased. This agrees with the trend that we find in our simulations. Furthermore, the values of the mean $X_{\rm{CO}}$ factor found by \citet{smith14}, averaged over the whole galaxy, overlap with the lowest values that we find for individual clouds, although we also find a large scatter in $X_{\rm{CO}}$ between clouds. 

\section{Conclusions}\label{conclusions} 

We have presented an analysis of GMCs identified in high-resolution ($750 \, \rm{M}_{\odot}$ per particle and a gravitational softening of $3.1 \, \rm{pc}$) SPH simulations of isolated, low-mass ($M_{\ast} \sim 10^{9} \, \rm{M}_{\odot}$) disc galaxies, with a particular emphasis on the evolution of molecular abundances and the implications for CO emission and the $X_{\rm{CO}}$ factor. Our simulations include a treatment for the non-equilibrium chemistry of $157$ species, including 20 molecules \citep{richings14a, richings14b}. 

We define dense clouds in our simulations using two different methods, one that is physically motivated and another that is observationally motivated. First, we define clouds to be regions with a density above a threshold of $n_{\rm{H}} = 10 \, \rm{cm}^{-3}$, which, depending on the metallicity and radiation field, is comparable to the density of the H\textsc{i}-to-H$_{2}$ transition \citep[e.g.][]{schaye01, gnedin09}. We group gas particles above this threshold into clouds using a Friends-of-Friends (FoF) algorithm, with a linking length of $l = 10 \, \rm{pc}$. 

The observationally motivated cloud definition is based on CO emission, where we restrict the cloud boundary to regions with a velocity-integrated CO intensity above a threshold of $0.25 \, \rm{K} \, \rm{km} \, \rm{s}^{-1}$. This allows for a fairer comparison with observations, as it excludes CO-faint regions that would be undetectable in typical CO surveys of molecular clouds, although it is still not a like-for-like comparison. 

To highlight the effects of metallicity and radiation field, we run our simulations at constant metallicity and with a uniform background UV radiation field, along with a prescription for local self-shielding by gas and dust \citep{richings14b}. Our three simulations cover two metallicities ($0.1 \, \rm{Z}_{\odot}$ and $\rm{Z}_{\odot}$) and two UV radiation fields (the ISRF of \citealt{black87}, measured in the local solar neighbourhood, and ten per cent of this ISRF). 

Our main results are as follows: 

\begin{enumerate}
\item Our simulated clouds have a median lifetime of $13 \, \rm{Myr}$ (Fig.~\ref{cloudAgeFig}), defined by the period over which at least half of the particles originally in the cloud when it was identified were in the main progenitor/descendant. This is consistent with observational estimates \citep[e.g.][]{bash77, kawamura09, murray11, miura12}, although see \citet{elmegreen00} and \citet{scoville79} for examples of shorter and longer observational estimates of GMC lifetimes, respectively. If we instead define the cloud lifetime by tracking the total mass of its main progenitor/descendant, rather than only the original particles, we find that clouds survive longer, with a median lifetime of $33 \, \rm{Myr}$, as new gas cycles through the cloud. 
\item Simulated clouds follow a mass-size relation $M \propto R^{2}$, as observed in molecular clouds \citep[e.g.][]{solomon87, romanduval10}. If we define clouds by a density threshold of $n_{\rm{H}} = 10 \, \rm{cm}^{-3}$ , then the normalisation is a factor $\approx 4$ below the observed relation (Fig.~\ref{massSizeFig}). However, if we restrict our cloud definition to CO-detectable regions, we find better agreement with observations, with a normalisation that is a factor $\approx 2$ below the observed relation (Fig.~\ref{massSizeThreshFig}). 
\item Clouds defined by a density threshold approximately follow the observed velocity dispersion-size relation if we account for the contribution of the pressure floor in the velocity dispersion (Fig.~\ref{velocityDispersionFig}), although there is a large scatter in the simulated relation. In particular, clouds within $1 \, \rm{kpc}$ of the galactic centre typically lie $0.5$ dex above the observed relation. We find better agreement with the observed relation when we use a CO-based cloud definition (Fig.~\ref{velocityDispersionThreshFig}). 
\item Most clouds in our simulations are gravitationally unbound, with a virial parameter $\alpha > 2$ (Fig.~\ref{virialFig}). While some clouds are marginally bound, with $\alpha \approx 1 - 2$, all clouds have $\alpha > 1$, i.e. no clouds are virialised. This is partially due to the pressure floor, at least in low-mass clouds ($\la 3 \times 10^{5} \, \rm{M}_{\odot}$). When we repeat the ref run with a lower pressure floor, reduced by a factor of 16 in terms of the Jeans mass, we do find some low-mass clouds with $\alpha < 1$ (Fig.~\ref{virialApp}). However, this may be an artifact of spurious fragmentation due to a poorly resolved Jeans scale, as we find no clouds with $\alpha < 1$ when we simultaneously increase the resolution and decrease the pressure floor (Fig.~\ref{virialRes}).
\item At ten per cent solar metallicity, young GMCs ($\la 10 - 15 \, \rm{Myr}$) are underabundant in H$_{2}$ and CO compared to chemical equilibrium, by factors $\approx 3$ and $1 - 2$ orders of magnitude respectively (Figs.~\ref{cloudH2Fraction} and \ref{cloudCOFraction}). These non-equilibrium effects are less apparent at solar metallicity. The H$_{2}$ fraction at solar metallicity reaches within a factor $2$ of equilibrium at $1 \, \rm{Myr}$, while the CO fraction at solar metallicity either remains close to equilibrium or becomes enhanced by up to two orders of magnitude compared to equilibrium. These non-equilibrium effects therefore depend strongly on metallicity, with no strong dependence on radiation field. 
\item If we restrict our analysis to CO-detectable regions (with $I_{\rm{CO}} > 0.25 \, \rm{K} \, \rm{km} \, \rm{s}^{-1}$), we find higher H$_{2}$ and CO fractions, as we exclude the atomic outer envelope of clouds (Figs.~\ref{cloudH2FractionThresh} and \ref{cloudCOFractionThresh}). The simulation with a low metallicity and a low UV radiation field (lowISRF) shows similar trends of $f_{\rm{H_{2}}}$ and $f_{\rm{CO}}$ with age as for our standard cloud definition, although the simulation at low metallicity and for a high UV radiation field (ref) shows weaker evolution of $f_{\rm{H_{2}}}$ and $f_{\rm{CO}}$ with age. 
\item The mean CO intensity, $I_{\rm{CO}}$, is strongly suppressed towards low dust extinction, $A_{\rm{v}}$, and may become saturated at high $A_{\rm{v}}$ (Fig.~\ref{co_Av_fig}), in agreement with observations \citep[e.g.][]{pineda08}. Our simulated $I_{\rm{CO}} - A_{\rm{v}}$ relation moves towards higher $A_{\rm{v}}$ at higher metallicities and, to a lesser extent, for stronger UV radiation fields. 
\item There is large scatter ($2-4$ orders of magnitude) in the mean $X_{\rm{CO}}$ factor of individual clouds (Fig.~\ref{co_Av_fig}). Clouds at high $A_{\rm{v}}$ show much less scatter in $X_{\rm{CO}}$ and are within a factor of $2$ of the Milky Way value (see also Fig.~\ref{IcoApp}). 
\item At ten per cent solar metallicity, we find weaker CO emission in young clouds, with ages $\la 10 - 15 \, \rm{Myr}$ (Fig.~\ref{co_Av_fig}), consistent with the trends we find for $f_{\rm{CO}}$. This is also reflected in the median $X_{\rm{CO}}$ factor in bins of cloud age, which decreases by more than an order of magnitude from $0$ to $15 \, \rm{Myr}$, although there is a large scatter in $X_{\rm{CO}}$ at fixed age. There are no strong trends with age at solar metallicity. 
\item By restricting our analysis to CO-detectable regions, we find less scatter in $X_{\rm{CO}}$ ($\approx 1 - 2$ orders of magnitude; Fig.~\ref{co_Av_thresh_fig}). We also find better agreement with observed GMC scaling relations (Figs.~\ref{massSizeThreshFig} and \ref{velocityDispersionThreshFig}). 
\end{enumerate} 

We have therefore shown that, at ten per cent solar metallicity, clouds younger than $\approx 10 - 15 \, \rm{Myr}$ are likely to be highly underabundant in H$_{2}$ and CO compared to chemical equilibrium. CO is more underabundant than H$_{2}$ in young clouds, which results in a trend of decreasing $X_{\rm{CO}}$ with increasing age from $0$ to $15 \, \rm{Myr}$, albeit with a large scatter. Clouds at solar metallicity reach chemical equilibrium faster (within $\approx 1 \, \rm{Myr}$). 

However, there are several caveats that we need to consider. Firstly, our simulations use a constant, uniform UV background. While we do include self-shielding by gas and dust, which varies both spatially and temporally, the interstellar radiation field itself should also vary, due to the birth and evolution of nearby stars and due to relative motions between the gas and the stars. If the fluctuations in the radiation field occur on time-scales shorter than the chemical time-scale, they can drive abundances out of equilibrium \citep[e.g.][]{oppenheimer13}. Additionally, radiative feedback from young, massive stars (due to photoionisation heating and/or radiation pressure) may disrupt their natal clouds and inhibit further formation of dense clouds \citep[e.g.][]{dale05, krumholz09, hopkins12, walch12, rosdahl15}. These effects are not included in our simulations. 

Secondly, the presence of turbulence can drive chemical abundances out of equilibrium,  if the time-scale associated with the turbulence is short compared to the chemical time-scale \citep[e.g.][]{gray15}. If we do not fully resolve small-scale turbulence in our simulations, we may therefore underestimate the effects of non-equilibrium chemistry. Conversely, the presence of unresolved turbulence will also create small-scale regions with higher densities than we resolve. The formation time-scales of molecules are shorter at higher densities, so they would reach equilibrium sooner in these unresolved high-density regions. This could lead us to overestimate the non-equilibrium chemistry. 

Furthermore, our simulations include a density-dependent pressure floor to ensure that we always resolve the Jeans mass by at least four times the kernel mass. However, in the lowest-mass clouds ($\la 3 \times 10^{5} \, \rm{M}_{\odot}$) this pressure floor prevents clouds from becoming strongly gravitationally bound ($\alpha < 1$). In Appendix~\ref{pfloor_appendix} we repeat our ref model with different pressure floors, to explore how this affects our results. As we lower the pressure floor, the lowest-mass, most poorly resolved clouds ($\la 2-3 \times 10^{5} \, \rm{M}_{\odot}$) become more compact, more strongly gravitationally bound, and can live longer (up to $\approx 50 \, \rm{Myr}$). However, higher-mass clouds are unaffected. This also means that the run with the lowest pressure floor no longer reproduces the observed slope of the mass-size relation. This is inconsistent with what we find when we increase the resolution (Appendix~\ref{res_appendix}). Therefore, it is likely that the increasingly compact and gravitationally bound low-mass clouds that we find when we lower the pressure floor are the result of artificial fragmentation and collapse that may arise when we do not fully resolve the Jeans scale \citep[e.g.][]{bate97, truelove97}. 

Despite these differences when we vary the pressure floor, we find that our main results for the median relations of molecule abundances with cloud age, and of CO intensity and $X_{\rm{CO}}$ factor with dust extinction, are insensitive to both the pressure floor and the resolution, although there is more scatter in these relations when we use a lower pressure floor or higher resolution.

A final caveat to note is that these conclusions depend on how the cloud age is defined. We have defined the cloud age from the time when half of the particles currently in the cloud were in a cloud progenitor. However, as we discussed in section~\ref{evolution_section}, alternative definitions can result in different ages. Furthermore, our definition requires that we trace individual gas particles back in time. This is trivial in SPH simulations, but is not possible in observations, for which we only have a single snapshot of the cloud at the present day. Observational estimates of GMC ages and lifetimes typically use nearby signatures of star formation, such as young stellar clusters and H\textsc{ii} regions \citep[e.g.][]{kawamura09}. We therefore need to be careful when comparing cloud ages from our simulations with observational estimates, as the two definitions may not be equivalent. 

\section*{Acknowledgments}
We are very grateful to Volker Springel for sharing \textsc{gadget3} and his inital conditions code, to Claudio Dalla Vecchia for allowing us to use \textsc{anarchy}, and to Benjamin Oppenheimer for code contributions. We thank Simon Glover and Ewine van Dishoeck for useful discussions. We gratefully acknowledge support from the European Research Council under the European Union's Seventh Framework Programme (FP7/2007-2013) / ERC Grant agreement 278594-GasAroundGalaxies. This work used the DiRAC Data Centric system at Durham University, operated by the Institute for Computational Cosmology on behalf of the STFC DiRAC HPC Facility (www.dirac.ac.uk). This equipment was funded by BIS National E-infrastructure capital grant ST/K00042X/1, STFC capital grant ST/H008519/1, and STFC DiRAC Operations grant ST/K003267/1 and Durham University. DiRAC is part of the National E-Infrastructure. This work also used computer resources provided by the Gauss Centre for Supercomputing/Leibniz Supercomputing Centre under grant:pr83le. We further acknowledge PRACE for awarding us access to resource Supermuc based in Germany at LRZ Garching (proposal number 2013091919). AJR is supported by the Lindheimer Fellowship at Northwestern University. 

{}

\appendix 

\section{Resolution tests}\label{res_appendix} 

In this paper we have focussed on the properties of GMCs in our simulations of isolated galaxies. These GMCs are among the smallest structures that we can resolve in these simulations, and thus are close to our resolution limit. We therefore need to test how sensitive our results are to the resolution of our simulations. 

We repeated our ref model ($Z = 0.1 \, \rm{Z}_{\odot}$ and the ISRF of \citealt{black87}) twice, with a factor four higher and lower mass resolution (runs hiRes and lowRes, respectively). We also decreased/increased the gravitational softening by a factor $4^{1/3}$, with $\epsilon_{\rm{soft}} = 2.0 \, \rm{pc}$ and $5.0 \, \rm{pc}$ in the hiRes and lowRes runs, respectively. For these resolution tests, we set the pressure floor such that $N_{\rm{J, \, m}} = 4$, as for our fiducial resolution (see section~\ref{models_section}). See Appendix~\ref{pfloor_appendix} for the effects of varying the pressure floor at fixed resolution. We ran the hiRes and lowRes simulations for $500 \, \rm{Myr}$, then we identified clumps of gas above a density threshold of $n_{\rm{H, \, min}} = 10 \, \rm{cm}^{-3}$ using a FoF algorithm as before (see section~\ref{clump_finder}), except that we used a FoF linking length of $6.3 \, \rm{pc}$ and $15.9 \, \rm{pc}$ in the hiRes and lowRes runs, respectively. These linking lengths correspond to the mean interparticle spacing at the density threshold for the given resolution. 

\begin{figure}
\centering
\mbox{
	\includegraphics[width=70mm]{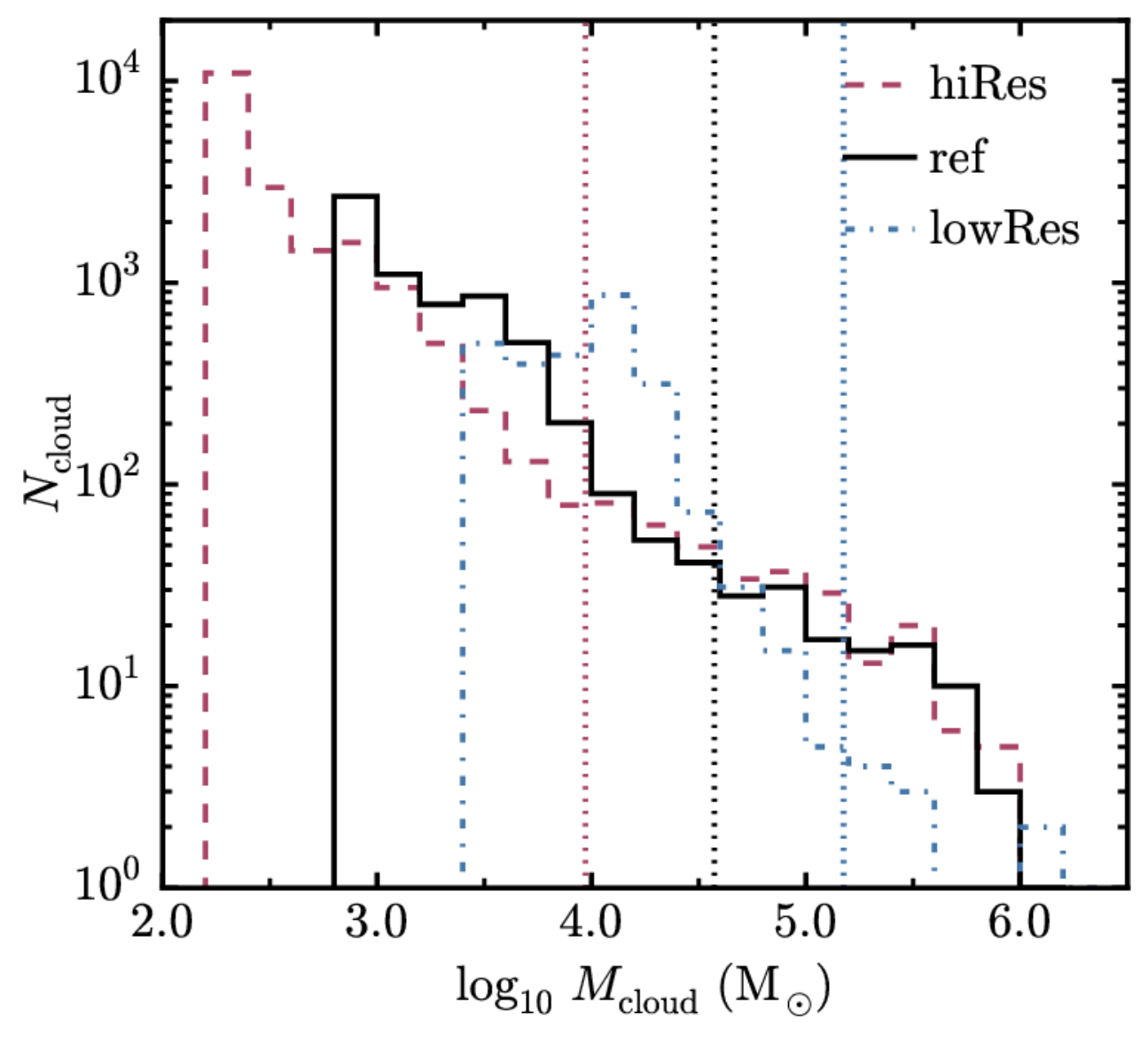}}
\caption{Cloud mass functions for the ref model with a mass of $187.5 \, \rm{M}_{\odot}$ \textit{(hiRes; red dashed curve)}, $750 \, \rm{M}_{\odot}$ \textit{(ref; black solid curve)} and $3000 \, \rm{M}_{\odot}$ \textit{(lowRes; blue dot-dashed curve)} per SPH particle. The vertical dotted lines indicate the cloud mass containing $50$ particles for each resolution. We include snapshots from $100$ to $400 \, \rm{Myr}$ at $100 \, \rm{Myr}$ intervals. The mass functions of the ref and hiRes runs agree at masses $\ga 10^{4} \, \rm{M}_{\odot}$.} 
\label{cloudFnRes}
\end{figure}

\begin{figure}
\centering
\mbox{
	\includegraphics[width=84mm]{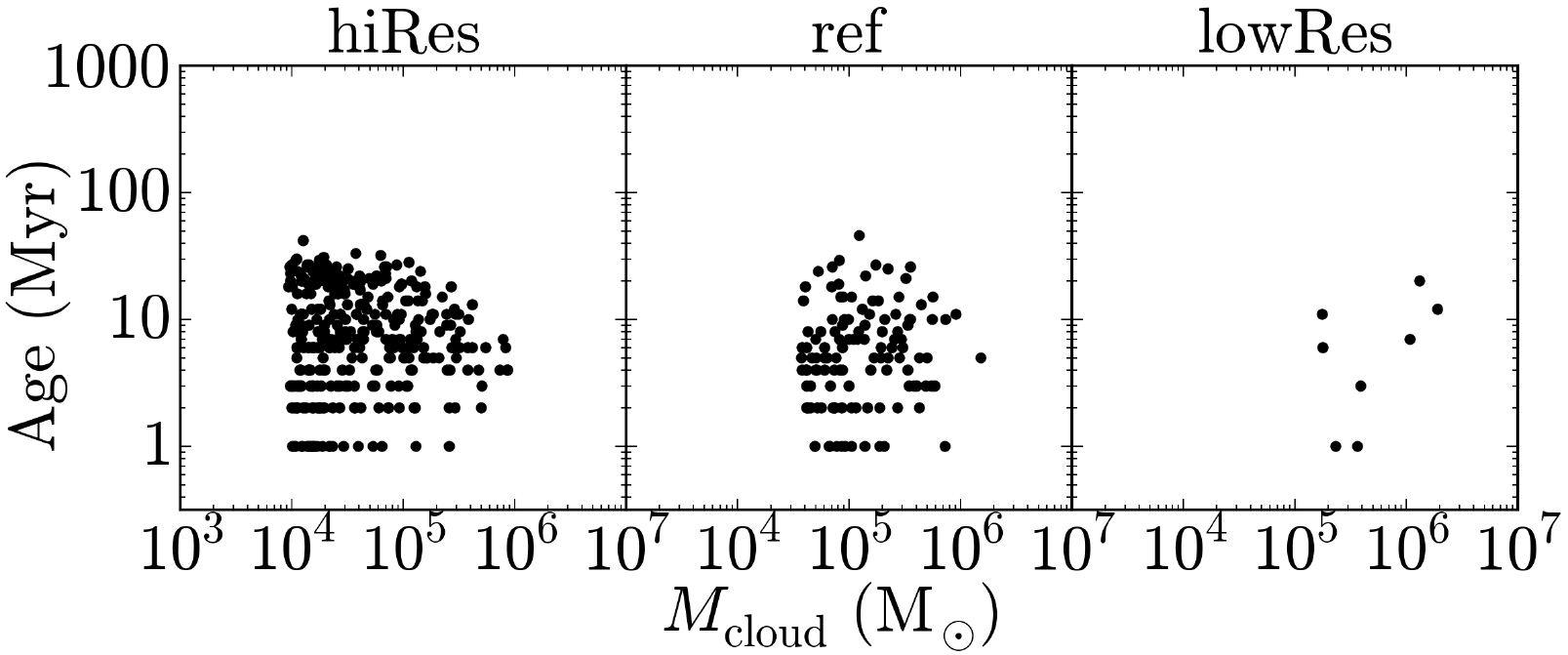}}
\caption{Cloud age, defined using the particles originally in the cloud in the current snapshot (see section~\ref{evolution_section}), versus mass for clouds in the ref model with a mass of $187.5 \, \rm{M}_{\odot}$, $750 \, \rm{M}_{\odot}$ and $3000 \, \rm{M}_{\odot}$ per SPH particle \textit{(left, centre and right panels, respectively)}. We include all clouds with at least 50 particles in snapshots at $100 \, \rm{Myr}$ intervals from $100$ to $400 \, \rm{Myr}$. The horizontal stripes of points at low ages are due to the finite interval ($1 \, \rm{Myr}$) between snapshots. The cloud ages do not depend strongly on the resolution of the simulation.} 
\label{cloudAgeRes}
\end{figure}

Fig~\ref{cloudFnRes} shows the cloud mass functions for the three resolutions. The vertical dotted lines indicate the cloud mass containing $50$ particles for each resolution. The mass function is steeper in the lowRes run, but the ref and hiRes runs agree at masses $\ga 10^{4} \, \rm{M}_{\odot}$. 

Fig.~\ref{cloudAgeRes} shows the cloud age as a function of cloud mass. The cloud age here uses our fiducial age definition, i.e. the time when half of the particles originally in the cloud in the current snapshot were in a progenitor of the cloud (see section~\ref{evolution_section}). We include all clouds in snapshots at intervals of $100 \, \rm{Myr}$ from $100$ to $400 \, \rm{Myr}$. The distribution of cloud ages is not strongly affected by resolution. 

\begin{figure}
\centering
\mbox{
	\includegraphics[width=84mm]{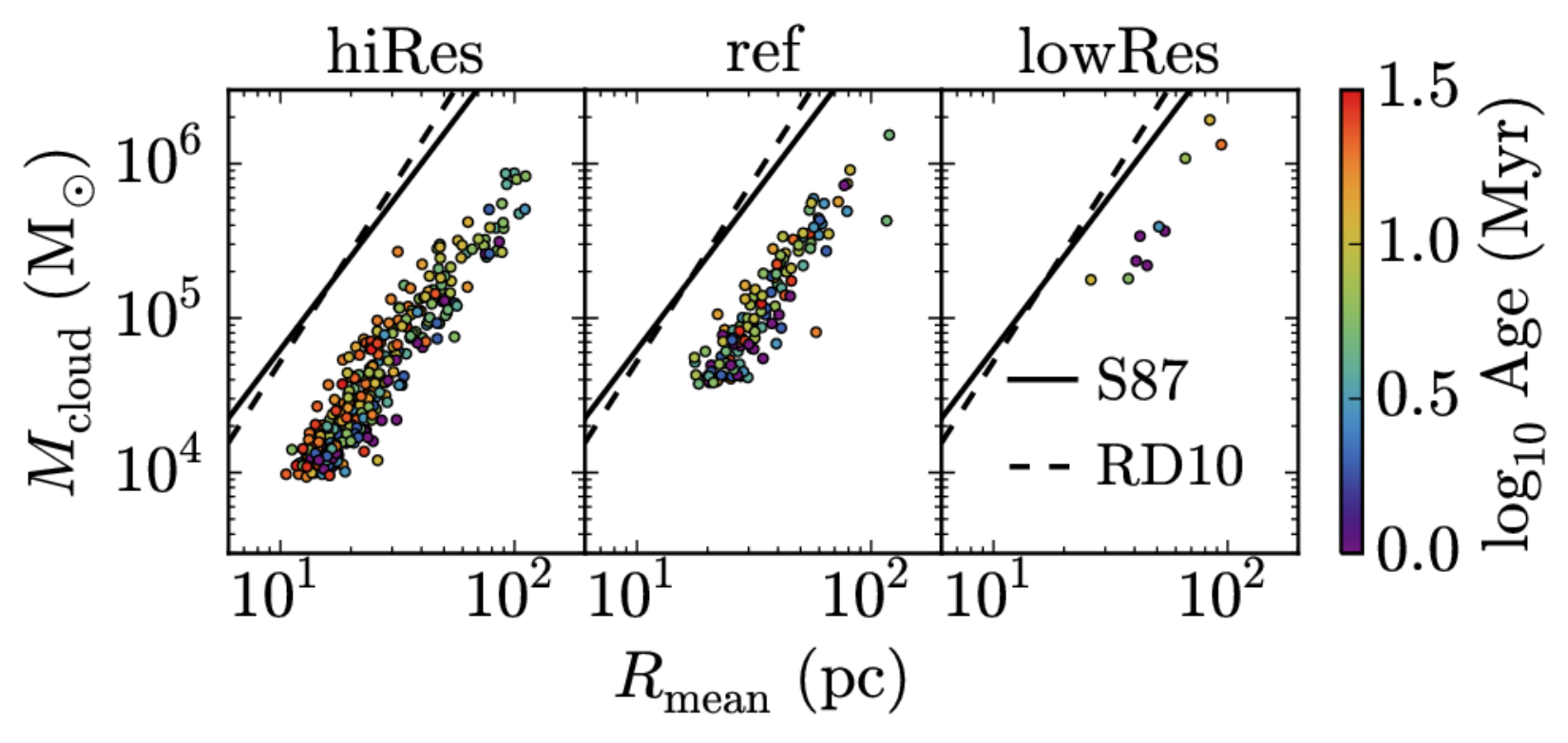}}
\caption{Mass-size relation for the ref model with three different particle masses: $187.5 \, \rm{M}_{\odot}$, $750 \, \rm{M}_{\odot}$ and $3000 \, \rm{M}_{\odot}$ \textit{(left, centre and right panels, respectively)}. The colour scale indicates the cloud age. We include all clouds with at least $50$ particles in snapshots at $100 \, \rm{Myr}$ intervals from $100$ to $400 \, \rm{Myr}$. We also show the observed relations from \citet{solomon87} \textit{(S87; solid cuves)} and \citet{romanduval10} \textit{(RD10; dashed curves}). The mass-size relation is insensitive to resolution.} 
\label{massSizeRes}
\end{figure}

Fig.~\ref{massSizeRes} shows the cloud mass-size relation for the three resolutions, from hiRes (left panel) to lowRes (right panel). The colour scale indicates the cloud age, and the solid and dashed lines show the observed relations of \citet{solomon87} and \citet{romanduval10}, respectively. All three simulations follow the same relation, with the same slope and normalisation. The hiRes simulation extends this relation to lower cloud masses and sizes, as it can resolve smaller clouds (and we include only clouds with at least $50$ particles). 

\begin{figure}
\centering
\mbox{
	\includegraphics[width=84mm]{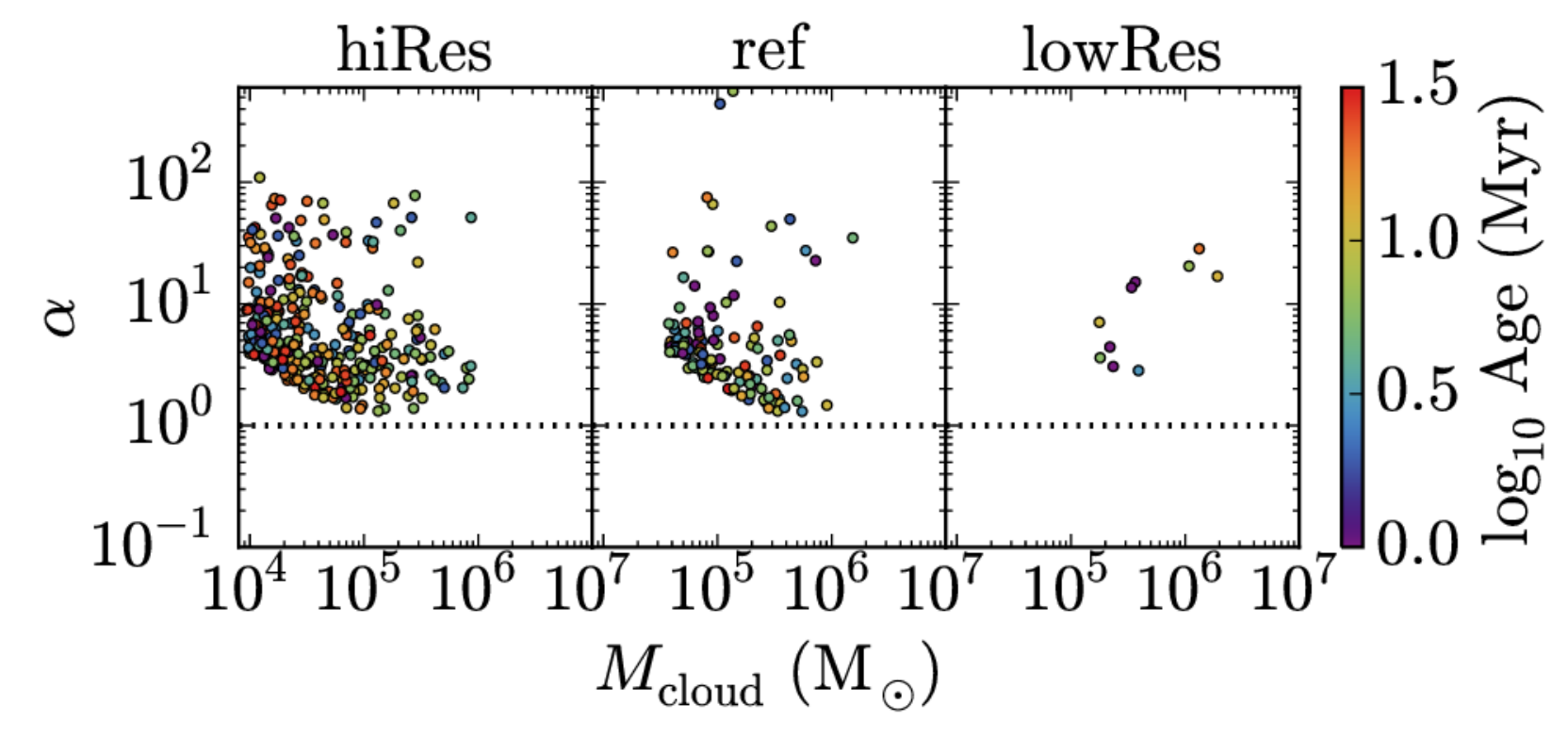}}
\caption{Cloud virial parameter, $\alpha$, versus mass, for the ref model with three different particle masses: $187.5 \, \rm{M}_{\odot}$, $750 \, \rm{M}_{\odot}$ and $3000 \, \rm{M}_{\odot}$ \textit{(left, centre and right panels, respectively)}. The colour scale indicates the cloud age, and the horizontal dotted lines show a value of $\alpha = 1$, which corresponds to virial equilibrium. As we increase the resolution, the lower envelope of $\alpha$ (which is influenced by the pressure floor) decreases. However, clouds with $M_{\rm{cloud}} \ga 10^{5} \, \rm{M}_{\odot}$ still have $\alpha > 1$, even in the hiRes run. These high-mass clouds are therefore genuinely unvirialised, and this is not an artifact of the pressure floor.} 
\label{virialRes}
\end{figure}

Fig.~\ref{virialRes} shows the virial parameter, $\alpha$, versus cloud mass for the three resolutions (different panels). The horizontal dotted line indicates $\alpha = 1$, which corresponds to virial equilibrium. In the ref and hiRes runs we clearly see a lower envelope of $\alpha \propto M^{-2/3}$. As discussed in section~\ref{scaling_relations_section}, this envelope is caused by the pressure floor, which prevents low-mass clouds from becoming gravitationally bound. As we increase the resolution, the envelope decreases, because the Jeans mass of the pressure floor (which we set to four times the kernel mass for these resolution tests) decreases. However, this envelope does not extend below $\alpha = 1$. For example, in the hiRes run, the lower envelope caused by the pressure floor reaches $\alpha = 1$ at $M_{\rm{cloud}} \approx 10^{5} \, \rm{M}_{\odot}$. Above this mass, the minimum $\alpha$ in our simulated clouds is just above unity, independent of cloud mass, and no longer follows the scaling that would arise from the pressure floor. This suggests that, in our hiRes simulation, clouds with $M_{\rm{cloud}} \ga 10^{5} \, \rm{M}_{\odot}$ are genuinely unvirialised, with $\alpha > 1$, and this is not an artifact of the pressure floor. 

\begin{figure}
\centering
\mbox{
	\includegraphics[width=84mm]{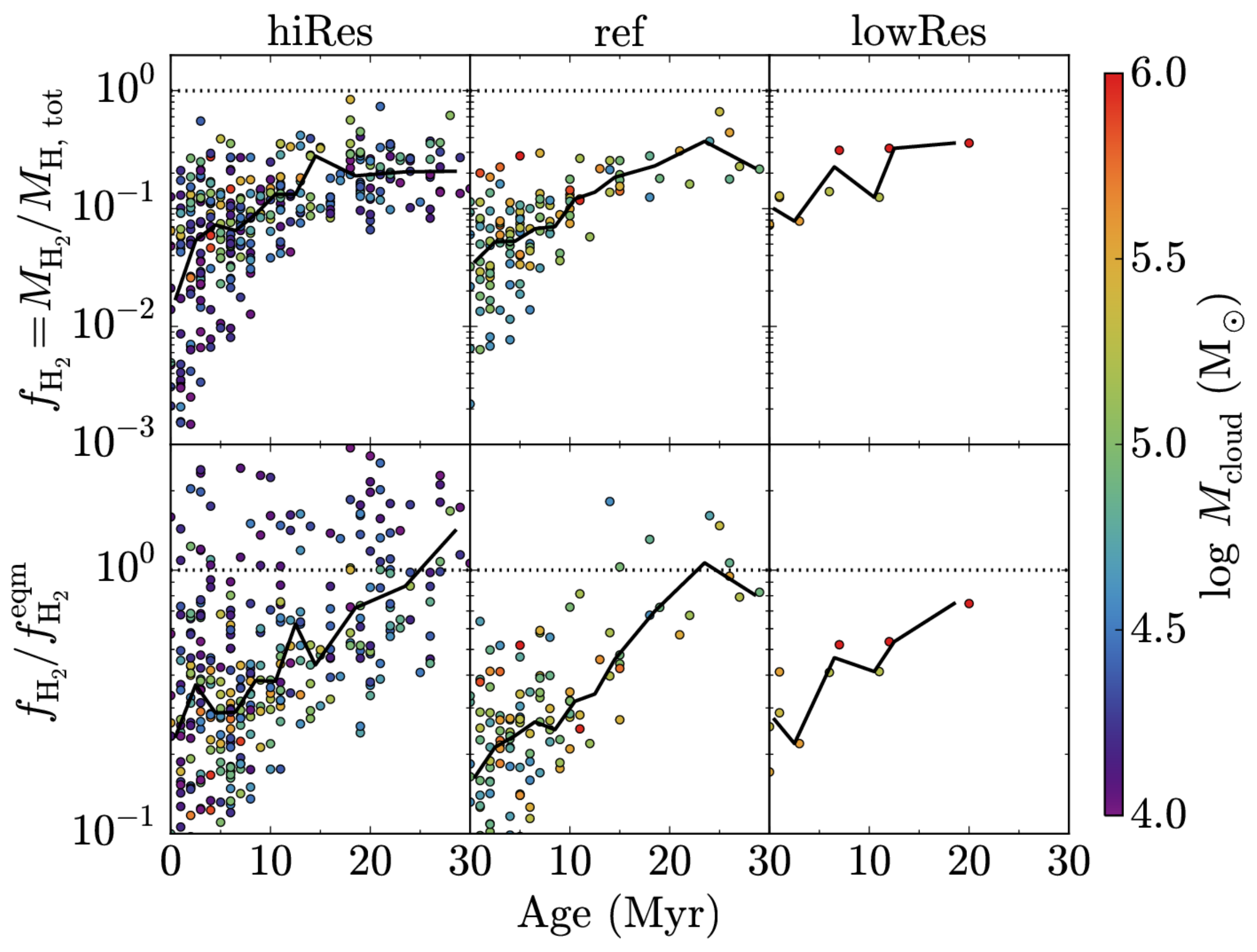}}
\caption{H$_{2}$ fraction, $f_{\rm{H_{2}}}$ \textit{(top row)}, and the ratio of $f_{\rm{H_{2}}}$ to the H$_{2}$ fraction in equilibrium, $f_{\rm{H_{2}}}^{\rm{eqm}}$, plotted against cloud age, for clouds in the ref model with three different particle masses: $187.5 \, \rm{M}_{\odot}$, $750 \, \rm{M}_{\odot}$ and $3000 \, \rm{M}_{\odot}$ \textit{(left, centre and right columns, respectively)}. The colour scale indicates the cloud mass, the solid curves indicate the median relation in bins of age, and the horizontal dotted lines indicate a value of unity. We include all clouds with at least $50$ particles in snapshots at $100 \, \rm{Myr}$ intervals from $100$ to $400 \, \rm{Myr}$. The median relations for the different resolutions agree, although there is more scatter in the hiRes run, which is primarily due to low-mass clouds ($\la 3 \times 10^{4} \, \rm{M}_{\odot}$) that are not present at lower resolutions.}
\label{H2Res}
\end{figure}

\begin{figure}
\centering
\mbox{
	\includegraphics[width=84mm]{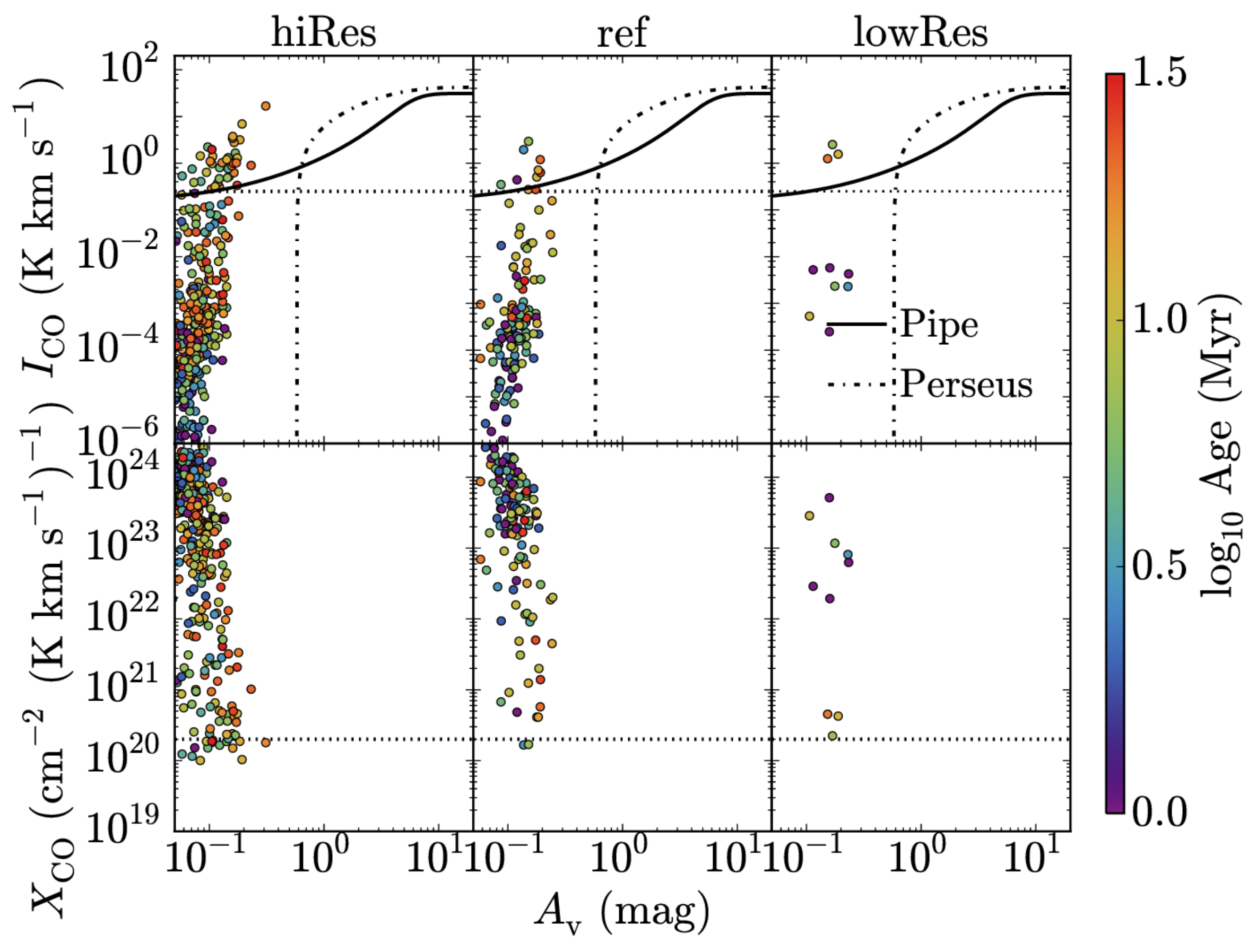}}
\caption{Mean CO intensity, $I_{\rm{CO}}$ \textit{(top row)}, and $X_{\rm{CO}}$ factor \textit{(bottom row)}, versus mean dust extinction, $A_{\rm{v}}$. The colour scale indicates the cloud age. In the top row, we also show the $I_{\rm{CO}} - A_{\rm{v}}$ relations observed in the Pipe nebula (\citealt{lombardi06}; solid curves) and the Perseus cloud (\citealt{pineda08}; dot-dashed curves). The horizontal dotted lines in the top row show a value of $I_{\rm{CO}} = 0.25 \, \rm{K} \, \rm{km} \, \rm{s}^{-1}$, which corresponds to the $3\sigma$ intensity threshold for the Small Magellanic Cloud in the observations of \citet{leroy11}. In the bottom row, the horizontal dotted lines show the typical value measured in the Milky Way, $X_{\rm{CO}} = 2 \times 10^{20} \, \rm{cm}^{-2} \, (\rm{K} \, \rm{km} \, \rm{s}^{-1})^{-1}$ \citep[e.g.][]{bolatto13}. The different resolutions show similar trends of $I_{\rm{CO}}$ and $X_{\rm{CO}}$ with $A_{\rm{v}}$.} 
\label{IcoRes}
\end{figure}

Fig.~\ref{H2Res} shows the H$_{2}$ fraction, $f_{\rm{H_{2}}}$ (top row), and the ratio of $f_{\rm{H_{2}}}$ to the equilibrium H$_{2}$ fraction, $f_{\rm{H_{2}}}^{\rm{eqm}}$ (bottom row), plotted against cloud age, for the three resolutions, from hiRes (left column) to lowRes (right column). The solid black curves show the median relations in bins of age, the horizontal dotted lines indicate a value of unity, and the colour scale indicates the cloud mass. All resolutions follow similar median relations, both for $f_{\rm{H_{2}}}$ versus age and $f_{\rm{H_{2}}} / f_{\rm{H_{2}}}^{\rm{eqm}}$ versus age. However, there is more scatter at higher resolution. In the hiRes run (left-hand column), the scatter is particularly large for low-mass ($\la 3 \times 10^{4} \, \rm{M}_{\odot}$) clouds (dark blue/purple circles), which are not present at lower resolutions because we include only clouds with at least $50$ particles. 

The CO fractions (not shown) also follow similar median relations with cloud age at different resolutions, albeit with more scatter at higher resolution due to the low-mass clouds. 

Fig.~\ref{IcoRes} shows the mean CO intensity, $I_{\rm{CO}}$ (top row), and the mean $X_{\rm{CO}}$ factor (bottom row) versus the mean dust extinction, $A_{\rm{v}}$, for the three resolutions. All three resolutions follow similar trends of $I_{\rm{CO}}$ and $X_{\rm{CO}}$ with $A_{\rm{v}}$, and the threshold dust extinction, $A_{\rm{V}}^{\rm{thresh}}$, below which $I_{\rm{CO}}$ is strongly suppressed, is insensitive to the resolution. 

\section{Effects of the pressure floor}\label{pfloor_appendix} 

In our simulations we impose a density-dependent pressure floor (equation~\ref{floor_mass}) to ensure that the Jeans mass is always at least a factor $N_{\rm{J, \, m}} = 4$ times the kernel mass. However, we saw in Fig.~\ref{virialFig} that this pressure floor prevents the lowest-mass clouds ($\la 3 \times 10^{5} \, \rm{M}_{\odot}$) from becoming strongly gravitationally bound ($\alpha < 1$). While the pressure floor could represent sources of pressure that are not explicitly included in our models, such as unresolved turbulence, its functional form was motivated by numerical reasons, so we may overestimate the true pressure. 

Conversely, if we remove the pressure floor entirely, so that we no longer resolve the Jeans mass in cold, dense gas, we may experience artificial fragmentation and collapse in gas that should be Jeans-stable. \citet{bate97} considered the resolution requirements in SPH simulations that include self-gravity, applied to the collapse and fragmentation of molecular clouds. In simulations where the gravitational softening length, $\epsilon_{\rm{soft}}$, is smaller than the SPH smoothing length, $h_{\rm{sml}}$, which is the case for nearly all gas particles in our simulations (we use $\epsilon_{\rm{soft}} = 3.1 \, \rm{pc}$ for the ref model), they found that, if the Jeans mass is not resolved by at least two times the kernel mass, gas can artificially undergo collapse when it should be Jeans-stable. The reason is that the thermal pressure is smoothed on scales of $h_{\rm{sml}}$, while the gravitational force is smoothed on scales of $\epsilon_{\rm{soft}}$. Therefore, once a gas cloud becomes unresolved, the pressure force will be smoothed out before the gravitational force. Hence it will artificially lose pressure support against gravity and will undergo gravitational collapse. 

A number of approaches have been proposed in the literature to alleviate these problems of artificial fragmentation that may occur when the Jeans scale is unresolved. \citet{booth07} developed a prescription for star formation and feedback in disc galaxies in which unresolved molecular gas is modelled by `sticky particles' that coagulate when they collide. \citet{narayanan11} also use a subgrid prescription to model the unresolved cold, molecular gas, which involves hybrid SPH particles that include both the warm and the cold ISM phases. \citet{schaye08} imposed a polytropic equation of state, $P \propto \rho^{\gamma_{\rm{eff}}}$, in their SPH simulations of disc galaxies. They used $\gamma_{\rm{eff}} = 4 / 3$ so that $m_{\rm{gas}} / M_{\rm{J}}$ and $h_{\rm{sml}} / L_{\rm{J}}$ are constant, where $m_{\rm{gas}}$ is the SPH particle mass, $h_{\rm{sml}}$ is the SPH smoothing length, and $M_{\rm{J}}$ and $L_{\rm{J}}$ are the Jeans mass and length, respectively. They set the normalisation of the equation of state such that $N_{\rm{J, \, m}} = 6$ in our nomenclature. \citet{robertson08} imposed a pressure floor in their SPH simulations of disc galaxies by setting a minimum internal energy (i.e. a temperature floor). The functional form of their temperature floor was equivalent to a polytropic equation of state with $P \propto \rho^{4/3}$. They explored a range of temperature floors, corresponding to $N_{\rm{J, \, m}} = 2 - 200$ (see their appendix A), and they used a fiducial floor with $N_{\rm{J, \, m}} = 30$. \citet{hopkins11} used the pressure floor of \citet{robertson08} in their SPH simulations of disk galaxies, with $N_{\rm{J, \, m}} = 10$ (although they also considered $N_{\rm{j, \, m}} = 4 - 15$, see their appendix A). 

However, not all simulation studies have imposed a Jeans limiter. For example, \citet{clark15} simply ended their SPH simulations of molecular clouds once the Jeans scale became unresolved. However, they used a very high resolution ($0.005 \, \rm{M}_{\odot}$ per particle and $50$ SPH neighbours), so they could follow the gravitational collapse up to a density of $\sim 10^{6} \, \rm{cm}^{-3}$. \citet{walch15} also did not include a Jeans limiter in their SPH simulations of supernova feedback in molecular clouds. The densities in some of their simulations extended up to $\sim 10^{6} \, \rm{cm}^{-3}$ (see their fig. 3), but they used a lower resolution than \citet{clark15}, with $0.1 \, \rm{M}_{\odot}$ per SPH particle, so the Jeans scale will be unresolved in some of their simulations. \citet{hu16} also did not include a Jeans limiter in their simulations of dwarf galaxies, although they did check that the resolution of their simulations ($4 \, \rm{M}_{\odot}$ per SPH particle, with 100 neighbours) is high enough that the Jeans scale is resolved in the majority of the gas. There are also other studies that do not explicitly include a Jeans limiter, for example \citet{gnedin11, glover11, shetty11a, shetty11b, dobbs13}. 

To explore how the pressure floor affects our results, we repeated the ref model twice, with the pressure floor reduced by factors of $4$ and $16$ in terms of the Jeans mass (i.e. with $N_{\rm{J, \, m}} = 1$ and $0.25$, respectively), using our fiducial resolution of $750 \, \rm{M}_{\odot}$ per SPH particle. These simulations were only run for $100 \, \rm{Myr}$, and we therefore compared clouds at $100 \, \rm{Myr}$ in these runs with the ref model, whereas previously we have been combining clouds from nine snapshots (or four snapshots for our resolution tests in Appendix~\ref{res_appendix}) at $100 \, \rm{Myr}$ intervals. 

\begin{figure}
\centering
\mbox{
	\includegraphics[width=70mm]{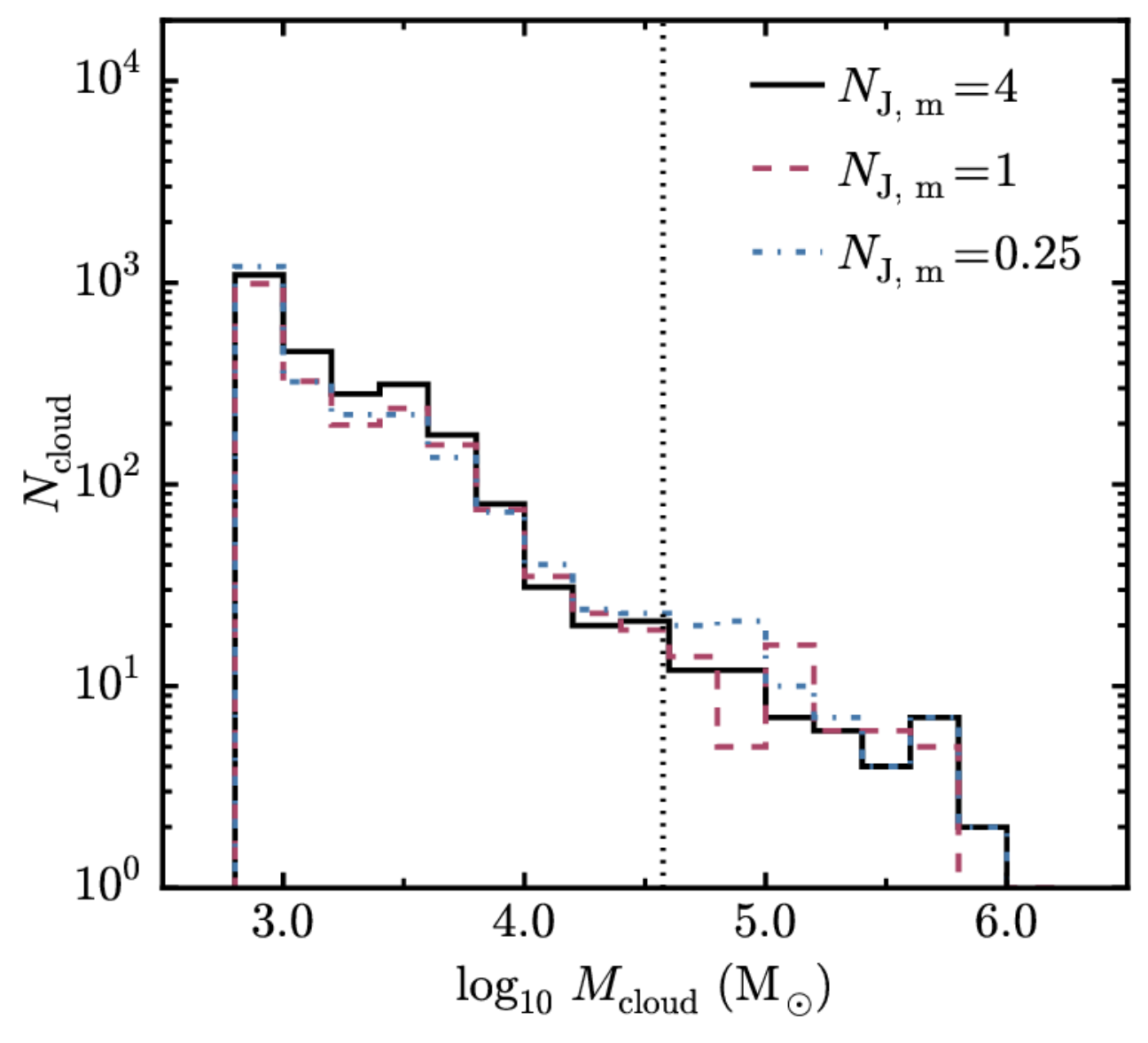}}
\caption{Cloud mass functions for the ref model, using a pressure floor corresponding to a constant Jeans mass with $N_{\rm{J, \, m}} = 4$ \textit{(black solid curve)}, $N_{\rm{J, \, m}} = 1$ \textit{(red dashed curve)} and $N_{\rm{J, \, m}} = 0.25$ \textit{(blue dot-dashed curve)}, where $N_{\rm{J, \, m}}$ is the ratio of the minimum Jeans mass to the mass within the SPH kernel. The cloud mass function is insensitive to the pressure floor.} 
\label{cloudFnApp}
\end{figure}

\begin{figure}
\centering
\mbox{
	\includegraphics[width=84mm]{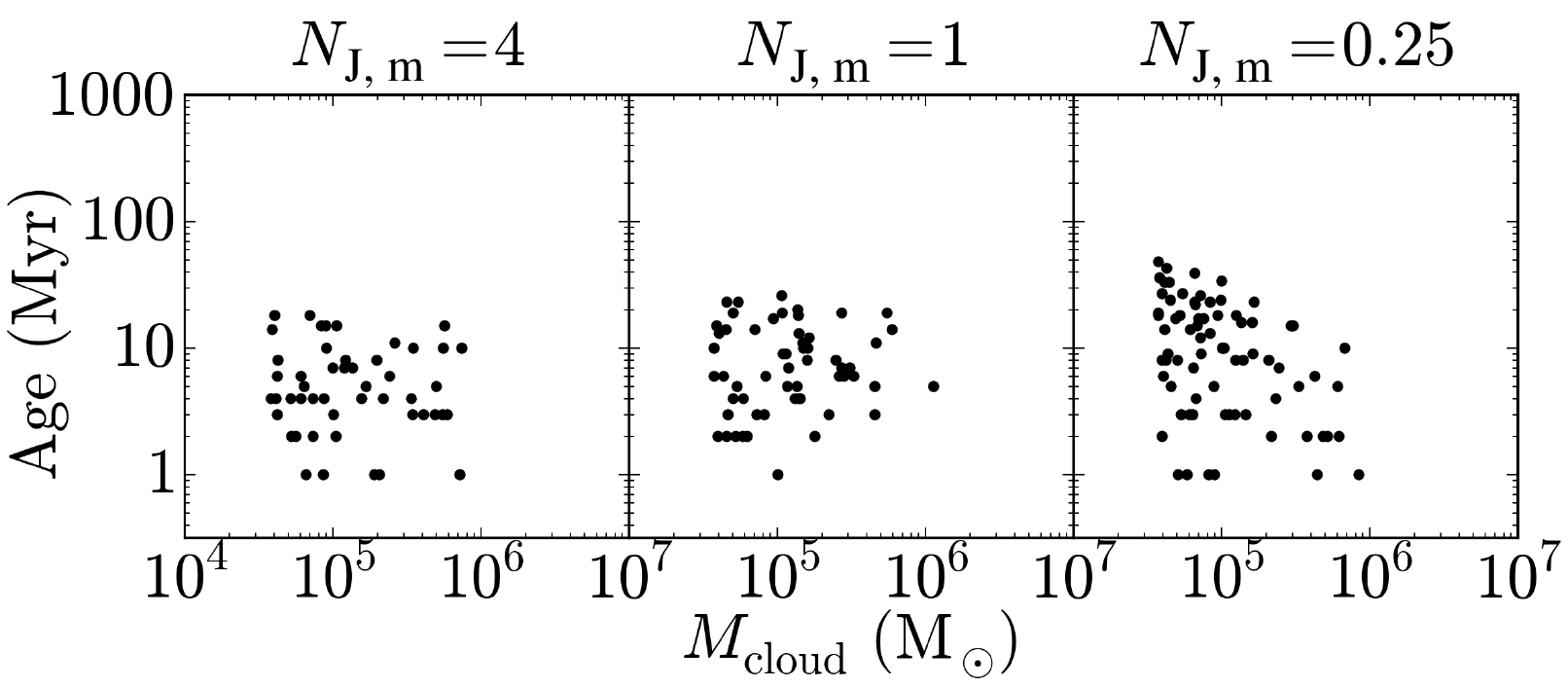}}
\caption{Cloud age, defined using the particles originally in the cloud in the current snapshot (see section~\ref{evolution_section}), versus mass for clouds in the snapshot at $100 \, \rm{Myr}$ in the ref model with three different pressure floors: $N_{\rm{J, \, m}} = 4$, $1$ and $0.25$ \textit{(left, centre and right panels, respectively)}. In the run with the lowest pressure floor (right panel), low-mass, poorly resolved clouds can survive for longer (up to $50 \, \rm{Myr}$) than in the run with our fiducial pressure floor (left panel).} 
\label{cloudAgeApp}
\end{figure}

\begin{figure}
\centering
\mbox{
	\includegraphics[width=84mm]{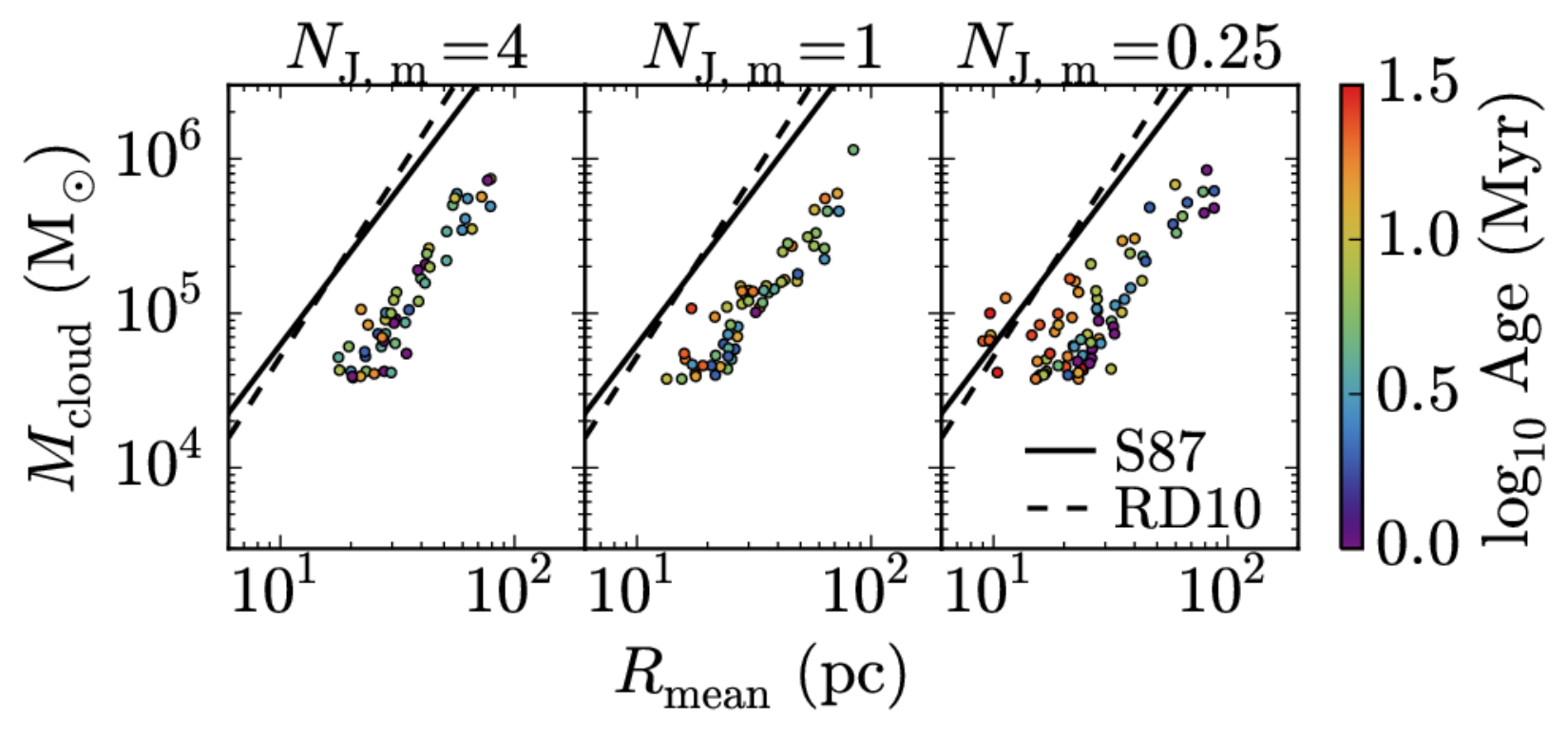}}
\caption{Mass-size relation for clouds in the snapshot at $100 \, \rm{Myr}$, from the ref model with three different pressure floors: $N_{\rm{J, \, m}} = 4$, $1$ and $0.25$ \textit{(left, centre and right panels, respectively)}. The colour scale indicates the cloud age. We also show the observed relations from \citet{solomon87} \textit{(S87; solid cuves)} and \citet{romanduval10} \textit{(RD10; dashed curves}). In the run with $N_{\rm{J, \, m}} = 0.25$, low-mass, poorly resolved clouds are more compact, but this run no longer reproduces the observed slope of this relation.} 
\label{massSizeApp}
\end{figure}

\begin{figure}
\centering
\mbox{
	\includegraphics[width=84mm]{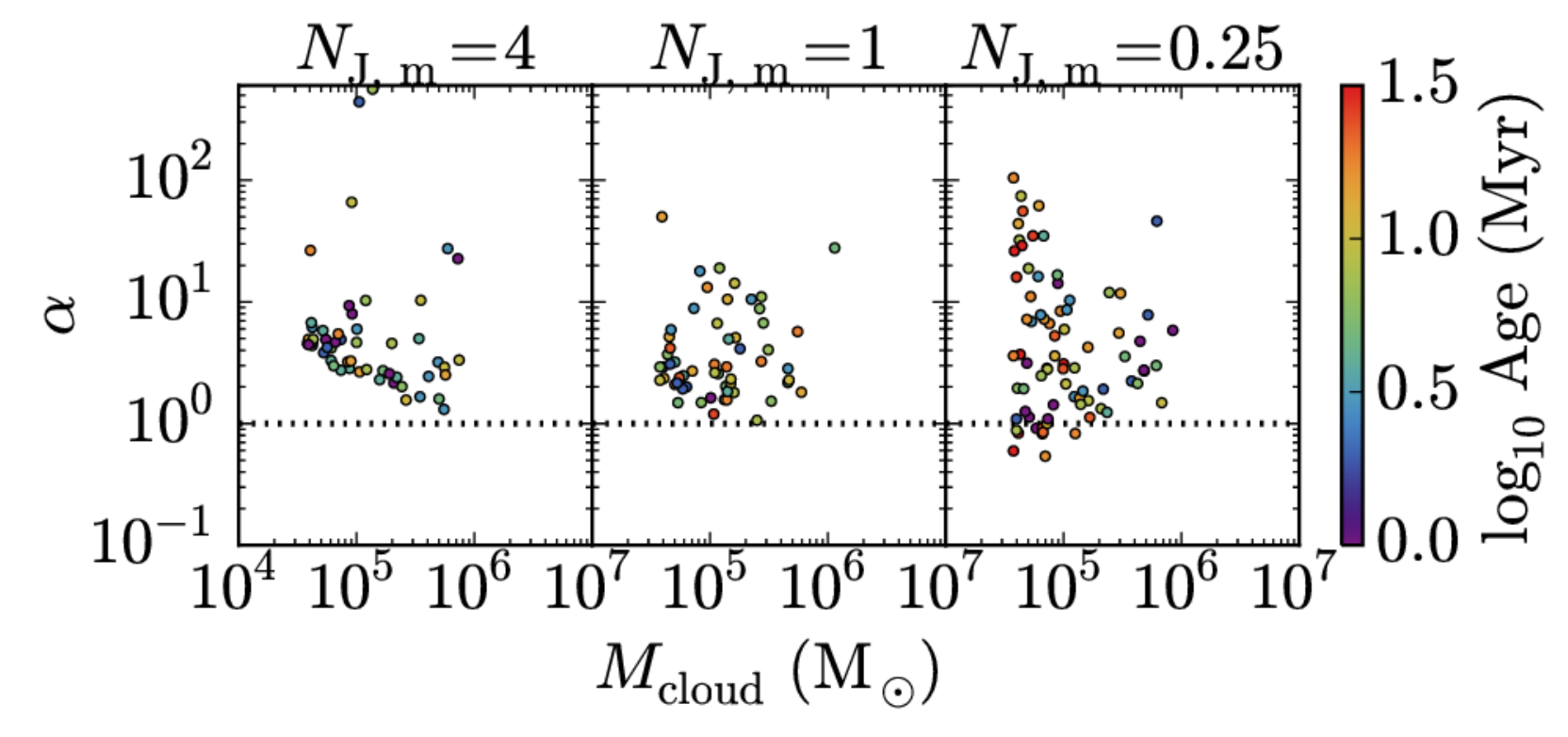}}
\caption{Cloud virial parameter, $\alpha$, versus mass, for the ref model with three different pressure floors: $N_{\rm{J, \, m}} = 4$, $1$ and $0.25$ \textit{(left, centre and right panels, respectively)}. The colour scale indicates the cloud age, and the horizontal dotted lines show a value of $\alpha = 1$, which corresponds to virial equilibrium. The lower envelope of $\alpha$ decreases as we lower the pressure floor, with some low-mass clouds in the $N_{\rm{J, \, m}} = 0.25$ run becoming strongly gravitationally bound. However, it is likely that the most compact clouds in the $N_{\rm{J, \, m}} = 0.25$ run are affected by spurious fragmentation and collapse that may arise when we do not fully resolve the Jeans scale (see text).} 
\label{virialApp}
\end{figure}

Fig.~\ref{cloudFnApp} shows the cloud mass functions, which are similar for the three pressure floors. In Fig.~\ref{cloudAgeApp} we show the cloud age as a function of mass, where the cloud age is defined by the time when half of the particles that were originally in the cloud in the current snapshot ($100 \, \rm{Myr}$) were in a progenitor of that cloud, i.e. our fiducial age definition (see section~\ref{evolution_section}). The left, centre and right panels show the results for a pressure floor with $N_{\rm{J, \, m}} = 4$, $1$ and $0.25$, respectively. In the run with the lowest pressure floor (right panel), we find clouds extending up to higher ages (up to $50 \, \rm{Myr}$). However, the longest-lived clouds ($\ga 30 \, \rm{Myr}$) in this simulation are only found in the lowest-mass, most poorly resolved clouds ($M_{\rm{cloud}} \la 10^{5} \, \rm{M}_{\odot}$), whereas we see no trends between cloud age and mass with higher pressure floors. 

Fig.~\ref{massSizeApp} shows the mass-size relation for clouds with the different pressure floors. The colour scale indicates the cloud age, and the black solid and dashed curves show the observed relations of \citet{solomon87} and \citet{romanduval10}, respectively. The run with the lowest pressure floor (right panel) shows more scatter in this relation, with clouds of the same mass generally being more compact than in the $N_{\rm{J, \, m}} = 4$ run. Additionally, the most compact clouds, i.e. those that lie to the left of the relation, tend to be older ($\ga 30 \, \rm{Myr}$). This is understandable, as more compact objects will be more strongly gravitationally bound, and thus can survive for longer. This explains the longer cloud ages that we saw in Fig.~\ref{cloudAgeApp} in the $N_{\rm{J, \, m}} = 0.25$ run. 

However, it is only the lowest-mass, most poorly resolved clouds that are more compact in the $N_{\rm{J, \, m}} = 0.25$ run than in the $N_{\rm{J, \, m}} = 4$ run. The highest-mass clouds ($\ga 3 \times 10^{5} \, \rm{M}_{\odot}$) appear unaffected by the pressure floor. This also means that the mass-size relation in the $N_{\rm{J, \, m}} = 0.25$ run is flatter than is observed. While the clouds in the $N_{\rm{J, \, m}} = 4$ run lie further from the observed relations, they do recover the same slope, and the difference in normalisation can be explained by the differences in cloud definition (as seen in Figs.~\ref{massSizeFig} and \ref{massSizeThreshFig}). Furthermore, in the hiRes run in Appendix~\ref{res_appendix}, the mass-size relation did not change when we increased the resolution, and we did not see low-mass clouds become more compact with longer ages. This suggests that the more compact clouds that we find when we lower the pressure floor may be an artifact of spurious fragmentation and collapse that may arise when we do not fully resolve the Jeans scale. 

\begin{figure}
\centering
\mbox{
	\includegraphics[width=84mm]{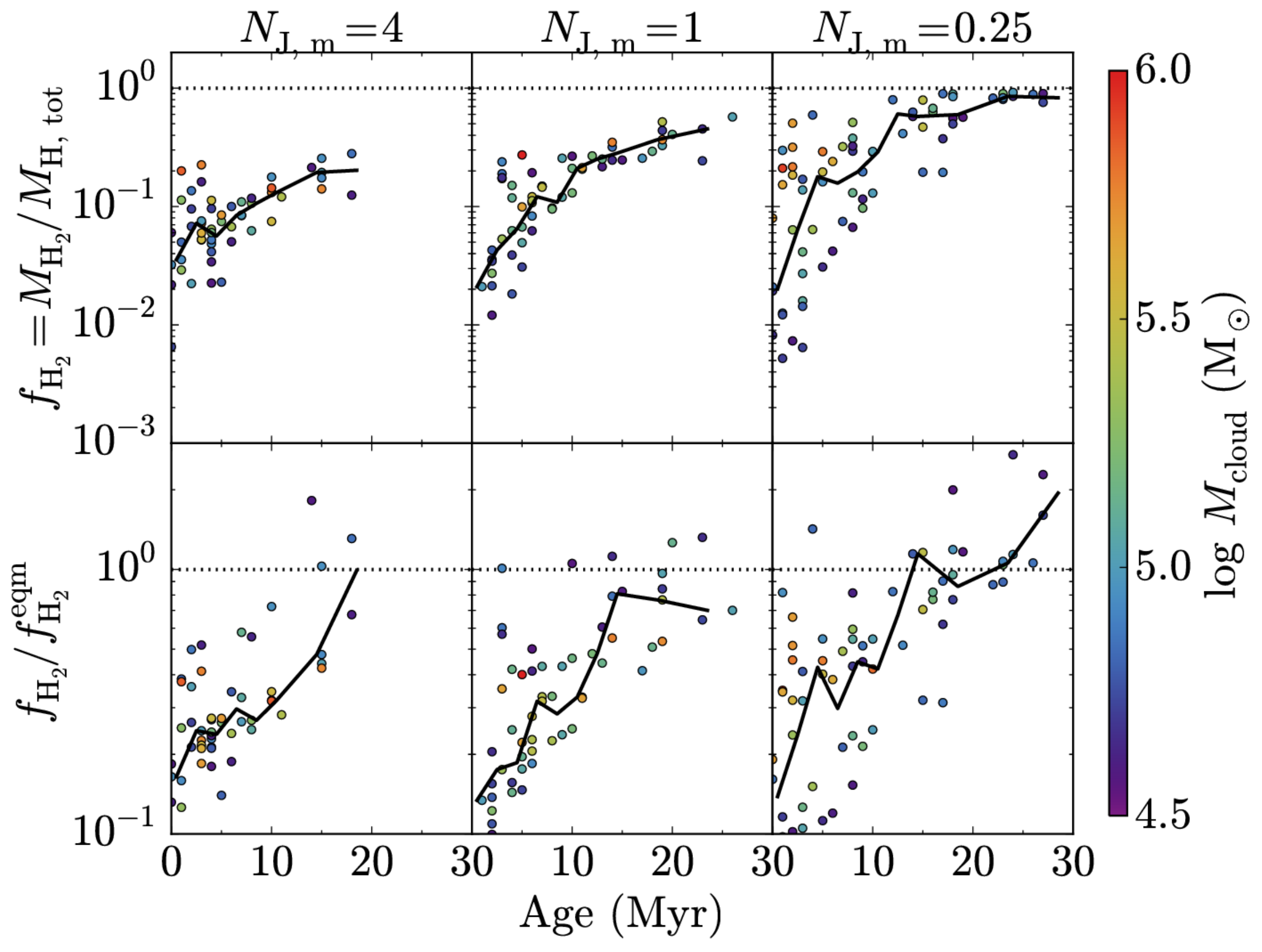}}
\caption{H$_{2}$ fraction, $f_{\rm{H_{2}}}$ \textit{(top row)}, and the ratio of $f_{\rm{H_{2}}}$ to the H$_{2}$ fraction in equilibrium, $f_{\rm{H_{2}}}^{\rm{eqm}}$, plotted against cloud age, for clouds in the snapshot at $100 \, \rm{Myr}$ from the ref model with three different pressure floors: $N_{\rm{J, \, m}} = 4$, $1$ and $0.25$ \textit{(left, centre and right columns, respectively)}. The colour scale indicates the cloud mass, the solid curves indicate the median relation in bins of age, and the horizontal dotted lines indicate a value of unity. As we lower the pressure floor, clouds extend to higher ages, but continue to follow similar median relations of $f_{\rm{H_{2}}}$ and $f_{\rm{H_{2}}} / f_{\rm{H_{2}}}^{\rm{eqm}}$ with age, albeit with more scatter.} 
\label{H2App}
\end{figure}

Fig.~\ref{virialApp} shows the virial parameter, $\alpha$, as a function of cloud mass, for the three pressure floors (in different panels). The horizontal dotted line indicates a value of $\alpha = 1$, which corresponds to virial equilibrium. In the left panel, for our fiducial pressure floor, $N_{\rm{J, \, m}} = 4$, we see a lower envelope in $\alpha$ that declines with cloud mass as $\alpha \propto M_{\rm{cloud}}^{-2/3}$. This scaling is due to the pressure floor, as discussed in section~\ref{scaling_relations_section}, and means that, in the $N_{\rm{J, \, m}} = 4$ run, the pressure floor prevents the lowest-mass clouds from becoming strongly gravitationally bound ($\alpha < 1$). 

As we lower the pressure floor, the low-mass clouds extend to lower values of $\alpha$, and can become strongly gravitationally bound ($\alpha < 1$) in the $N_{\rm{J, \, m}} = 0.25$ run (although there are still clouds with high values of $\alpha$ as well). This is consistent with the more compact low-mass clouds that we saw in this run in Fig.~\ref{massSizeApp}. However, as noted above, the more compact low-mass clouds that we find when we lower the pressure floor are likely to be an artifact of spurious fragmentation and collapse. Furthermore, there were no clouds with $\alpha < 1$ in the hiRes run in Appendix~\ref{res_appendix}. 

We now need to consider whether varying the pressure floor affects the evolution of molecular abundances, the CO emission or the $X_{\rm{CO}}$ factors of individual clouds. Fig.~\ref{H2App} shows the H$_{2}$ fraction, $f_{\rm{H_{2}}}$ (top row), and the ratio of $f_{\rm{H_{2}}}$ to the equilibrium H$_{2}$ fraction, $f_{\rm{H_{2}}}^{\rm{eqm}}$ (bottom row), plotted against cloud age, for the three pressure floors (in different columns). As we lower the pressure floor (left to right), clouds extend to higher ages, as we saw in Fig.~\ref{cloudAgeApp}. However, the median values of $f_{\rm{H_{2}}}$ and $f_{\rm{H_{2}}} / f_{\rm{H_{2}}}^{\rm{eqm}}$ at fixed cloud age are similar, regardless of the pressure floor. Thus, by lowering the pressure floor, we simply extend the same median relations of $f_{\rm{H_{2}}}$ and $f_{\rm{H_{2}}} / f_{\rm{H_{2}}}^{\rm{eqm}}$ versus cloud age to higher ages, although the scatter also increases. In particular, in all three runs it takes $\approx 10 - 15 \, \rm{Myr}$ for the median $f_{\rm{H_{2}}}$ to reach within a factor two of its equilibrium value. 

When we lower the pressure floor, the CO fractions (not shown) also follow similar median relations with cloud age as for our fiducial pressure floor, albeit with more scatter. 

\begin{figure}
\centering
\mbox{
	\includegraphics[width=84mm]{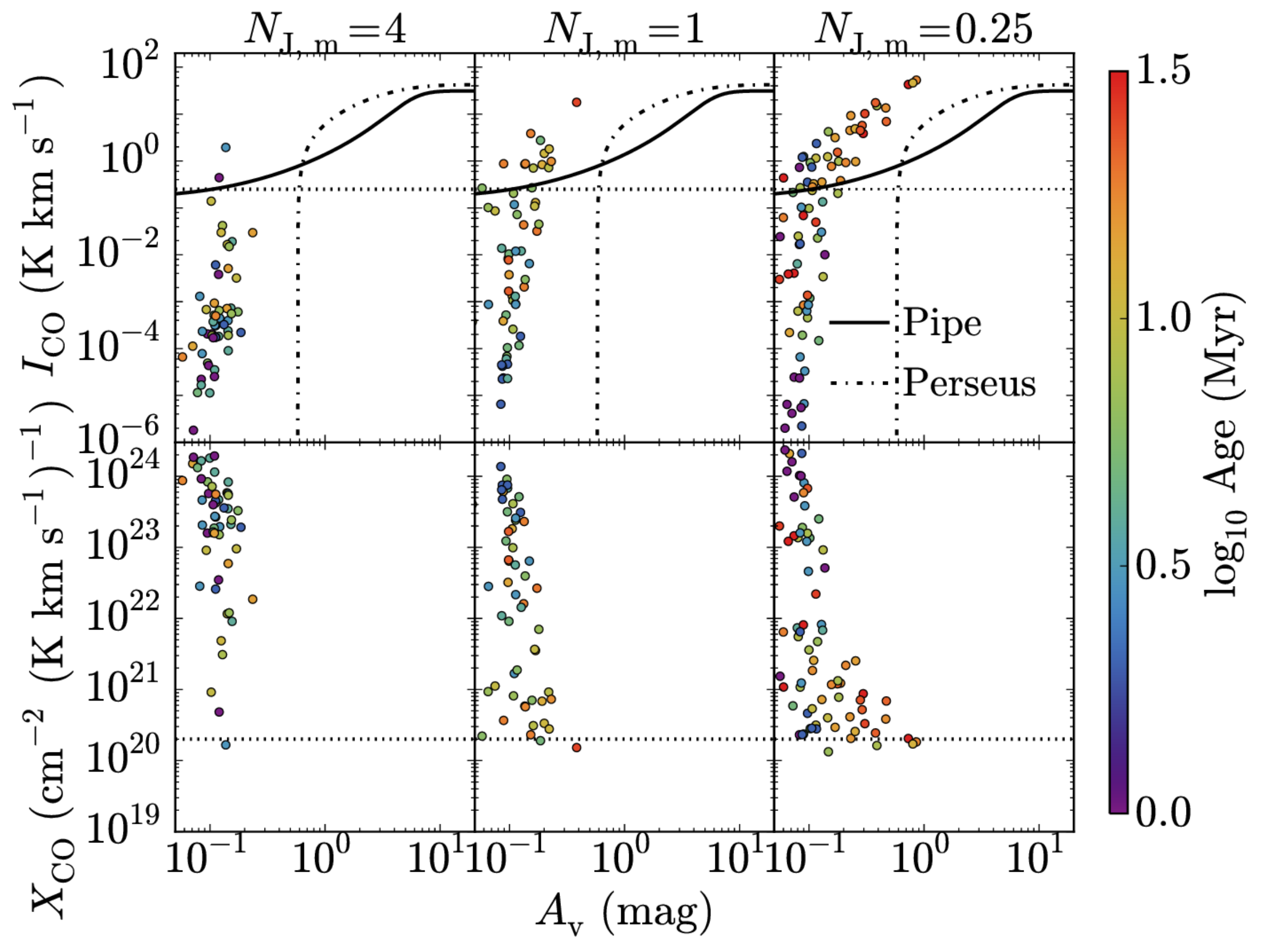}}
\caption{Mean CO intensity, $I_{\rm{CO}}$ \textit{(top row)}, and $X_{\rm{CO}}$ factor \textit{(bottom row)}, plotted against mean dust extinction, $A_{\rm{v}}$. The colour scale indicates the cloud age. In the top row, we also show the $I_{\rm{CO}} - A_{\rm{v}}$ relations observed in the Pipe nebula (\citealt{lombardi06}; solid curves) and the Perseus cloud (\citealt{pineda08}; dot-dashed curves). The horizontal dotted lines in the top row indicate a value of $I_{\rm{CO}} = 0.25 \, \rm{K} \, \rm{km} \, \rm{s}^{-1}$, which corresponds to the $3\sigma$ intensity threshold for the Small Magellanic Cloud in the observations of \citet{leroy11}. In the bottom row, the horizontal dotted lines indicate the typical value measured in the Milky Way, $X_{\rm{CO}} = 2 \times 10^{20} \, \rm{cm}^{-2} \, (\rm{K} \, \rm{km} \, \rm{s}^{-1})^{-1}$ \citep[e.g.][]{bolatto13}. As we lower the pressure floor, clouds extend to higher $A_{\rm{v}}$, as they become more compact. However, for the values of $A_{\rm{v}}$ where the different runs overlap, the $I_{\rm{CO}} - A_{\rm{v}}$ relations are consistent, albeit with more scatter, particularly in $I_{\rm{CO}}$, and with a shallower cut-off in $I_{\rm{CO}}$ towards low $A_{\rm{v}}$.} 
\label{IcoApp}
\end{figure}

Fig.~\ref{IcoApp} shows the mean CO intensity, $I_{\rm{CO}}$ (top row), and the mean $X_{\rm{CO}}$ factor (bottom row) of each cloud for the three pressure floors as a function of mean dust extinction, $A_{\rm{v}}$. As we lower the pressure floor, clouds extend to higher $A_{\rm{v}}$, as they become more compact. However, they continue to follow a similar trend of $I_{\rm{CO}}$ with $A_{\rm{v}}$, although there is more scatter in this relation with a lower pressure floor. In particular, the threshold $A_{\rm{v}}$ below which $I_{\rm{CO}}$ cuts off does not appear to be strongly affected by the pressure floor, i.e. the $I_{\rm{CO}} - A_{\rm{v}}$ relation does not move horizontally in this plot. However, the increased scatter as we reduce the pressure floor means that we find some low-$A_{\rm{v}}$ clouds ($A_{\rm{v}} \approx 0.06$) with higher CO intensities ($I_{\rm{CO}} \approx 0.3 \, \rm{K} \, \rm{km} \, \rm{s}^{-1}$), so the cut-off in $I_{\rm{CO}}$ at low $A_{\rm{v}}$ is less steep than in the $N_{\rm{J, \, m}} = 4$ run. 

The $X_{\rm{CO}}$ factors all show similar scatter of four orders of magnitude, regardless of pressure floor. In the $N_{\rm{J, \, m}} = 0.25$ run, the high-$A_{\rm{v}}$ clouds ($\ga 0.3$) show less scatter in $X_{\rm{CO}}$ (one order of magnitude) and suggest a trend of decreasing $X_{\rm{CO}}$ with increasing $A_{\rm{v}}$. This trend cannot be verified in the $N_{\rm{J, \, m}} = 4$ run, because it does not include these high-$A_{\rm{v}}$ clouds. However, for the range of $A_{\rm{v}}$ where these three runs overlap, they show consistent $X_{\rm{CO}} - A_{\rm{v}}$ relations. 

To conclude, lowering the pressure floor results in low mass clouds ($\la 3 \times 10^{5} \, \rm{M}_{\odot}$) becoming more compact, more strongly gravitationally bound, and able to survive for longer. However, this is inconsistent with what we found when we increased the resolution in Appendix~\ref{res_appendix}, so it is likely that this is an artifact of spurious fragmentation and collapse that may arise when we do not fully resolve the Jeans scale.  In particular, the run with the lowest pressure floor ($N_{\rm{J, \, m}} = 0.25$) no longer reproduces the observed slope of the cloud mass-size relation \citep{solomon87, romanduval10}, as it becomes flatter when the pressure floor is lowered, as the low-mass clouds become too compact. This flattening of the mass-size relation may be a characteristic signature that we can use in the future to determine whether a given simulation suffers from spurious fragmentation due to an unresolved Jeans scale. 

Furthermore, we find similar median trends of $f_{\rm{H_{2}}}$ and $f_{\rm{H_{2}}} / f_{\rm{H_{2}}}^{\rm{eqm}}$ versus cloud age, and $I_{\rm{CO}}$ and $X_{\rm{CO}}$ versus $A_{\rm{v}}$, regardless of pressure floor. Lowering the pressure floor simply extends these relations to higher ages and $A_{\rm{v}}$, albeit with more scatter. 

\label{lastpage}

\end{document}